\documentclass[trackchanges]{aastex701}

\usepackage{subfig}
\usepackage{tabularx}      
\usepackage{booktabs}     
\usepackage{threeparttable} 
\usepackage{amsmath}
\usepackage{multirow}   
\usepackage{booktabs}   
\usepackage{graphicx}
\usepackage{subfig}

\begin{document}

\title{Slitless Spectroscopy Source Detection Using YOLO Deep Neural Network}

\author[0009-0007-2538-5102]{Xiaohan Chen}
\affiliation{National Astronomical Observatories, Chinese Academy of Sciences, Beijing 100101,  People's Republic of China}
\affiliation{School of Physics and Astronomy, China West Normal University, Nanchong 637009, People's Republic of China}
\email{xhchen@bao.ac.cn}

\author[]{Man I Lam\textsuperscript{*,$\dagger$}}
\affiliation{National Astronomical Observatories, Chinese Academy of Sciences, Beijing 100101,  People's Republic of China} 
\email[show]{mlam@nao.cas.cn (Lead Corresponding Author)}

\author[0000-0002-8030-4383]{Yingying Zhou}
\affiliation{Key Laboratory of Space Astronomy and Technology, National Astronomical Observatories, Chinese Academy of Sciences, Beijing 100101, P.R. China}
\email{zhouyingying@bao.ac.cn}

\author[0009-0007-5610-6495]{Hongrui Gu}
\affiliation{National Astronomical Observatories, Chinese Academy of Sciences, Beijing 100101,  People's Republic of China}
\affiliation{School of Astronomy and Space Science, University of Chinese Academy of Sciences, Beijing 100049, People's Republic of China}
\email{guhr@bao.ac.cn}

\author[0009-0003-7031-9498]{Jinzhi Lai}
\affiliation{National Astronomical Observatories, Chinese Academy of Sciences, Beijing 100101,  People's Republic of China}
\affiliation{School of Astronomy and Space Science, University of Chinese Academy of Sciences, Beijing 100049, People's Republic of China}
\email{laijz@bao.ac.cn}

\author[0000-0002-6790-2397]{Zhou Fan\textsuperscript{*}}
\affiliation{National Astronomical Observatories, Chinese Academy of Sciences, Beijing 100101,  People's Republic of China}
\affiliation{School of Astronomy and Space Science, University of Chinese Academy of Sciences, Beijing 100049, People's Republic of China}
\email[show]{zfan@nao.cas.cn}

\author[0000-0002-4953-1545]{Jing Li\textsuperscript{*}}
\affiliation{School of Physics and Astronomy, China West Normal University, Nanchong 637009, People's Republic of China}
\email[show]{lijing@bao.ac.cn}

\author{Xin Zhang}
\affiliation{National Astronomical Observatories, Chinese Academy of Sciences, Beijing 100101,  People's Republic of China} 
\email{zhangx@nao.cas.cn}

\author[0000-0003-3347-7596]{Hao Tian}
\affiliation{National Astronomical Observatories, Chinese Academy of Sciences, Beijing 100101,  People's Republic of China} 
\email{tianhao@nao.cas.cn}


\begin{abstract}

     Slitless spectroscopy eliminates the need for slits, allowing light to pass directly through a prism or grism to generate a spectral dispersion image that encompasses all celestial objects within a specified area. 
     This technique enables highly efficient spectral acquisition. 
     However, when processing CSST slitless spectroscopy data, the unique design of its focal plane introduces a challenge: photometric and slitless spectroscopic images do not have a one-to-one correspondence. 
     As a result, it becomes essential to first identify and count the sources in the slitless spectroscopic images before extracting spectra.
     To address this challenge, we employed the You Only Look Once (YOLO) object detection algorithm to develop a model for detecting targets in slitless spectroscopy images. 
     This model was trained on 1,560 simulated CSST slitless spectroscopic images. 
     These simulations were generated from the CSST Cycle 6 and Cycle 9 main survey data products, representing the Galactic and nearby galaxy regions and the high galactic latitude regions, respectively.
     On the validation set, the model achieved a precision of 88.6\% and recall of 90.4\% for spectral lines, and 87.0\% and 80.8\% for zeroth-order images. 
     In testing, it maintained a detection rate $>$80\% for targets brighter than 21 mag (medium-density regions) and 20 mag (low-density regions) in the Galactic and nearby galaxies regions, 
     and $>$70\% for targets brighter than 18 mag in high galactic latitude regions.

\end{abstract}

\keywords{Astronomical object identification (87) --- Deep learning (1938) --- Spectroscopy (1558)}



\section{Introduction} \label{sec:intro}

Slitless spectroscopy, produced by a prism or grism without the use of a slit, is an efficient survey tool because it records the spectrum of all objects within the field of view. 
It is also more straightforward than traditional methods, as it eliminates the need for pre-imaging or slit alignment. 
Observations can be conducted simply by positioning the target within the field of view, greatly simplifying the observational process and reducing the need for precise telescope pointing. 
Additionally, traditional slit spectroscopy can result in light loss due to parts of the object not passing through the slit, which reduces the signal-to-noise ratio (SNR) of the spectrum.  
In contrast, slitless spectroscopy can minimize light loss by allowing more photon to reach the detector, resulting in more accurate flux calibration \citep{Willott_2022}.

Due to its operational simplicity and high efficiency in spectral acquisition, slitless spectroscopy has become a widely used technique in both ground-based and space-based observational astronomy. 
In ground-based surveys, \cite{1980MNRAS.193..415C} employed objective prism plates to perform low-resolution spectroscopic measurements of quasars, a technique that has been crucial for studying the statistical properties, redshift distribution, and spectral characteristics of quasars. 
This method allows researchers to efficiently obtain spectral data for large numbers of quasars without the need for massive telescopes. 
The efficiency of this technique paved the way for several major quasar surveys. 
Earlier, wide-field, objective prism-based slitless spectroscopic surveys, such as the Hamburg Quasar Survey (HQS) \citep{1988ASPC....2..143E, 1995A&AS..111..195H, 1998A&AS..128..507E} 
and the Hamburg/ESO Survey (HES) \citep{1994IAUS..161..723W, 1996A&AS..115..227W, 1996A&AS..115..235R, 1997Msngr..88...14R}, discovered a large number of bright quasars. 
Later, \cite{Worseck_2008} employed a deeper, targeted, CCD-based slitless spectroscopic survey — the `Quasars near Quasars' (QNQ) survey — to discover a significant number of fainter quasars at intermediate and high redshifts.

Slitless spectroscopy is particularly efficient in space-based surveys due to the reduced background noise in space \citep{K_mmel_2009}. 
Space telescopes such as the Hubble Space Telescope (HST) \citep{22b385bc-245f-39bd-a48e-ccaaab6d04eb}, the James Webb Space Telescope (JWST) \citep{2016SPIE.9904E..0EG, Willott_2022, 2023PASP..135i8001D}, and Euclid \citep{2025A&A...697A...1E} are all equipped with slitless spectroscopic modules. 
Leveraging the near-infrared slitless spectrographs (G102 and G141) on its Wide Field Camera 3 (WFC3), 
the HST observes the central regions of galaxy clusters to obtain spectral data for high-redshift galaxies. 
These observations advance studies in galaxy formation and evolution, star formation history, chemical composition, and the early universe.
For example, observations of the Spiderweb protocluster core using the G141 grism on HST WFC3 revealed the important role of active galactic nuclei (AGN) in influencing the evolution of quiescent galaxies \citep{2024ApJ...977...58N}. 
Moreover, by using HST WFC3/IR imaging and spectroscopic data, researchers successfully identified galaxies at redshifts as high as z$\approx$11.09, providing key constraints on galaxy assembly processes in the early universe \citep{2016ApJ...819..129O}. 
Another study combined deep slitless spectroscopic and imaging data from multiple HST surveys to investigate the star formation history and metallicity evolution of massive galaxies that had already quenched when the universe was only $\sim$2 Gyr \citep{2019ApJ...877..141M}.
Similarly, the Wide Field Slitless Spectroscopy (WFSS) mode of the Near Infrared Imager and Slitless Spectrograph (NIRISS), together with the Near Infrared Camera (NIRCam) on JWST, enables slitless spectroscopic observations across the 0.8-2.2 $\mu$m and 2.4-5.0 $\mu$m wavelength ranges, respectively.
With its powerful capability, JWST/NIRCam can observe high-redshift galaxies to study their metallicity, ionizing photon production efficiency, physical conditions of the interstellar medium, and the evolutionary characteristics of galaxy luminosity functions in the early universe, thereby revealing the formation and evolution mechanisms of galaxies during the Epoch of Reionization \citep{2023ApJ...953...53S}.
And JWST/NIRISS achieves a depth and spatial resolution higher than ever before, 
which is particularly crucial for resolving individual stars or galaxies in crowded cluster environments \citep{2022ApJ...940L..52B}.
Furthermore, JWST/NIRISS is capable of providing high-quality data for research in the field of exoplanets, assisting scientists in interpreting the atmospheric composition and physical conditions of exoplanets, as well as inferring their origins and evolutionary pathways \citep{2024ApJ...974L..10P}.
Following in the footsteps of HST and JWST, Euclid is also set to contribute significantly to our understanding of the universe. 
Using data from the Euclid Quick Data Release (Q1) \citep{2025arXiv250315302E}, the Near-Infrared Spectrometer and Photometer (NISP) aboard the Euclid satellite has enabled significant findings across numerous fields such as nearby galaxies, galaxy morphology, and strong gravitational lensing. 
A notable demonstration of NISP's capability is the discovery of a high-redshift quasar ($z = 5.404$), EUCL J181530.01+652054.0, using its slitless spectrograph (R$\gtrsim$450 across 950-2020 nm) \citep{2025A&A...697A...3E, article, 2025MNRAS.tmp.1224B}.

Although slitless spectroscopy significantly enhances observational efficiency by eliminating the need for a slit, 
it also introduces contamination effects that complicate spectrum extraction.
Currently, there are several specialized data processing pipelines designed for slitless spectroscopy. 
For example, GRIZLI \citep{2019ascl.soft05001B} focuses on comprehensive modeling and fitting of slitless spectroscopic observations; 
LINEAR \citep{2018PASP..130c4501R,2018wfc..rept...13R} reconstructs high SNR one-dimensional spectra by solving linear equations while effectively handling contamination issues; 
and aXe \citep{K_mmel_2009} is a versatile slitless spectroscopy data processing tool specifically developed for the HST, supporting multiple instruments and featuring advanced spectral extraction and correction capabilities.
Contamination is a common challenge in slitless spectroscopy. 
In cases with a high source density, contamination effects become more frequent and severe, making spectral extraction more difficult. 
In such scenarios, the scientific value of the data can be compromised. 
Therefore, it is crucial to detect and count the sources in slitless spectroscopic images, excluding data with excessively high source densities (i.e., data with significant contamination) before proceeding with spectral extraction.

\begin{figure*}[ht!]
\gridline{
    \fig{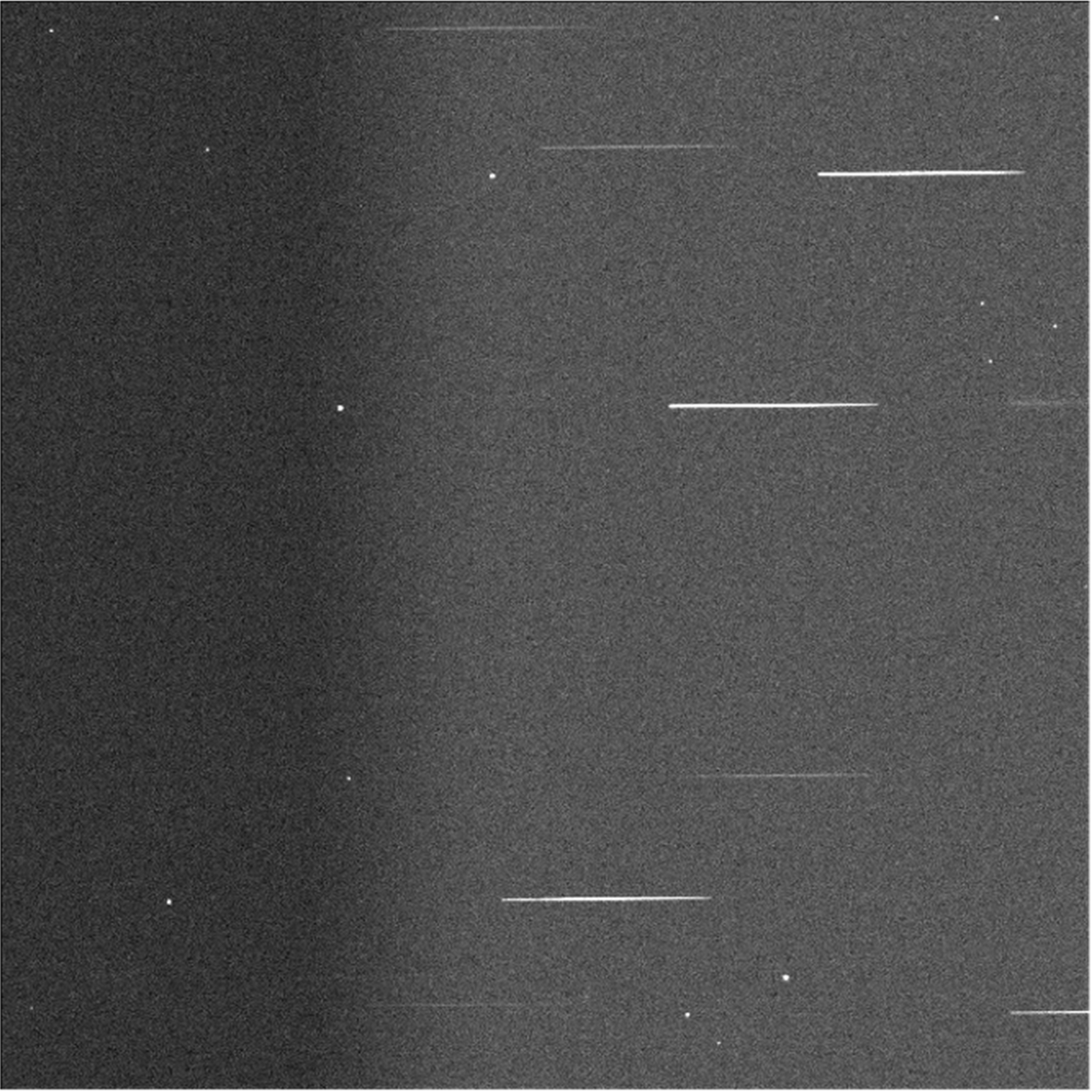}{0.32\textwidth}{(a)}
    \fig{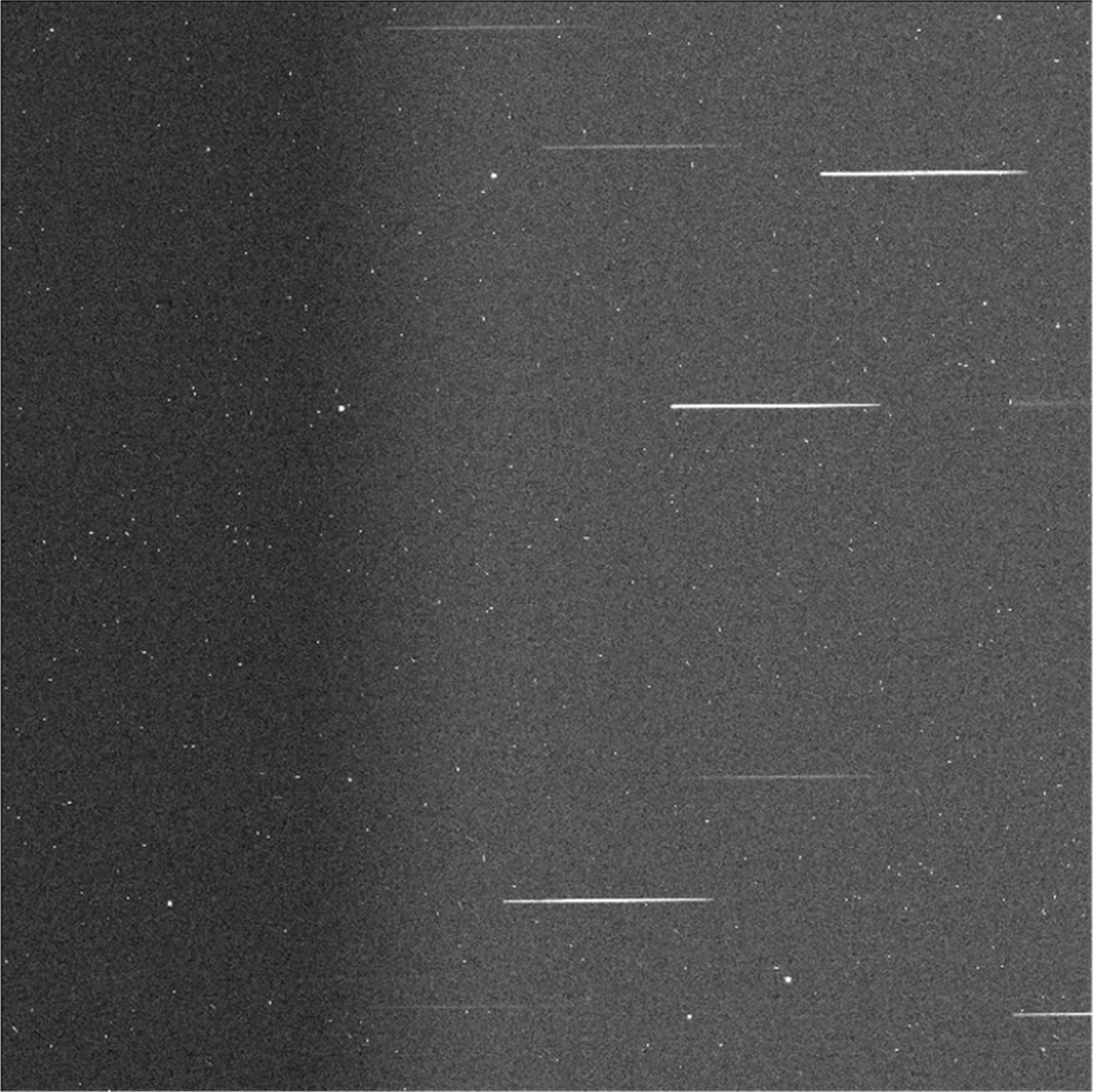}{0.32\textwidth}{(b)}
    \fig{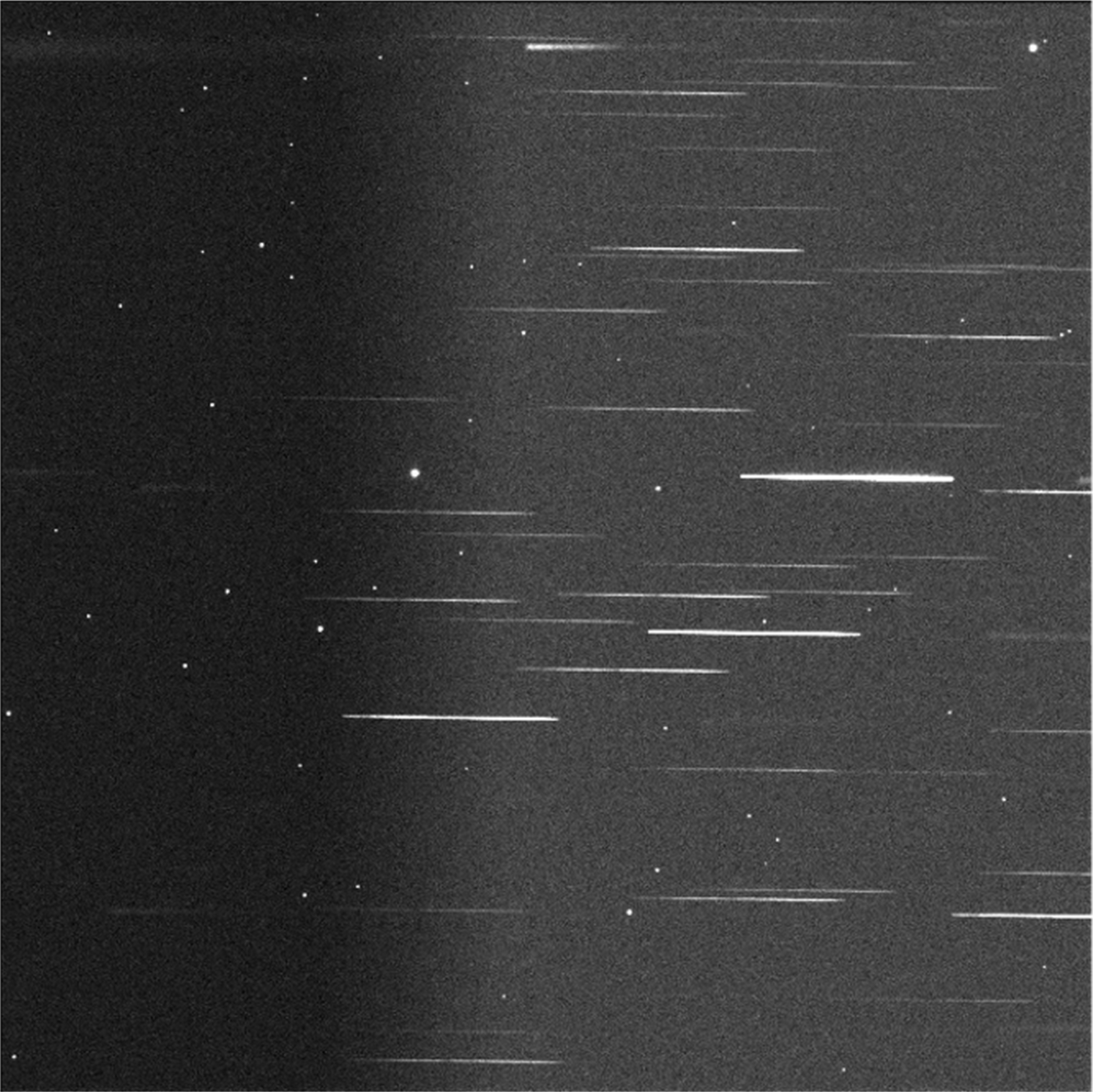}{0.32\textwidth}{(c)}
}
\gridline{
    \fig{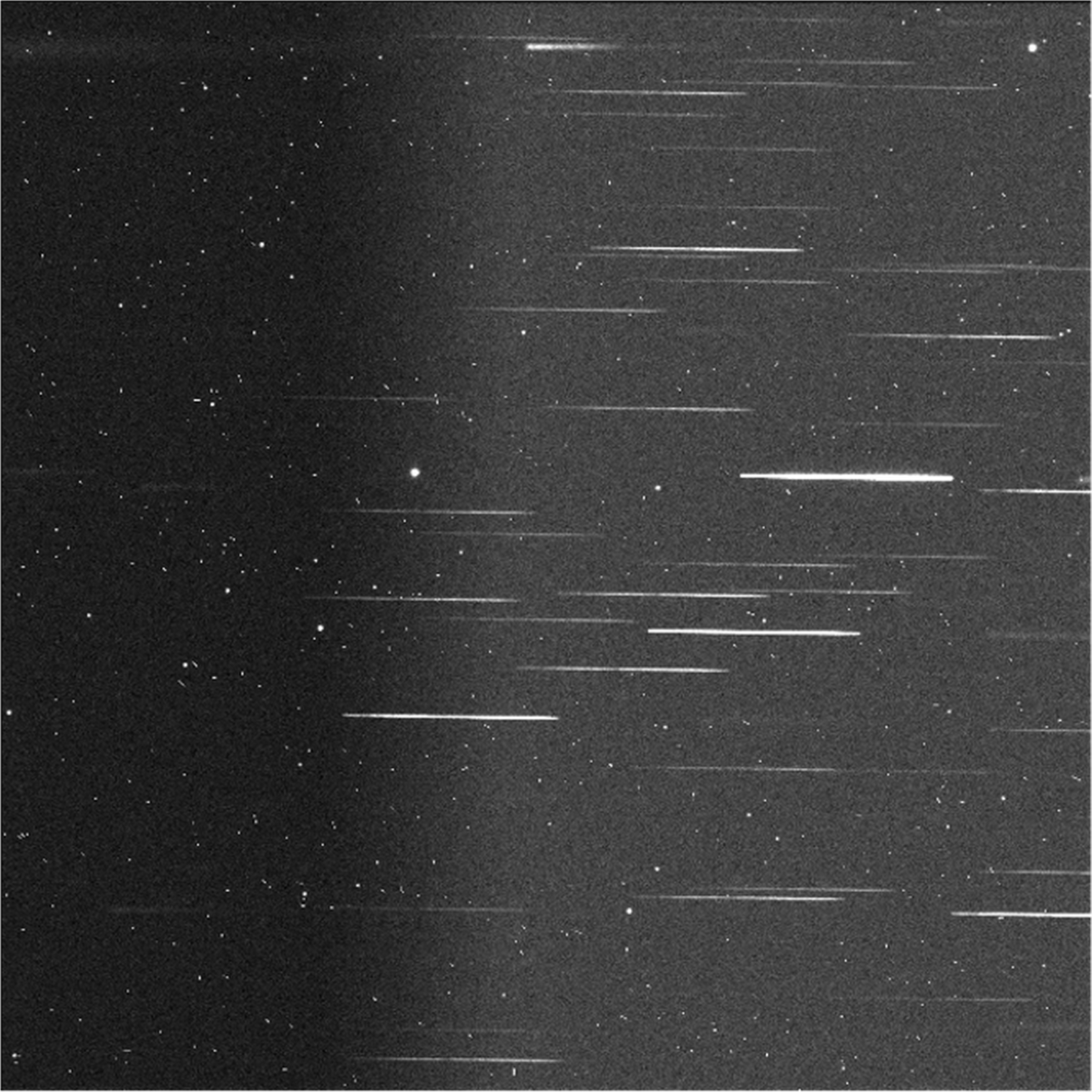}{0.32\textwidth}{(d)}
    \fig{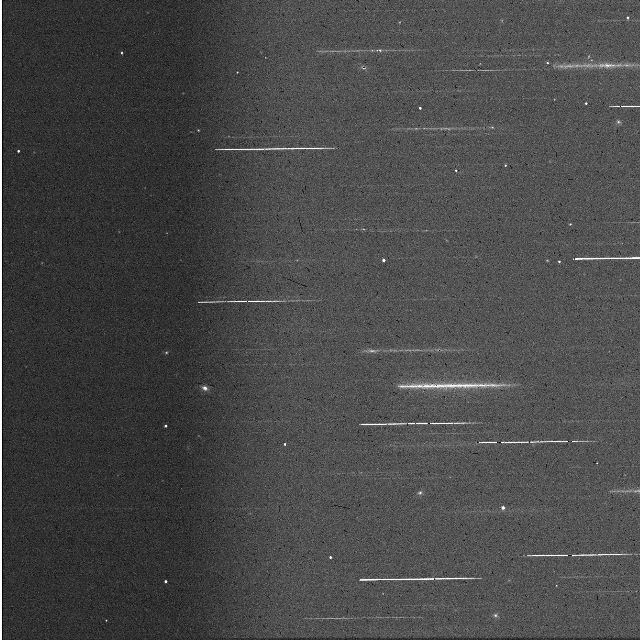}{0.32\textwidth}{(e)}
}
\caption{The simulated CSST slitless spectroscopy images, showing 1/16 of the full-size image:
(a) the low-density region without cosmic rays;
(b) the low-density region with cosmic rays;
(c) the medium-density region without cosmic rays;
(d) the medium-density region with cosmic rays;
(e) the high-Galactic latitude region.
The full slitless spectroscopy image has a size of approximately 9k $\times$ 9k pixels. For improved detection performance, each image was divided into 16 equal sections by resampling them to 640 $\times$ 640 pixels. 
Compared to the simulated images of the Milky Way and nearby galaxy regions, the high-Galactic latitude field contains a larger number of galaxies.}
\label{fig:simulation_images}
\end{figure*}

The China Space Station Telescope (CSST) is equipped with a slitless spectroscopy module and is set to conduct a large-scale slitless spectroscopic survey. 
This survey will cover approximately 17,500 square degrees of celestial area with a wavelength range of 255-1000 nm, aiming to observe celestial objects with a limiting magnitude of about 23 (5$\sigma$, point source) \citep{Zhan2021,10.1093/mnras/stae157}. 
The goal of these observations is to study the large-scale structure of the universe, dark energy, and the formation and evolution of galaxies. 
CSST's main focal plane consists of 30 detectors—18 dedicated to multiband imaging and 12 to slitless spectroscopy. 
Unlike HST, where detectors are switched between bands using filter and grating wheels, each detector on CSST is fixed to observe a specific band. 
This design eliminates the need for moving parts, reducing technical risks and improving reliability. 
However, due to this unique design, CSST cannot obtain photometric-slitless spectroscopy data pairs as HST does. 
Therefore, before processing the slitless spectroscopic data, it is essential to detect and count the sources in the slitless images, 
providing critical information for the subsequent spectral extraction. 

Object detection is a fundamental task in computer vision, focused on identifying objects within images and determining their locations. 
Traditional non-deep-learning approaches rely on hand-crafted features, such as HOG \citep{1467360}, SIFT \citep{2004Distinctive}, and Haar-like \citep{1038171}, and shallow models, such as the Supported Vector Machine (SVM) \citep{Cortes1995SupportVectorN}, AdaBoost \citep{FREUND1997119} and Deformable Part-based Model (DPM) \citep{5255236}. 
These methods suffer from poor robustness to complex scenes, multi-scale variations, occlusions, and lighting changes, resulting in evident performance bottlenecks. 
In contrast, deep-learning-based approaches demonstrate clear advantages. 
Deep-learning methods can be broadly categorized into tow classes: Two-Stage detectors and One-Stage detectors\citep{8825470}.
Two-Stage detectors first generate region proposals and then perform classification and regression on these proposals. 
The advent of deep learning, particularly the development of Convolutional Neural Networks (CNNs), has led to significant breakthroughs in this area \citep{8627998}. 
CNNs automatically learn high-level semantic features from images by combining convolutional layers, pooling layers, and fully connected layers, enabling both object classification and localization. 
One of the earliest frameworks to apply deep learning to object detection is R-CNN, which introduced CNNs into the field and laid the foundation for modern object-detection techniques \citep{6909475}. 
Subsequent developments include Fast R-CNN \citep{10.1109/ICCV.2015.169}, Faster R-CNN \citep{7485869}, and Mask R-CNN \citep{8237584}. 
Two-Stage methods achieve high accuracy and are well-suited to complex scenes; however, their multi-stage processing pipeline leads to slower inference speeds.
One-Stage detectors directly predict objects from images, streamlining the computational process and thereby improving detection efficiency. 
They are particularly well-suited for real-time applications. 
YOLO (You Only Look Once) is a prominent example of One-Stage Detectors. 
In 2016, \cite{7780460} introduced YOLO, treating object detection as a regression problem that directly predicts bounding boxes and class probabilities from image pixels. 
In addition to YOLO, notable One-Stage methods include SSD \citep{10.1007/978-3-319-46448-0_2}, RetinaNet \citep{8237586},  and M2Det \citep{10.1609/aaai.v33i01.33019259}. 
Nonetheless, these approaches generally exhibit lower accuracy on small and densely packed objects-a limitation that is particularly relevant to astronomical images, where targets are often small, closely spaced, and numerous.

Object detection in wide-field astronomical surveys presents unique challenges due to the vast areas of the sky that need to be rapidly scanned to capture a large number of celestial objects. 
The key to success in these surveys is the ability to quickly process massive amounts of image data. 
YOLO, with its ability to perform efficient and rapid object detection, is particularly well-suited for high-throughput data environments, meeting the real-time performance and resource utilization demands of such survey projects. 
As a result, the application of YOLO in astronomical object detection has been increasing. 
Researchers have successfully applied YOLO models to detect galaxies \citep{Xing_2023, 2023SPIE12511E..1CW}, galaxy clusters \citep{2025A&A...695A.246G}, solar radio bursts (SRBs) \citep{2020EGUGA..22.5109C}, and streaks caused by moving objects like satellites or meteorites \citep{2019amos.confE..89V}. 
Additionally, YOLO has shown strong performance across various observational datasets, including photometric, slitless spectroscopic and radio data \citep{9624090, Cornu_2024, Zhou_2025, 2024eas..conf.1350A}. 
Traditionally, object detection in astronomy often relied on supplementary data, such as photometric catalogs or photometric redshift information, to detect galaxy clusters. 
In contrast, YOLO's ability to work directly on images allows for faster, more convenient detection while avoiding the systematic errors often associated with traditional methods \citep{2025A&A...695A.246G}. 

This study focuses on identifying zero-order images and first-order spectral lines of stars and galaxies in slitless spectroscopic data using the YOLO algorithm, addressing the unique challenges posed by high target densities in astronomical images.
The structure of this paper is organized as follows. 
Section \ref{sec:data} provides an overview of the training and test samples used to construct our object detection model. 
The methods employed in this study are detailed in Section \ref{sec:methods}. 
Section \ref{sec:results} presents the detection results and a detailed discussion is presented in Section \ref{sec:discussion}. 
Finally, we summarize and conclude our study in Section \ref{sec:conclution}. 

\section{DATA} \label{sec:data}

\begin{figure*}
\gridline{\fig{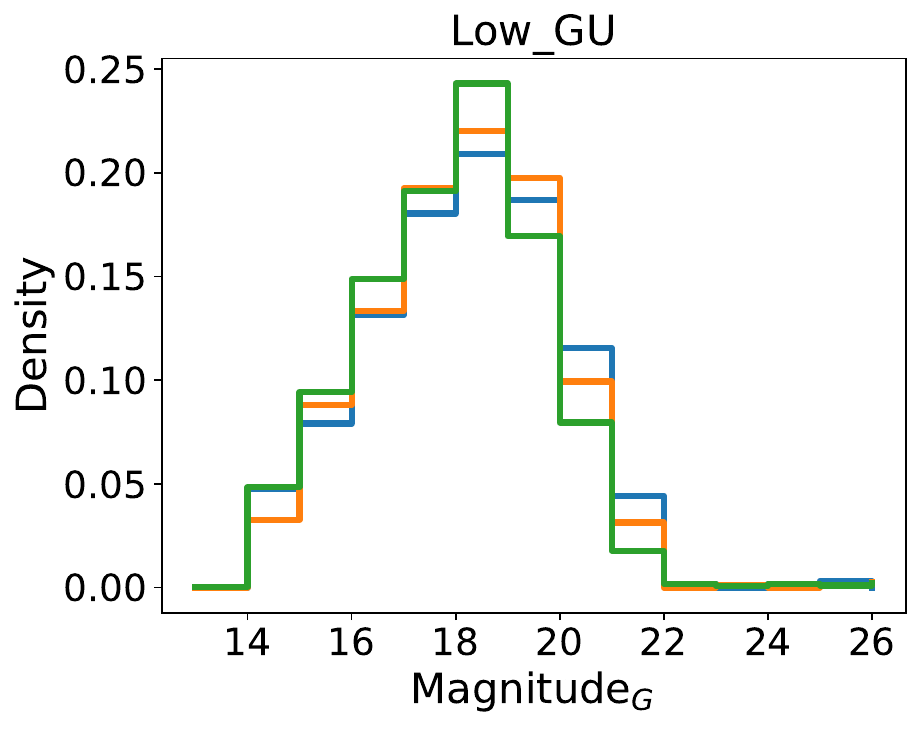}{0.32\textwidth}{(a)}
          \hspace{-5pt}
          \fig{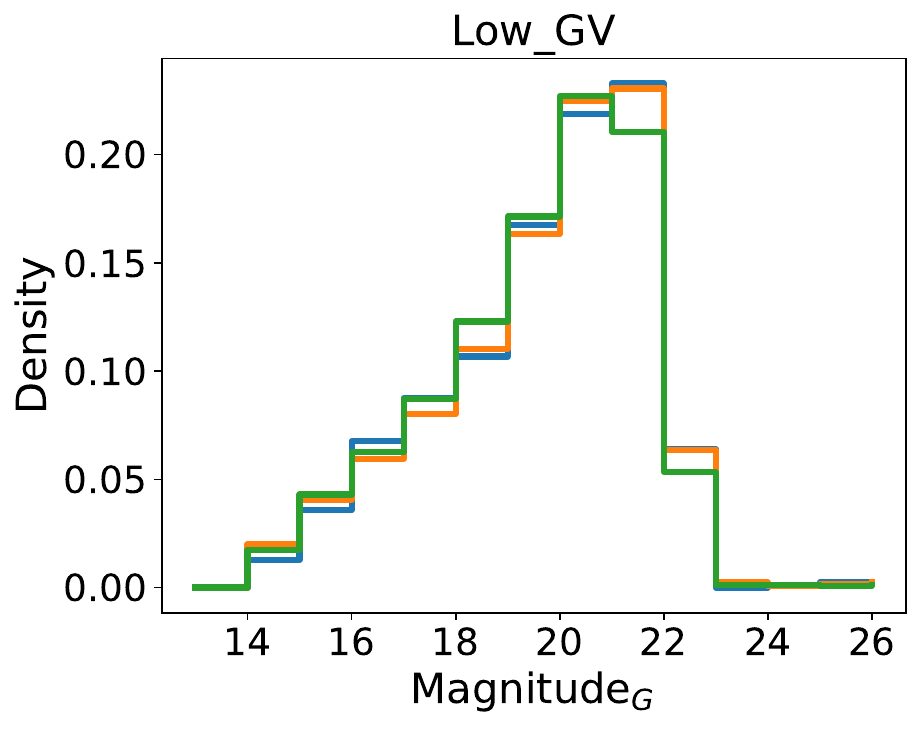}{0.32\textwidth}{(b)}
          \hspace{-5pt}
          \fig{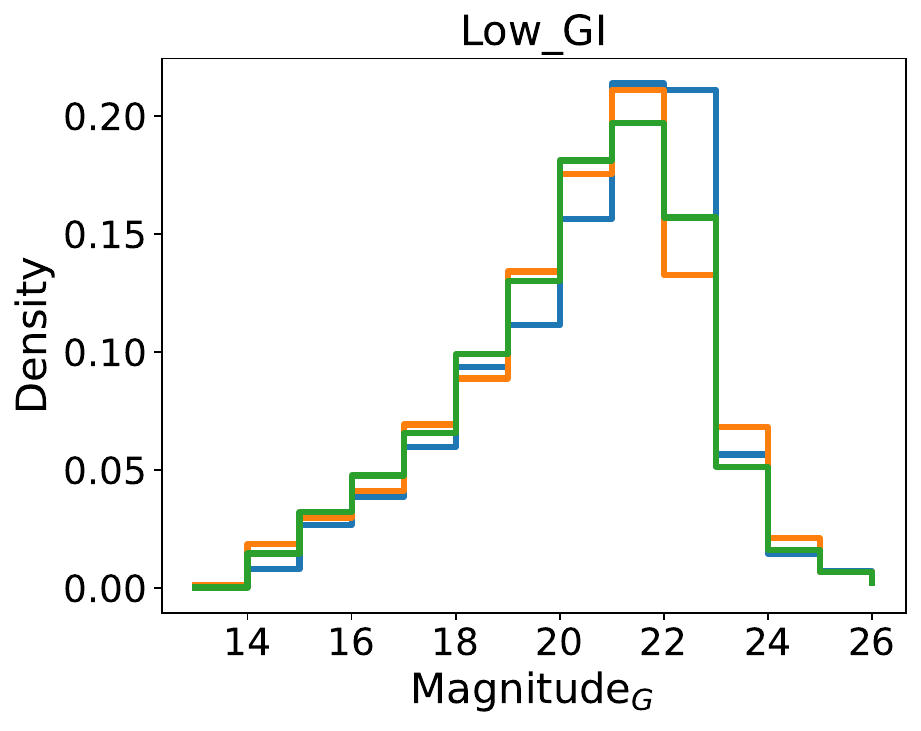}{0.32\textwidth}{(c)}
          }
\gridline{\fig{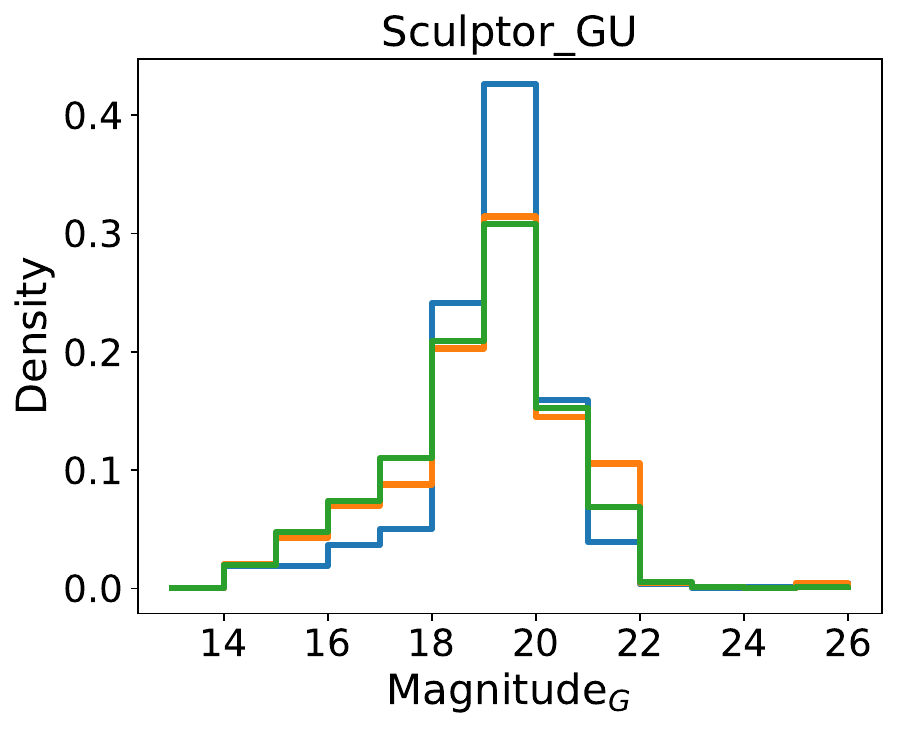}{0.32\textwidth}{(d)}
          \hspace{-5pt}
          \fig{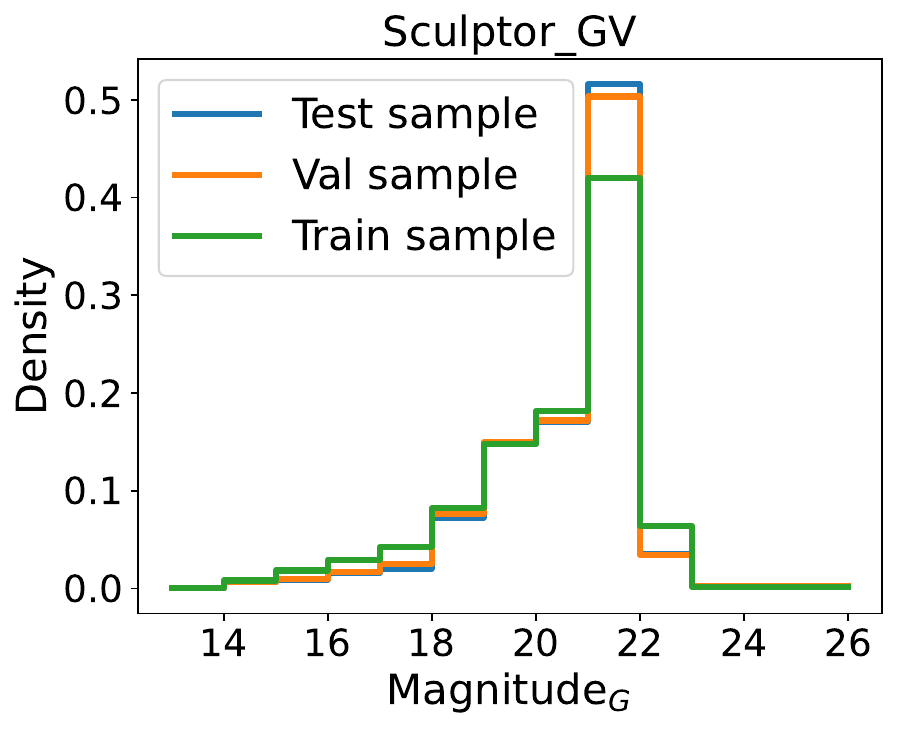}{0.32\textwidth}{(e)}
          \hspace{-5pt}
          \fig{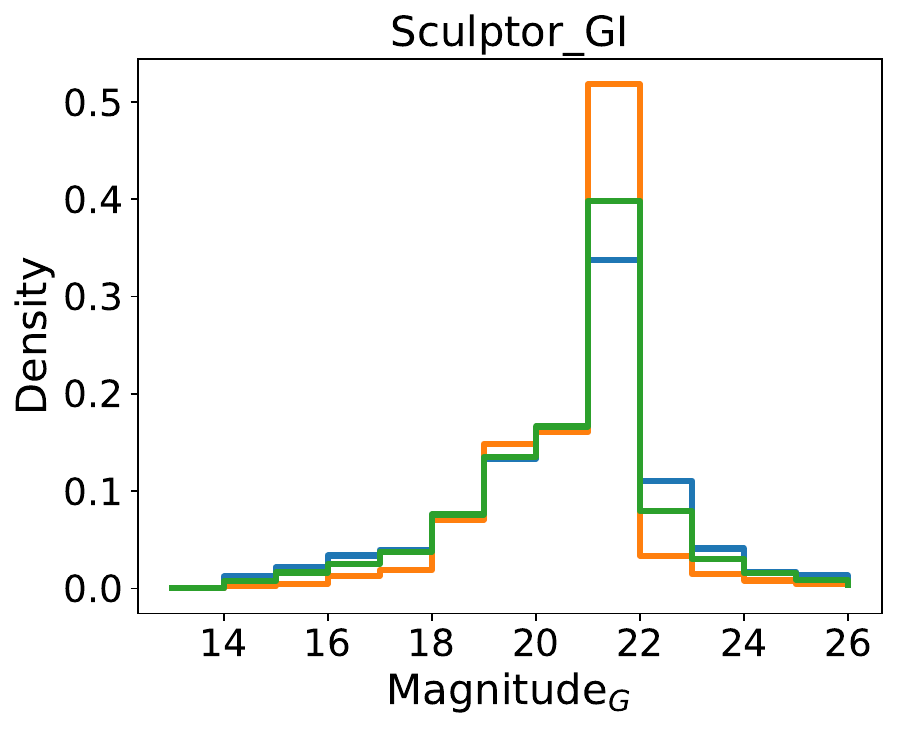}{0.32\textwidth}{(f)}
          }
\gridline{\fig{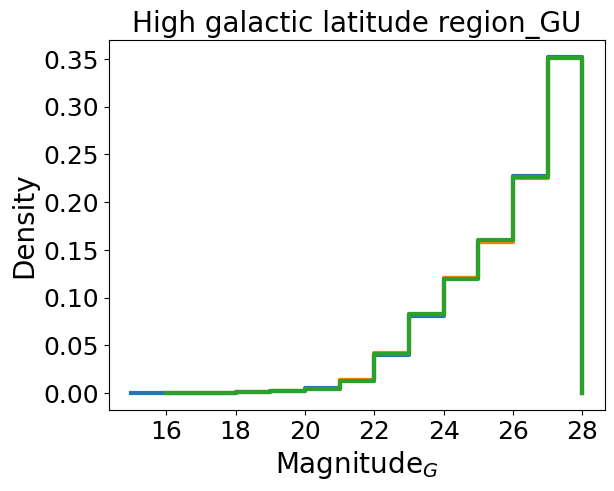}{0.32\textwidth}{(g)}
          \hspace{-5pt}
          \fig{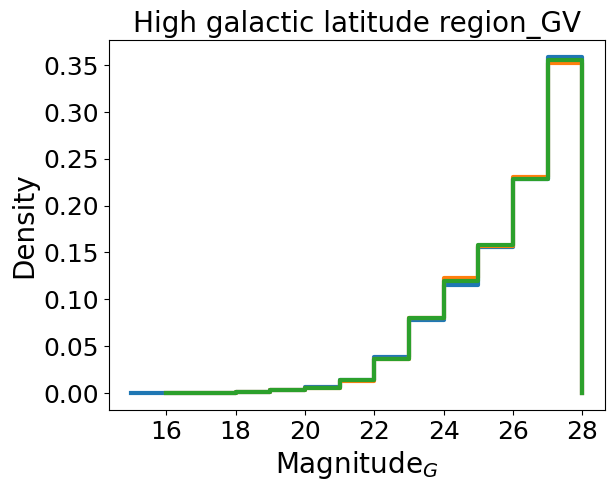}{0.32\textwidth}{(h)}
          \hspace{-5pt}
          \fig{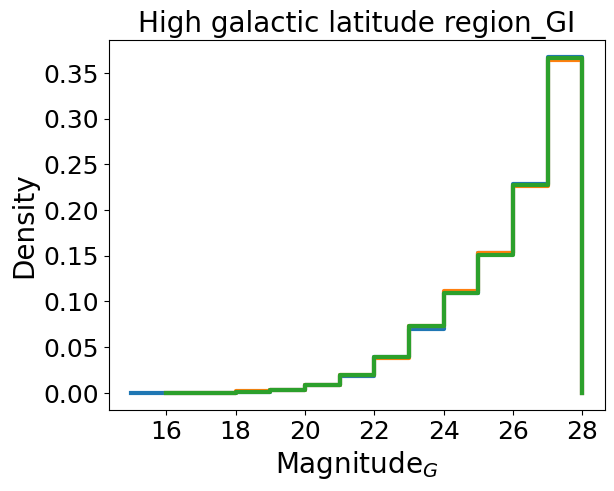}{0.32\textwidth}{(i)}
          }
\caption{The normalized distribution of datasets in the GU, GV, and GI bands for high galactic latitude regions, as well as for low-density and medium-density regions of the Galactic and nearby galaxies, where the vertical axis "Density" represents the proportion of sources within that magnitude interval relative to the total number. 
For the Milky Way and nearby galaxy regions, the simulated images use g-band magnitudes, whereas for the high-Galactic latitude region, the slitless spectroscopy images use white-light magnitudes corresponding to each chip.
The distributions for the training, validation, and test sets within each chip and region are generally consistent, ensuring balanced training performance. \label{fig:datasets}
}\end{figure*}

\subsection{Slitless Spectroscopic Data of CSST} \label{subsec:sls_spec}

The CSST \citep{Zhan2021} is a large space-based optical telescope with a 2-meter aperture, equipped with a variety of advanced observational instruments, 
including a wide-field optical survey module, a terahertz module, a multi-channel imager, an integral field spectrograph, and an exoplanet imaging coronagraph. 
Among these instruments, the wide-field optical survey module is one of the most critical observational devices of the CSST. 
The main focal plane of the wide-field optical survey module covers an area of approximately 1.1$^\circ \times$1.2$^\circ$ at the center of the field of view and consists of 30 detectors. 
Among these 30 detectors, 18 are dedicated to multi-band imaging observations (NUV, u, g, r, i, z, y), while 12 are assigned to slitless spectroscopic observations (GU, GV, GI).
Each exposure generates 30 observational images, corresponding to both the multi-band imaging and the slitless spectroscopic images. 
Each detector used for slitless spectroscopic observation by the CSST corresponds to two gratings, meaning each grating unit in the GU, GV, and GI bands consists of two gratings. 
The spectra from these two gratings, which operate at different orders (+1 or -1), disperse in opposite directions. 
In the final slitless spectroscopic image, the demarcation line is located at the 2/5 position of the detector, with sources on the left dispersing towards the right and sources on the right dispersing towards the left. 
In slitless spectroscopic images, our primary detection targets are the zero-order images and first-order spectra of stars and galaxies.
The zero-order spectrum is undispersed and exhibits better image quality, accounting for approximately 10\% of the total incident energy. It serves as a reference point for wavelength calibration. The first-order spectrum, which is the primary dispersed spectral signal, contains wavelength-dependent information.

\subsection{Data Simulation} \label{subsec:data_sim}

To ensure the timeliness and reliability of scientific outputs from the CSST, researchers have developed a comprehensive data simulation software designed to evaluate data processing pipelines and scientific analysis tools. Within this framework, the slitless spectral simulation module is built upon a detailed model of the optical system, incorporating the diffraction characteristics of the grating to achieve realistic and accurate results.
The simulated slitless spectroscopic data incorporate realistic observational characteristics by modeling the relationship between spectral traces and wavelength through polynomial fitting, while accounting for instrumental effects such as the point spread function (PSF), detector noise, and charge transfer efficiency. The simulations are generated using astrophysical source catalogs combined with optical response functions and detector characteristics to produce realistic slitless spectral images \citep{Zhang2025slitless,Wei2025overview}.

In this study, we used Cycle 6 \citep{csst_cycle6_sim}, the software for simulating CSST Main Sky Survey data, to generate synthetic CSST slitless spectroscopy images of the Galactic and nearby galaxies. 
The simulated data products for Cycle 6 are centered at Ra $=$ 244.9727 degrees, Dec $=$ 39.8959 degrees, covering approximately 1.53 square degrees. They include 137 continuously overlapping exposure pointings.
There is a shift of approximately one CCD size between consecutive exposures, covering a total sky area of about 2.65 square degrees. The central 1.53 square degree region constitutes the complete survey area.
These simulated data primarily represent stars, with galaxies excluded for this simulation. 
For practical applications, it is assumed that the data input into the model has already undergone correction for instrumental effects. 
However, it cannot be guaranteed that cosmic ray contamination has been fully removed. 
To account for this, cosmic ray effects were added to half of the simulation data, while the other half only included sky background noise. 
For the data with cosmic ray effects, different cosmic ray maps were generated for various outputs (e.g., calibrated output CAL and main science output MS) during the simulation process. 
This approach helps to realistically simulate the impact of cosmic rays on the detector, as their distribution is random, and different observational stages may be affected by different cosmic ray events. 
Two input star catalogs were selected: one for a low-density celestial region and the other for a medium-density region, specifically the Sculptor region. 
These two input catalogs are stellar catalogs, consisting of stars selected from the same sky region in Gaia DR3. 
They retain the spatial coordinates provided in Gaia DR3, along with the three Gaia magnitudes: G, BP, and RP. 
A star-by-star correspondence is established by comparing the G, BP, and RP magnitudes of stars in Gaia DR3 with those generated by the Galaxia software. 
In Galaxia, stellar magnitudes are computed via synthetic photometry: the Spectral Energy Distributions (SEDs), produced using the BT-Settl library based on stellar parameters ($T_{eff}$, log$g$, [Fe/H]), are convolved with the CSST filter transmission curves to derive magnitudes in each band \citep{Wei2025overview}. 
For each Gaia DR3 star S\_GDR3, a corresponding Galaxia star S\_Gal with consistent G, BP, and RP magnitudes is then selected.
The parameters and ugriz magnitudes of S\_Gal are then assigned to S\_GDR3 to facilitate matching with model spectra. 
Since all stars in the Gaia DR3 sample are brighter than G$=$21 mag, the stars in Gaia DR3 are used to replace those with G$<$21 mag in the simulated Galaxia catalog. 
These star catalogs are used for the development of the pipeline in the process of positional and flux calibration, as well as for other CSST data processing tasks \citep{Zhou_2025}.
These catalogs contain detailed information on stars within their respective regions.
For stars with a Gaia G magnitude less than 22, the density in the low-density region is 5,197.76 objects per square degree, while in the medium-density region, it is 7,328.82 objects per square degree.
The simulation program generates simulated images based on a pointing file (pointing.data). 
The survey module’s focal plane consists of 30 detectors, and each pointing generates 30 images per exposure (150 seconds). 
Among these, 12 are slitless spectroscopy images. 
The dataset includes 30 exposure pointings for both the low-density and medium-density celestial regions, with each pointing consisting of two types of images: those with cosmic rays and those without. 
Additionally, each pointing includes 12 slitless spectroscopy images in the GU, GV, and GI bands. 
As a result, this sample set contains a total of 1,440 slitless spectroscopy images, with 720 images from the low-density celestial region and 720 from the medium-density celestial region. 
Figure \ref{fig:simulation_images} (a)-(d) presents four simulated CSST slitless spectroscopy images of the Galactic and nearby galaxies. 
To intuitively demonstrate the contamination of cosmic rays on astronomical images and their distinction from real astrophysical signals, Figure 1 provides a comparative set of simulated image pairs.
Panels (a) and (b) show slitless spectroscopic images of the same low-density region in the Galactic and nearby galaxies. Specifically, (a) is free of cosmic rays, while (b) includes cosmic rays—all other content remains identical.
Similarly, panels (c) and (d) present slitless spectroscopic images of the same medium-density region in the Galactic and nearby galaxies, where (c) is without cosmic rays and (d) includes them—again, all other elements are unchanged.
Cosmic rays typically appear as sharp, bright peaks confined to one or a very small cluster of pixels, exhibiting a morphology inconsistent with the telescope’s point spread function (PSF).
In contrast, genuine astrophysical point sources display smooth, symmetric profiles that conform to the expected PSF shape.

For the high galactic latitude cases, we used the Cycle 9 simulation data \citep{csst_cycle9_sim}, which includes a larger number of galaxies. 
The simulation data products for Cycle 9 comprise five simulated sky regions, centered at [Ra, Dec] = [180.0, 60.0], [240.0, 30.0], [170.0, -23.0], [30.0, -48.0], and [300.0, -60.0], respectively. 
Each region includes simulated images and slitless spectral data for galaxies and stars, covering an area of 5 square degrees, with a total of 1,221 exposure pointings. 
Furthermore, to account for the effect of intrinsic ellipticity in current weak gravitational lensing shear measurements, the intrinsic ellipticity of each galaxy within the regions was rotated by 90 degrees in the Cycle 9 simulations before the imaging process was repeated. 
In total, image simulations for 50 square degrees were completed.
In each continuous sky region, there is a shift of roughly one CCD size between consecutive exposures. The central 5 square degree area represents the complete survey region.
We selected 10 exposure pointings from the Cycle 9 simulation data, and Figure \ref{fig:simulation_images} (e) shows a simulated CSST slitless spectroscopy image of the high galactic latitude region. 

Finally, the complete dataset consists of 1,560 slitless spectroscopy images: 1,440 from Cycle 6 contain only stars and no galaxies, while the remaining 120 from Cycle 9 include both stars and a substantial population of galaxies.
Figure \ref{fig:datasets} illustrates the distribution of datasets across the GU, GI, and GV bands for the low-density and medium-density regions of the Galaxy and nearby galaxies, as well as for the high galactic latitude regions. 
The distributions for the training, validation, and test sets are generally consistent. 
And the observation numbers for the GU, GV, and GI bands are shown in Table \ref{tab:observation_number}.

Our dataset contains only ordinary stars and galaxies, and does not contain special types such as AGNs. Although AGNs are not explicitly included, this does not affect our results, as the aim of this work is to estimate the number and counts of sources. In slitless spectroscopy, both AGNs and typical galaxies appear as extended-source spectra, and thus their distinction is not critical for our model’s learning process.

\begin{deluxetable}{lccc}
\tablecaption{The observation numbers for the GU, GV, and GI bands. \label{tab:observation_number}}
\tablewidth{0pt}
\tablehead{
\colhead{Region} & \colhead{GU bands} & \colhead{GV bands} & \colhead{GI bands} 
}
\startdata
the low-density regions & 35,266 & 34,440 & 34,122 \\
the medium-density regions & 54,416 & 50,588 & 51,516 \\
the high galactic latitude regions & 25,369 & 27,756 & 46,033 \\
\enddata
\begin{tablenotes}
  \item Note: The table displays the count of sources with magnitudes brighter than 22 in the output star catalog. The output catalog is a source information file produced alongside the simulated images by the simulation software Cycle 6 and Cycle 9.
\end{tablenotes}
\end{deluxetable}

\section{METHODOLOGY} \label{sec:methods}
\subsection{YOLO} \label{subsec:yolo}

In this work, we developed a model to detect zeroth-order images and first-order spectra in simulated CSST slitless spectroscopy images using the YOLOv8 algorithm, a member of the YOLO family \citep{7780460}. 
The YOLO algorithm is a groundbreaking deep learning approach for real-time object detection. 
It reformulates the object detection task as a regression problem, directly predicting bounding box coordinates and class probabilities from image pixels, which enables fast and efficient detection. 
YOLO divides the input image into an $S\times S$ grid, where each grid cell predicts $B$ bounding boxes along with their confidence scores. 
Each grid cell also predicts $C$ conditional class probabilities. 
The final output is a tensor of size $S\times S\times (B\times 5+C)$, where $B\times 5$ corresponds to the bounding box coordinates ($x$, $y$, width $w$, height $h$) and confidence scores, while $C$ represents the class probabilities. 
The YOLO loss function consists of three components: bounding box coordinate loss, confidence loss, and class probability loss \citep{7780460}. 
These losses are combined in a weighted sum for multi-objective optimization. 
YOLO processes images in real-time, and because it observes the entire image during both training and detection, it can use global contextual information to minimize false positives from background noise. 
However, YOLO may encounter difficulties in localizing small or densely packed objects. 

YOLOv8, released by Ultralytics on January 10, 2023, introduces significant advancements over its predecessors. 
It features a deeper network architecture with additional convolutional layers and residual connections, which enhance the network's capacity for feature extraction and improve detection performance. 
YOLOv8 also offers models of varying sizes, including YOLOv8n, YOLOv8s, YOLOv8m, YOLOv8l, and YOLOv8x, allowing for flexible applications based on computational requirements. 
For this project, we selected the large-scale architecture, YOLOv8l, from the Ultralytics YOLOv8 library to develop an object detection model aimed at identifying zeroth-order images and first-order spectra in simulated CSST slitless spectroscopy images. 

\subsection{Data Preprocessing} \label{subsec:Preprocessing}

\begin{figure*}
\gridline{
  \fig{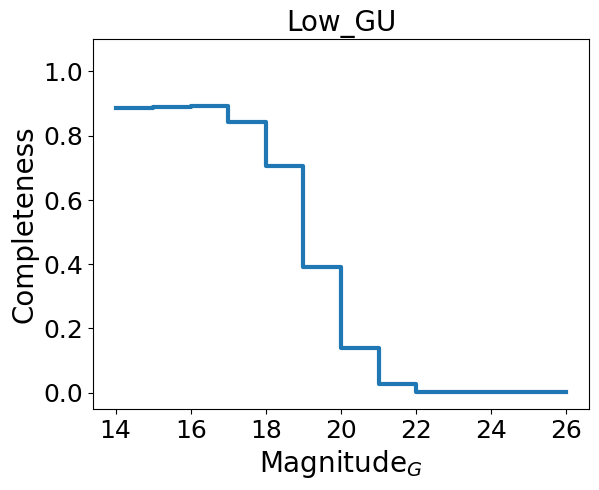}{0.32\textwidth}{(a)}
  \hspace{-5pt} 
  \fig{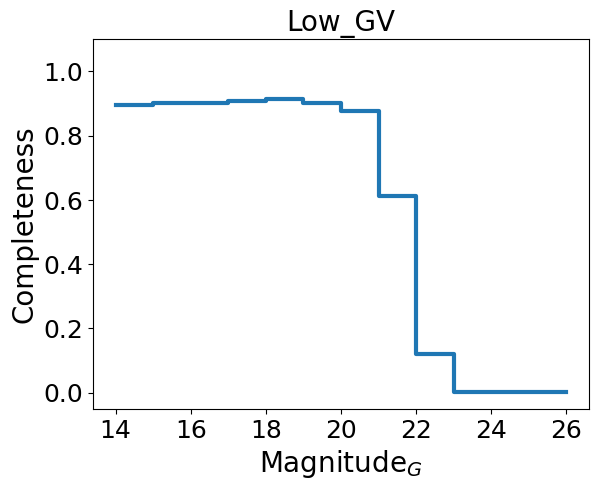}{0.32\textwidth}{(b)}
  \hspace{-5pt}
  \fig{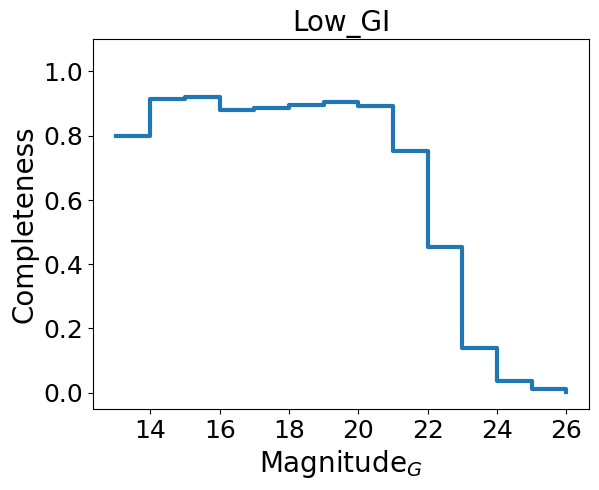}{0.32\textwidth}{(c)}
}
\gridline{
  \fig{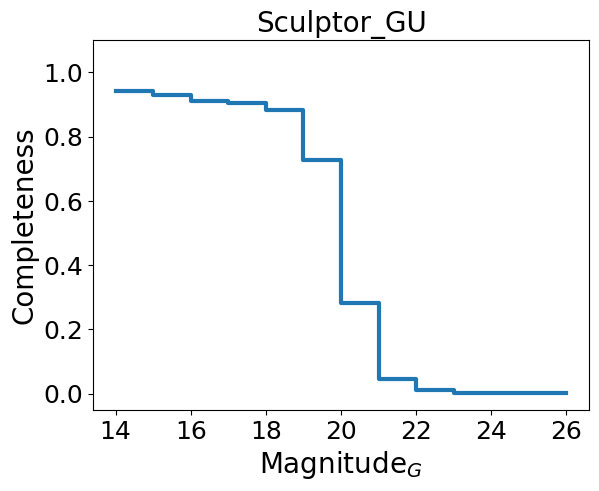}{0.32\textwidth}{(d)}
  \hspace{-5pt}
  \fig{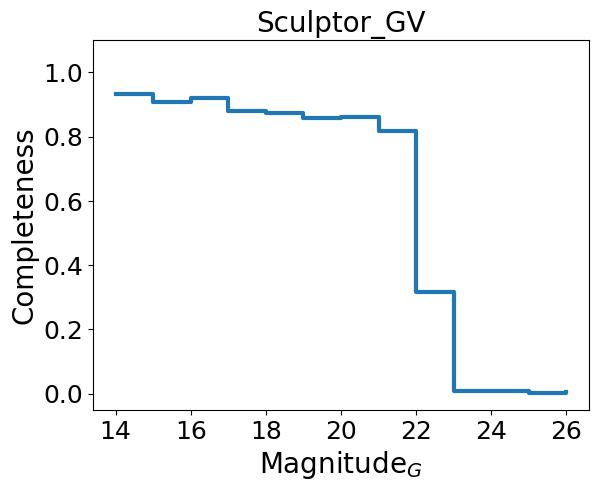}{0.32\textwidth}{(e)}
  \hspace{-5pt}
  \fig{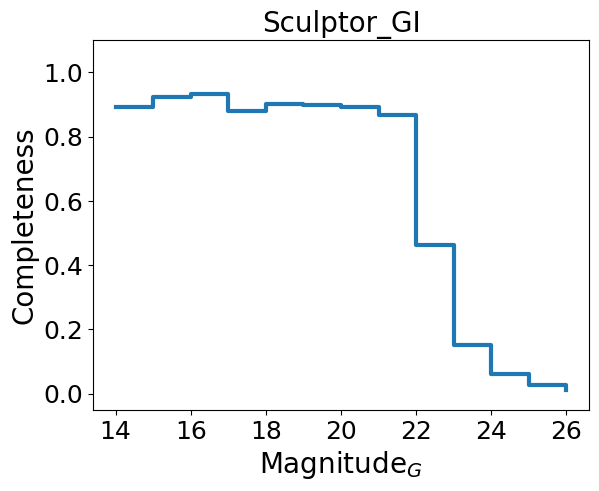}{0.32\textwidth}{(f)}
}
\gridline{\fig{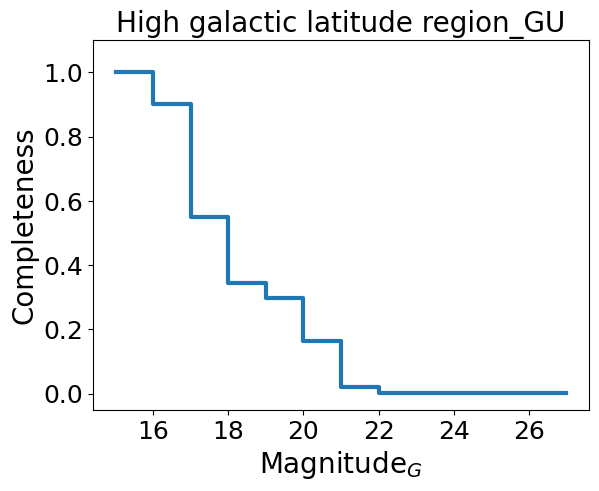}{0.32\textwidth}{(g)}
          \hspace{-5pt}
          \fig{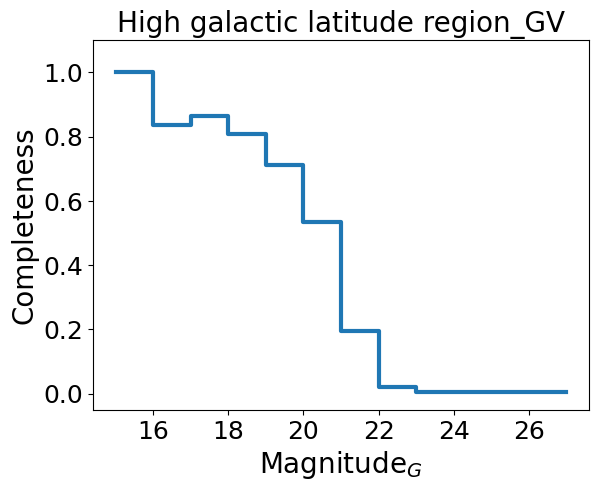}{0.32\textwidth}{(h)}
          \hspace{-5pt}
          \fig{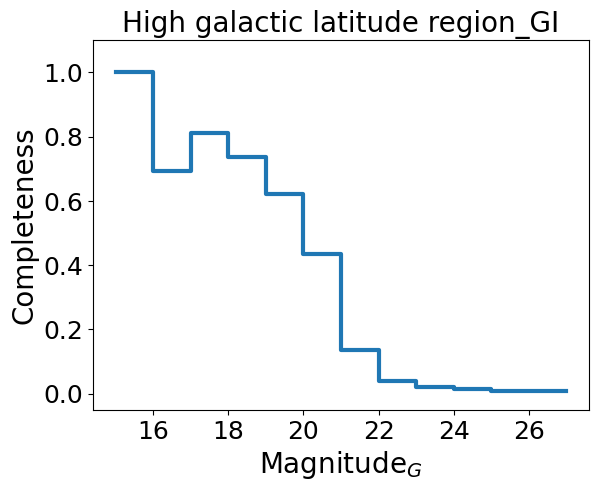}{0.32\textwidth}{(i)}
}
\caption{The completeness of visual annotation. The completeness of manual annotation is lowest in the high galactic latitude regions, primarily because these regions contain a large number of galaxies, which are extended sources. At a given magnitude, galaxies appear visually fainter than point sources, reducing their detectability. Moreover, the completeness varies across different bands due to differences in SNRs. In particular, the GU band images exhibit relatively low SNR, which limited the number of sources that could be reliably identified through visual inspection.\label{fig:completeness}}
\end{figure*}

Due to the high density of sources in some of the slitless spectroscopy images for the medium-density celestial region (as shown in Figure \ref{fig:simulation_images} (c) and (d)), detecting objects using deep learning poses significant challenges. 
To address this, we divided each slitless spectroscopy image into 16 equal parts by resampling each section to 640$\times$640 pixels, as the YOLOv8 model requires input images of this size. 
This approach resulted in a total of 24,960 segmented slitless spectroscopy images, each of 640$\times$640 pixels, for training and testing. 
These images were annotated using CVAT \citep{2019zndo...3497106S}, a interactive video and image annotation tool for computer vision, to accurately define the bounding box dimensions, which enclose the detected objects. 
Each label includes the object's central coordinates ($x$, $y$) and the width and height of the bounding box ($w$, $h$).
The labels were visually annotated, and the completeness of visual annotation is shown in Figure \ref{fig:completeness}.
The final dataset of 24,960 segmented images was split into training, validation, and test sets at a ratio of 8:1:1. 
Specifically, 19,968 images were used for training, 2,496 images for validation, and 2,496 images for testing. 

\subsection{Model Training} \label{subsec:training}

During model training, a batch size of 16 was used, and the Stochastic Gradient Descent (SGD) optimizer was employed to fine-tune the model parameters. 
The parameters for image translation, image scaling, and image mosaicking were set to 0.1, 0.5, and 1.0, respectively. 
The initial learning rate was configured to 0.01, with the Intersection over Union (IoU) training threshold set to 0.7. 
After 920 epochs of training, completed in 206.911 hours, the model's loss showed stable convergence. 
As a result, we successfully developed an object detection model capable of identifying zeroth-order images and first-order spectra in the simulated CSST slitless spectroscopy images. 

\begin{figure}[ht!]
\centering
\includegraphics[width=0.5\textwidth]{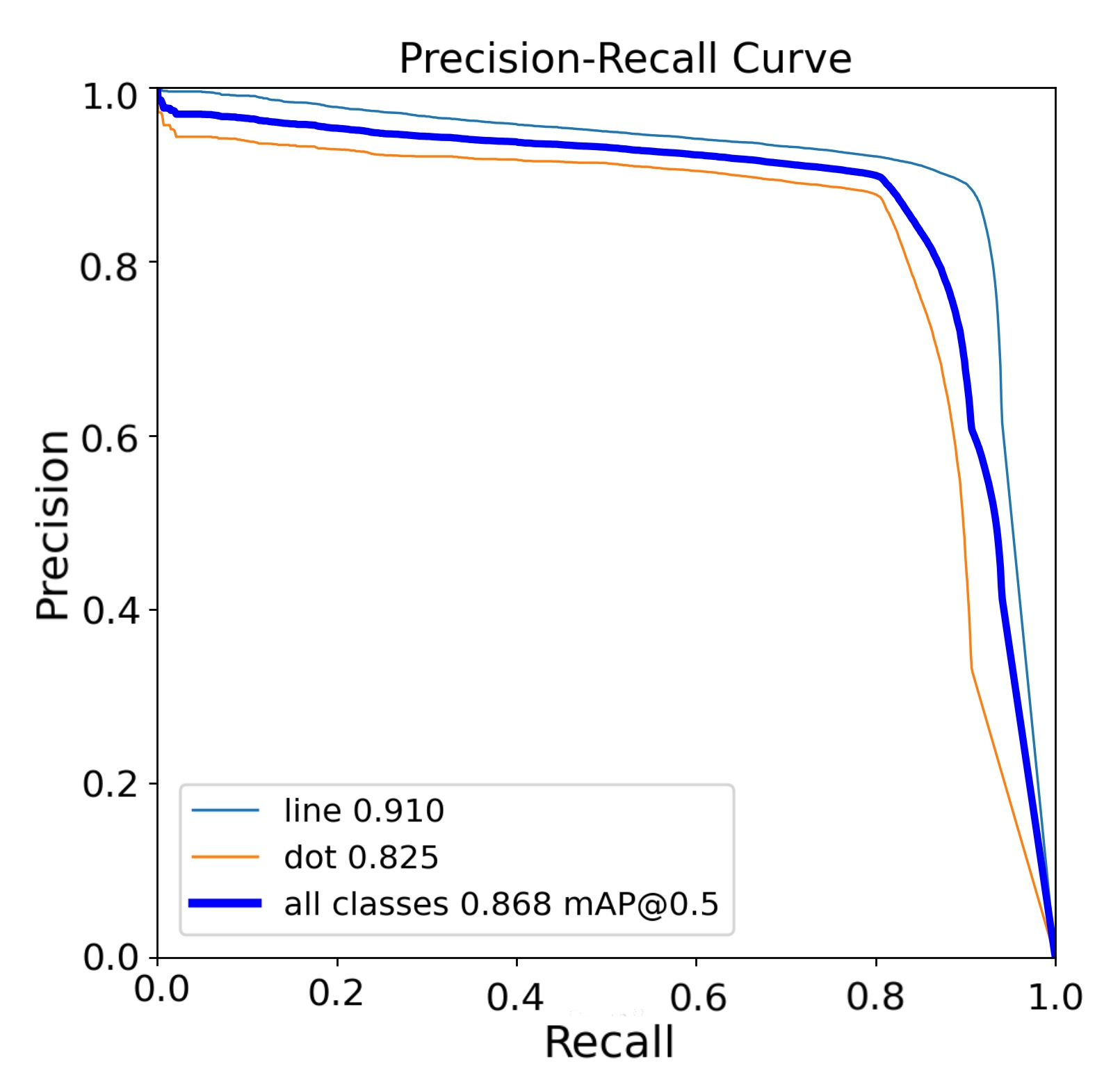}
\caption{The precision-recall curve of the source detection model, evaluated on the validation dataset. The model achieves a mean Average Precision (mAP@0.5) of 0.868 across all classes, with the ``line'' class (spectral lines) performing best at 0.910 precision and the ``dot'' class (zero-order image) at 0.825. \label{fig:pr_curve}}
\end{figure}

To evaluate the model's performance in detecting sources, we computed the precision and recall for the test dataset images.  
These metrics were then used to calculate the model's mAP score. 
The formulas for computing precision, recall, average precision (AP), and IoU are as follows:
\begin{equation}
  precision = \frac{TP}{TP + FP}
\end{equation}
\begin{equation}
  recall = \frac{TP}{TP + FN}
\end{equation}
\begin{equation}
  AP = \int_0^1 P(R)dR
\end{equation}
\begin{equation}
  IOU = \frac{A\cap B}{A\cup B}
\end{equation}
where TP (True Positives) refers to the number of targets correctly identified, FP (False Positives) denotes the number of targets incorrectly identified, and FN (False Negatives) represents the number of targets that were missed or incorrectly rejected. 
$A$ represents the predicted bounding box, while $B$ represents the ground-truth bounding box. 
Precision and recall are key metrics for evaluating the performance of machine learning models. 
Precision measures the ratio of true positives (correctly detected objects) to the total number of detected objects, reflecting the accuracy of the detection results, 
while recall measures the ratio of true positives to the total number of actual existing objects, reflecting the completeness of the detection method \citep{8825470}. 
Typically, precision and recall exhibit a trade-off: improving one often leads to a reduction in the other. 
To balance these metrics, the AP is used, which represents the area under the Precision-Recall curve. 
The AP value ranges from 0 to 1, with a higher value indicating better model performance. 
The mAP is a metric used to evaluate the performance of object detection models by calculating the average of the AP scores across multiple object categories.
mAP@50 (mAP50) is a composite metric that averages the AP scores across multiple object categories, providing an overall performance evaluation of the detection model. mAP@50 (also known as mAP50) considers only those detections where the IoU is at least 50\%, while mAP50-95 calculates the mAP considering IoU thresholds ranging from 50\% to 95\%.
\begin{deluxetable}{lcccc}
\tablecaption{Performance Metrics of Slitless Spectroscopy Source Detection Model Across Different Classes \label{tab:performance}}
\tablewidth{0pt}
\tablehead{
\colhead{Class} & \colhead{Precision} & \colhead{Recall} & \colhead{mAP50} & \colhead{mAP50-95}
}
\startdata
all & 0.878 & 0.856 & 0.868 & 0.435 \\
spectral lines & 0.886 & 0.904 & 0.910 & 0.515 \\
zeroth-order images & 0.870 & 0.808 & 0.825 & 0.354 \\
\enddata
\end{deluxetable}

\begin{figure*}
\gridline{\fig{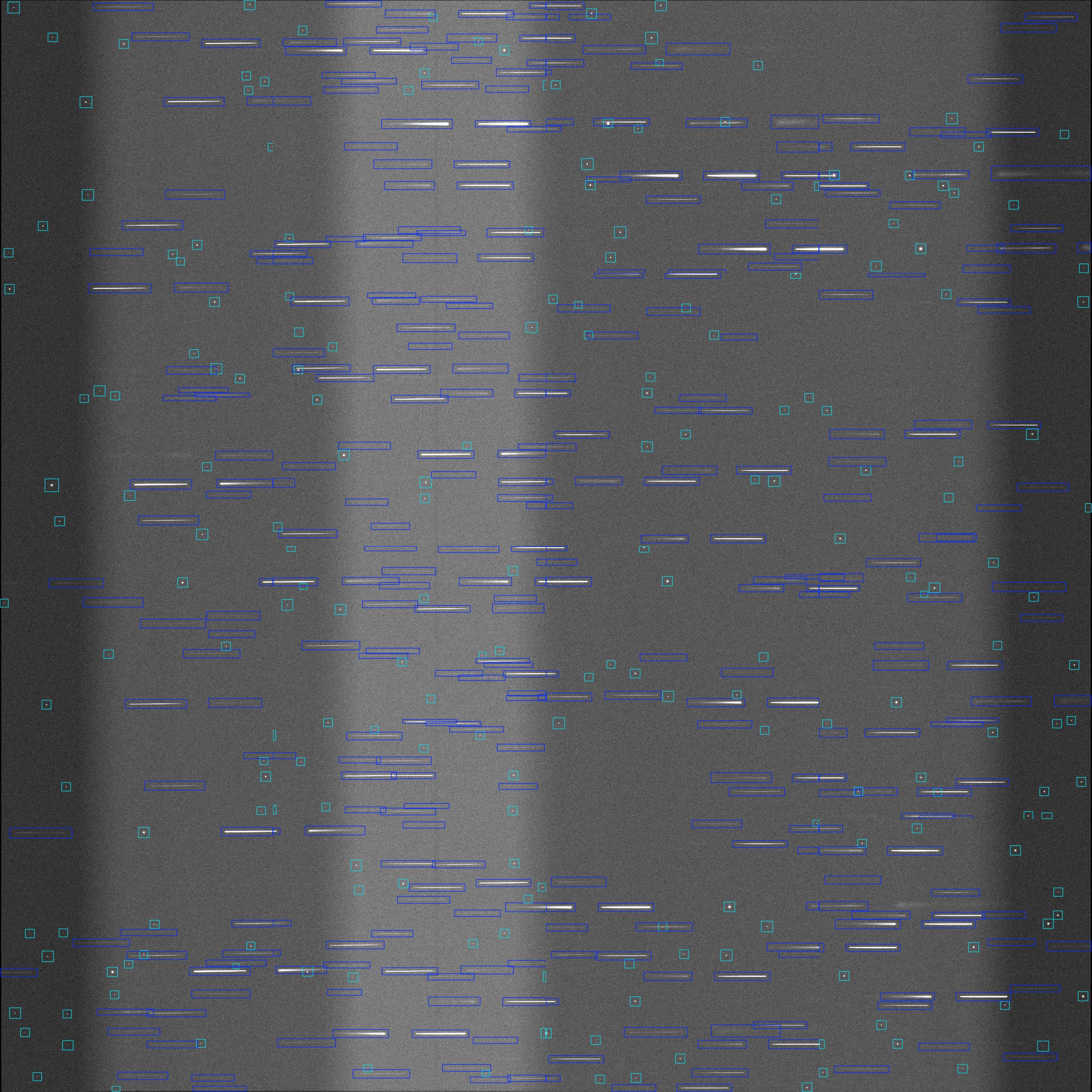}{0.33\textwidth}{(a)}
          \fig{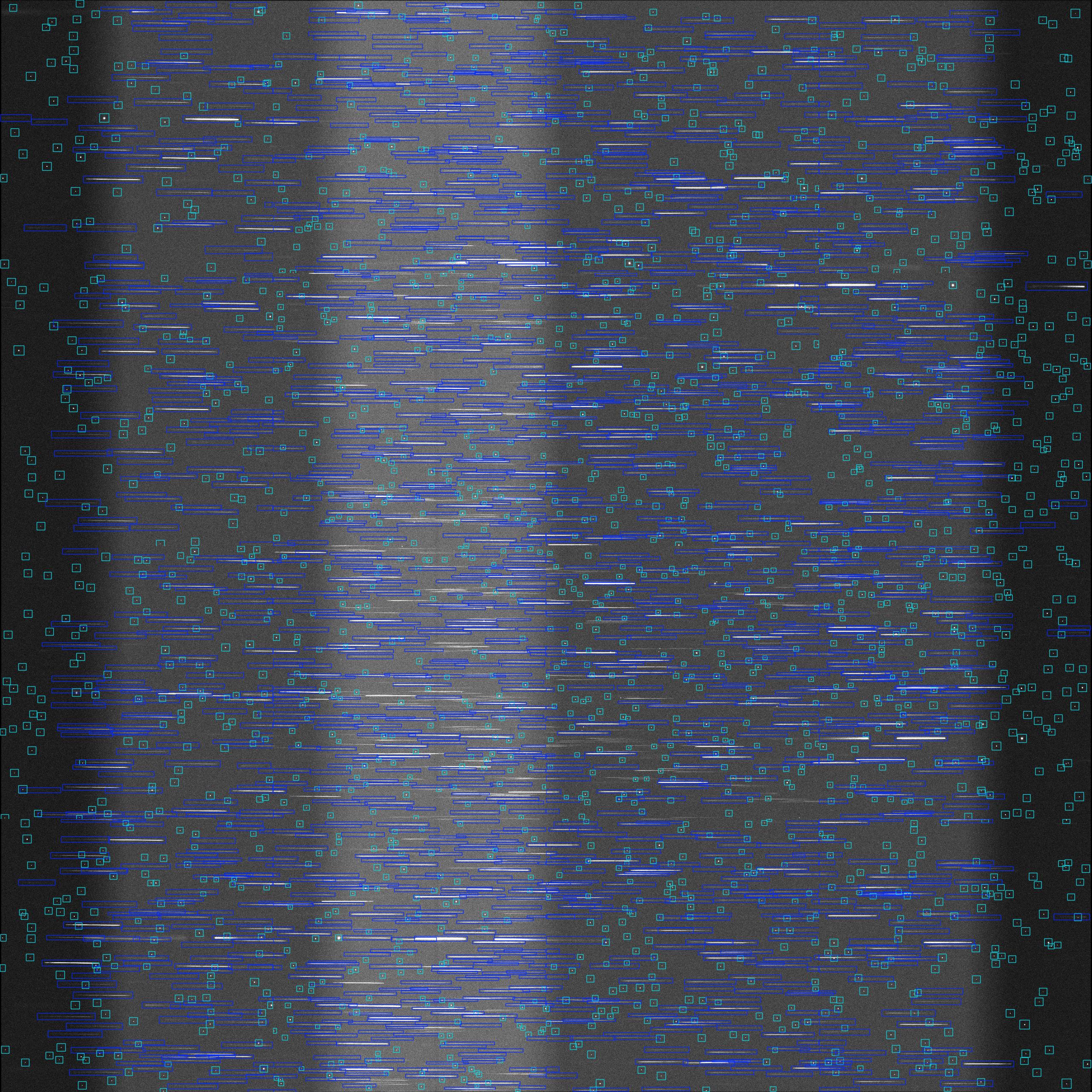}{0.33\textwidth}{(b)}
          \fig{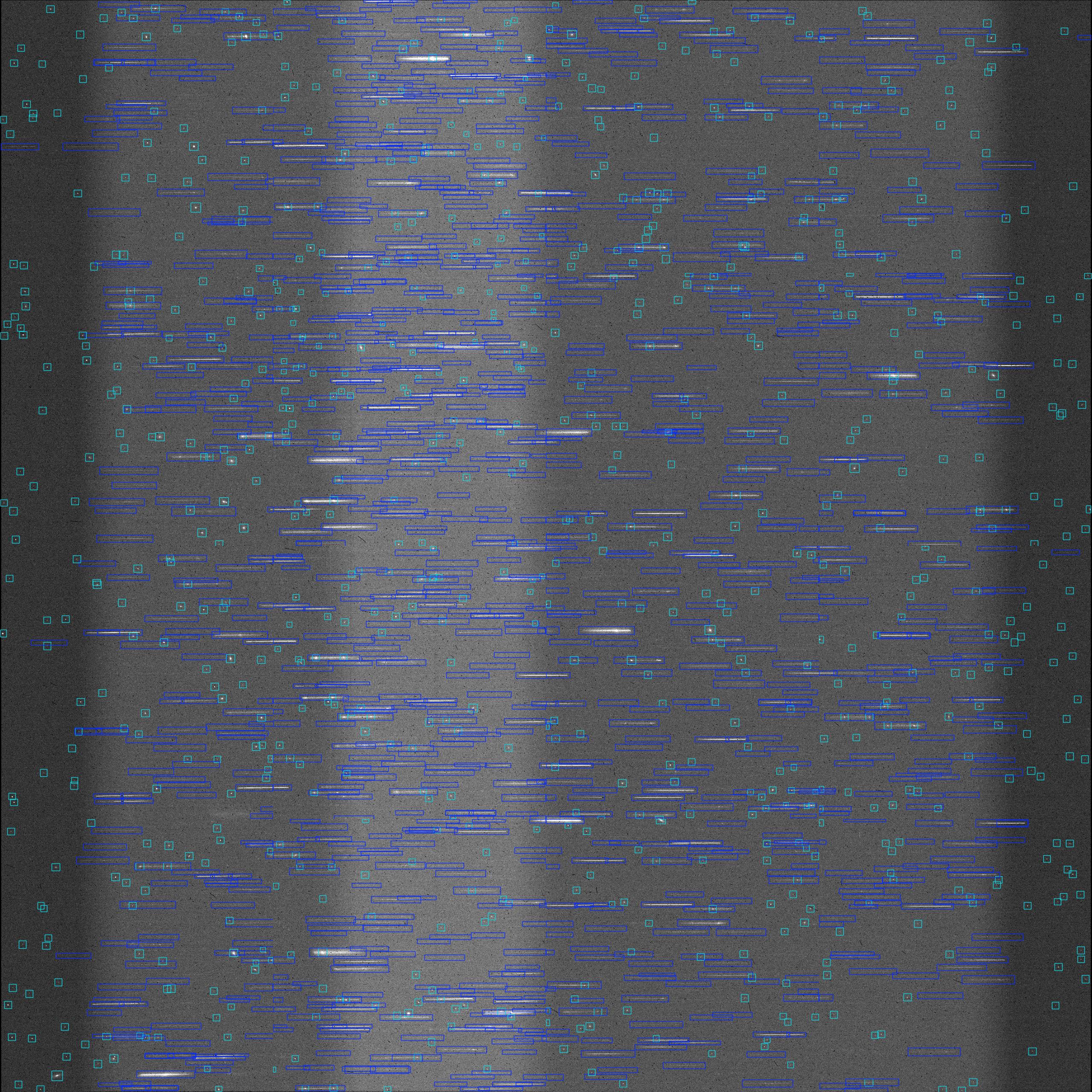}{0.33\textwidth}{(c)}
          }
\gridline{\fig{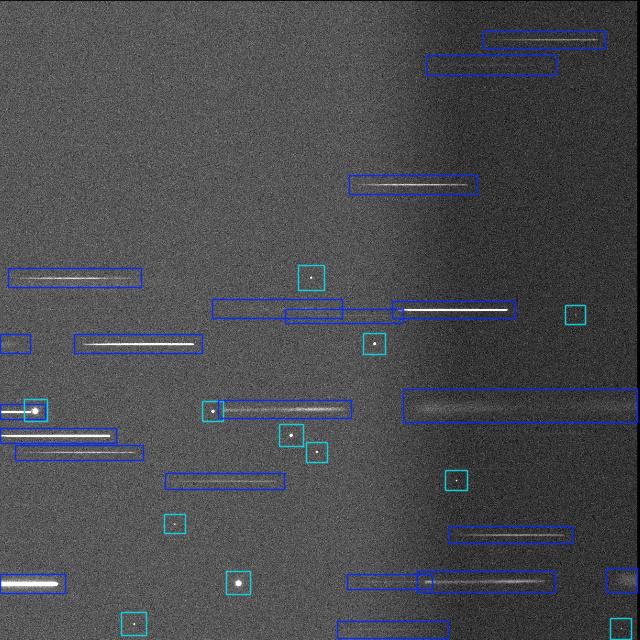}{0.33\textwidth}{(d)}
          \fig{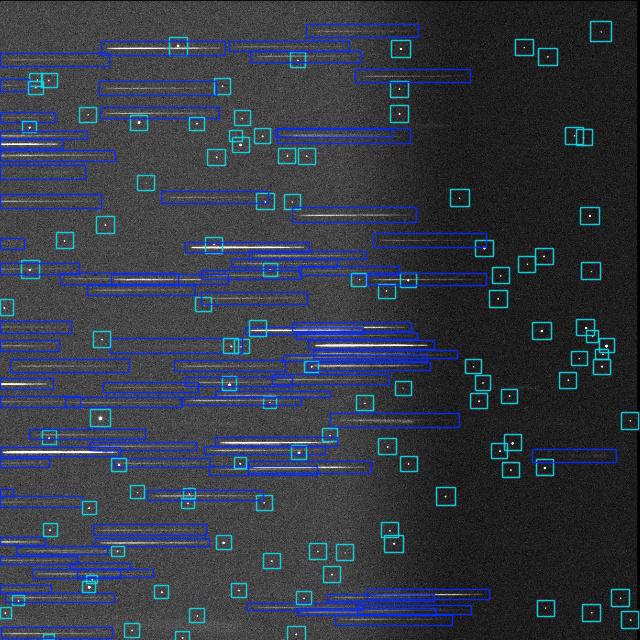}{0.33\textwidth}{(e)}
          \fig{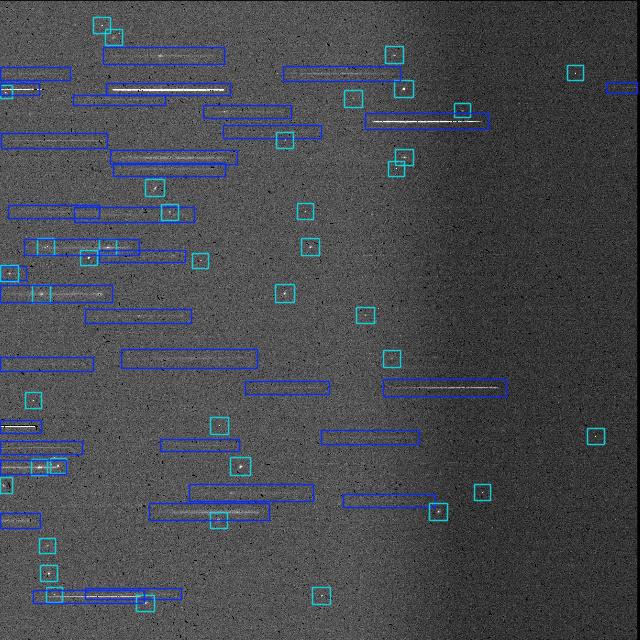}{0.33\textwidth}{(f)}
          }
\caption{Schematic diagram of slitless spectroscopy image detection:
(a) Illustration of target detection in a low-density celestial region of the Galactic and nearby galaxies,
(b) Illustration of target detection in a medium-density celestial region of the Galactic and nearby galaxies,
(c) Illustration of target detection in a high galactic latitude region. 
The figure shows the final CSST slitless spectral simulation image, assembled by stitching together 16 sub-images after object detection.
Panel (d), (e), and (f) are local images magnified 16 times from panel (a), (b), and (c) respectively, to better demonstrate the target detection effect.
\label{fig:detected_example}}
\end{figure*}

\section{RESULTS} \label{sec:results}

The detection model outputs the center coordinates of the object, along with the width and height of the bounding box, and the confidence score. 
The confidence level represents the probability that an object is present within a given bounding box. 
In our study, we used a standard confidence threshold of 0.001 during validation. 
This resulted in precision values of 87.8\%, 88.6\%, and 87.0\%, and recall rates of 85.6\%, 90.4\%, and 80.8\% for the total, spectral lines, and zeroth-order images, respectively. 
The detailed test results are presented in Table \ref{tab:performance}. 
Figure \ref{fig:pr_curve} shows the precision-recall curve for the detection model. 
The performance for detecting spectral lines is superior to that for zeroth-order images. 
This may be attributed to the influence of cosmic rays, which can be confused with zeroth-order images. 
Additionally, the SNR for zeroth-order images is lower, as their energy contributes only about 10\% of the total incident energy \citep{Zhan2021}.

In the test set, we used a standard confidence threshold of 0.07. 
Each image was processed in approximately 9.3 ms, resulting in a total processing time of 0.387 minutes for all 2,496 images. 

\begin{figure*}
\gridline{
  \fig{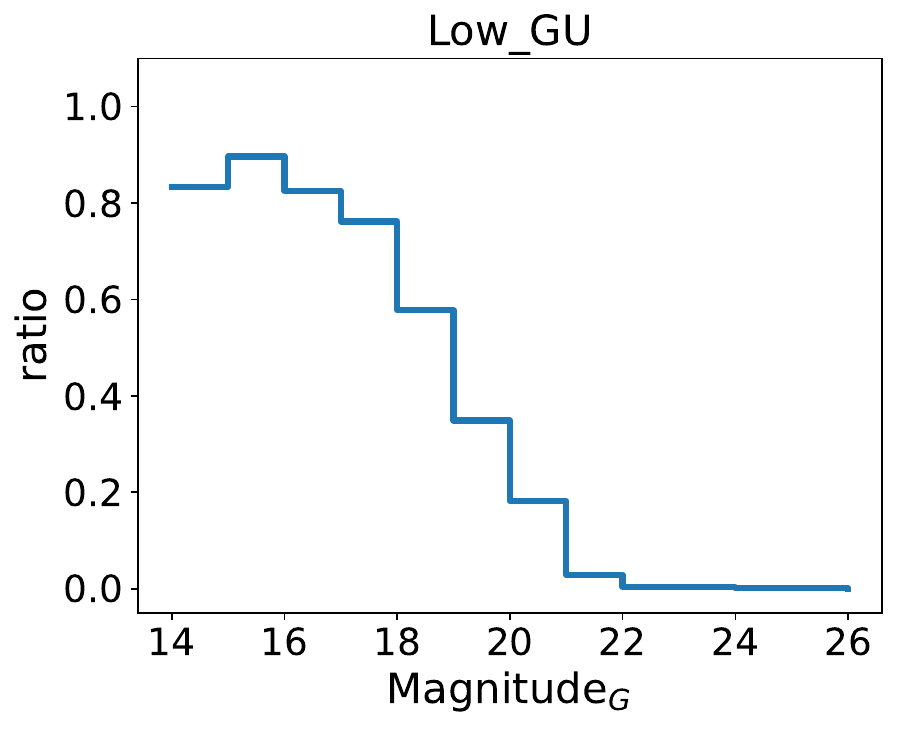}{0.32\textwidth}{(a)}
  \hspace{-5pt} 
  \fig{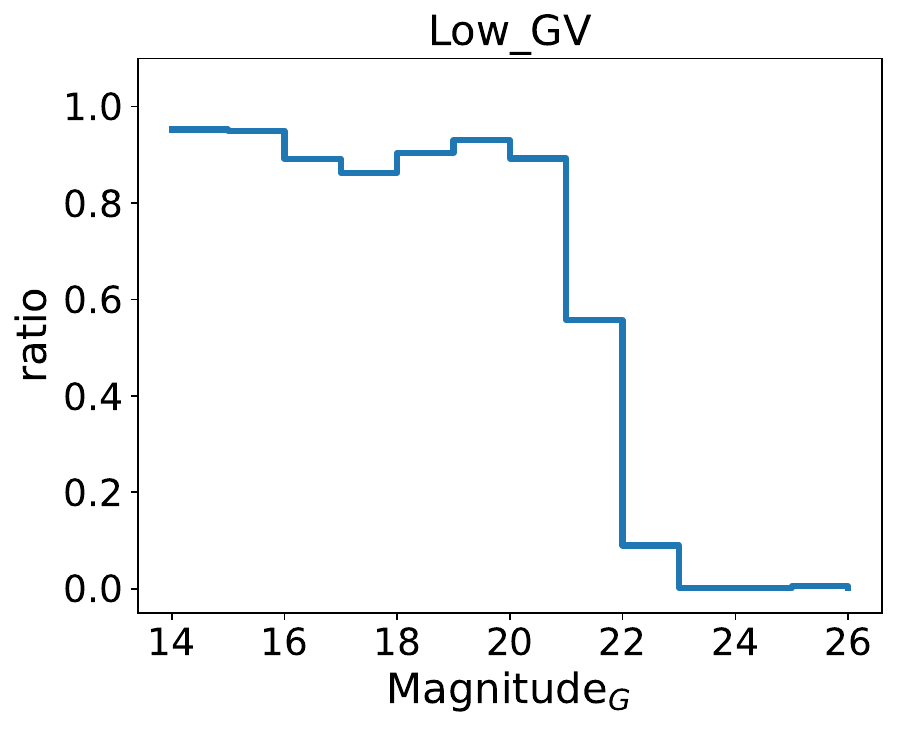}{0.32\textwidth}{(b)}
  \hspace{-5pt}
  \fig{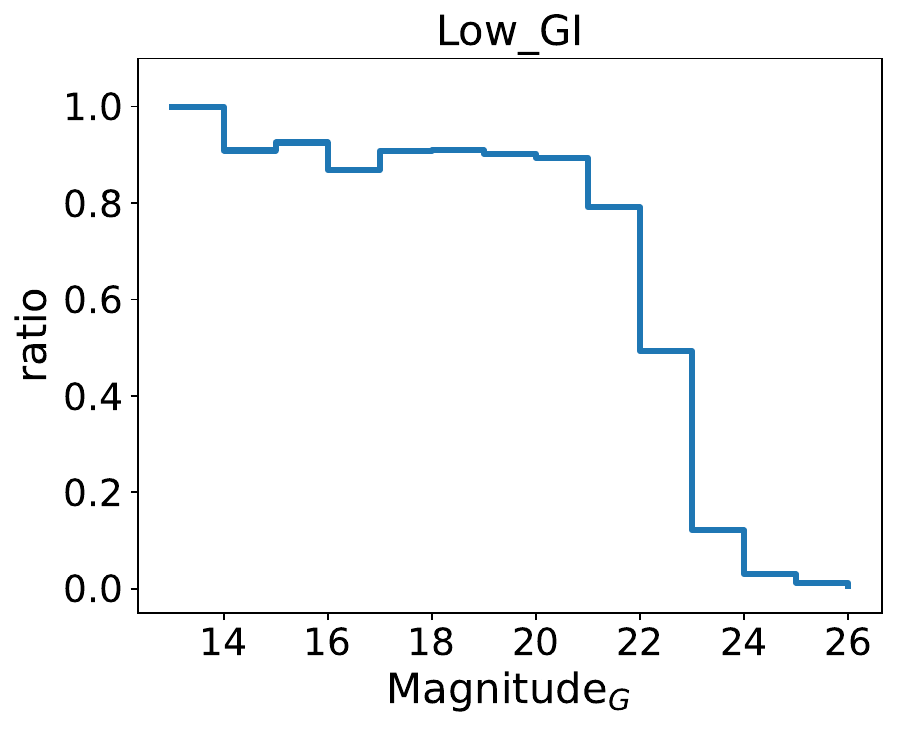}{0.32\textwidth}{(c)}
}
\gridline{
  \fig{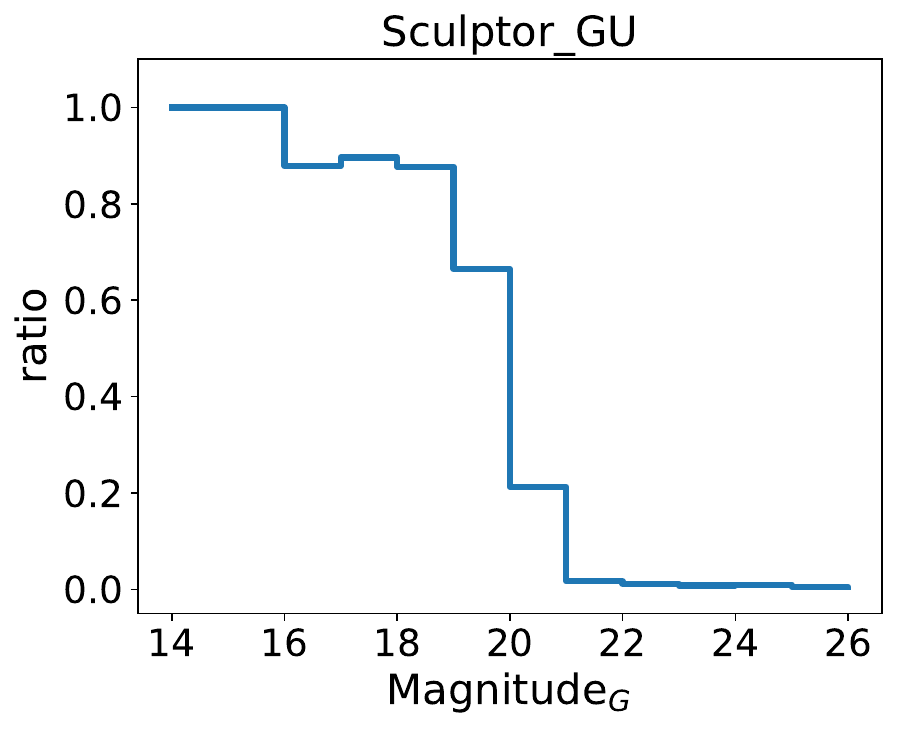}{0.32\textwidth}{(d)}
  \hspace{-5pt}
  \fig{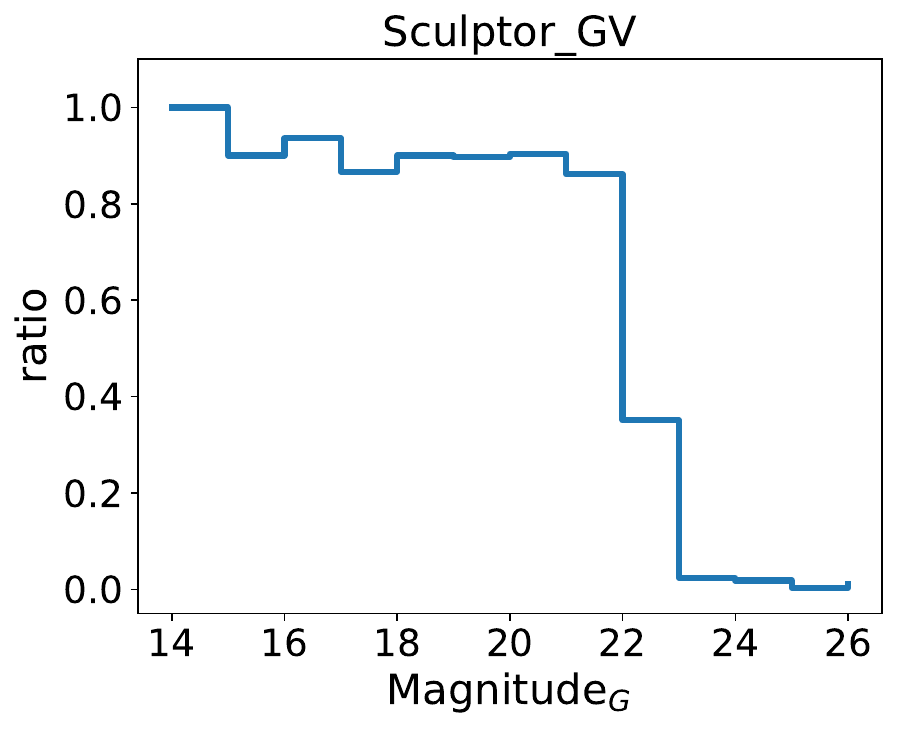}{0.32\textwidth}{(e)}
  \hspace{-5pt}
  \fig{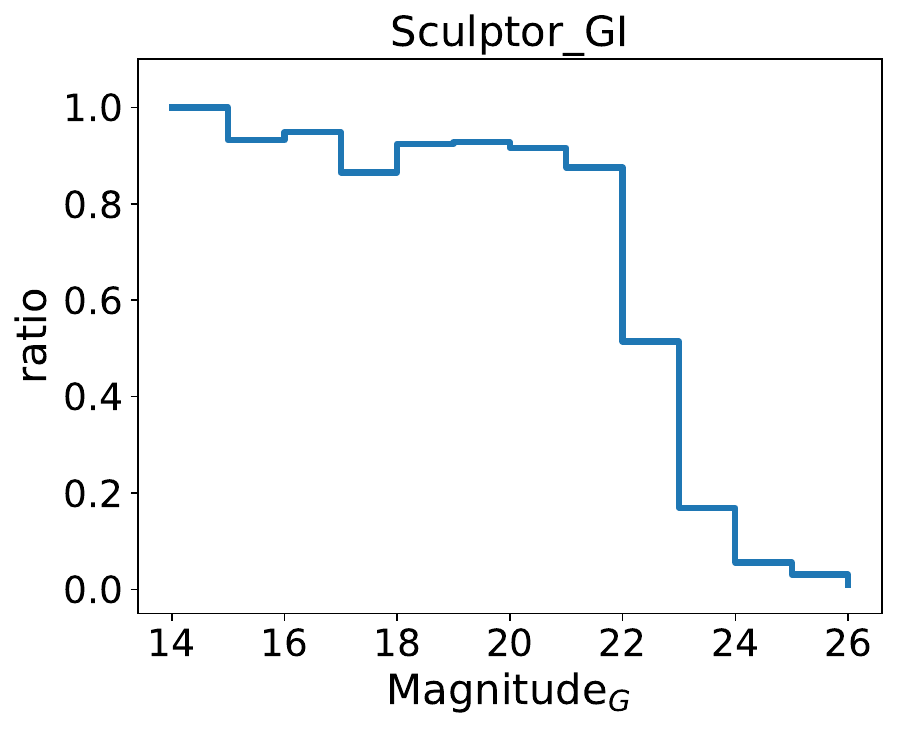}{0.32\textwidth}{(f)}
}
\gridline{\fig{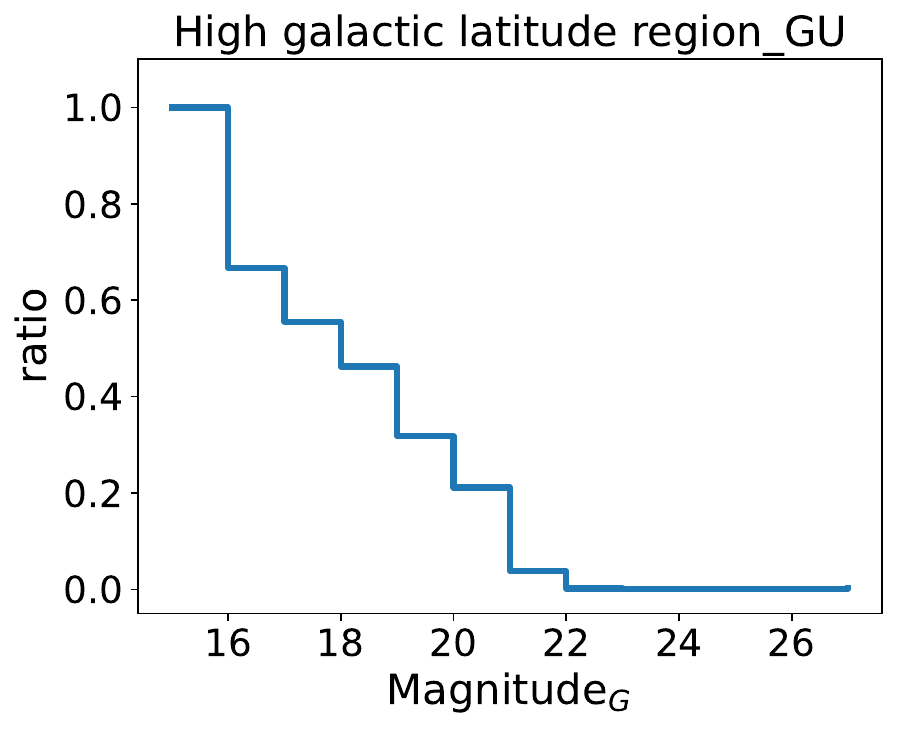}{0.32\textwidth}{(g)}
          \hspace{-5pt}
          \fig{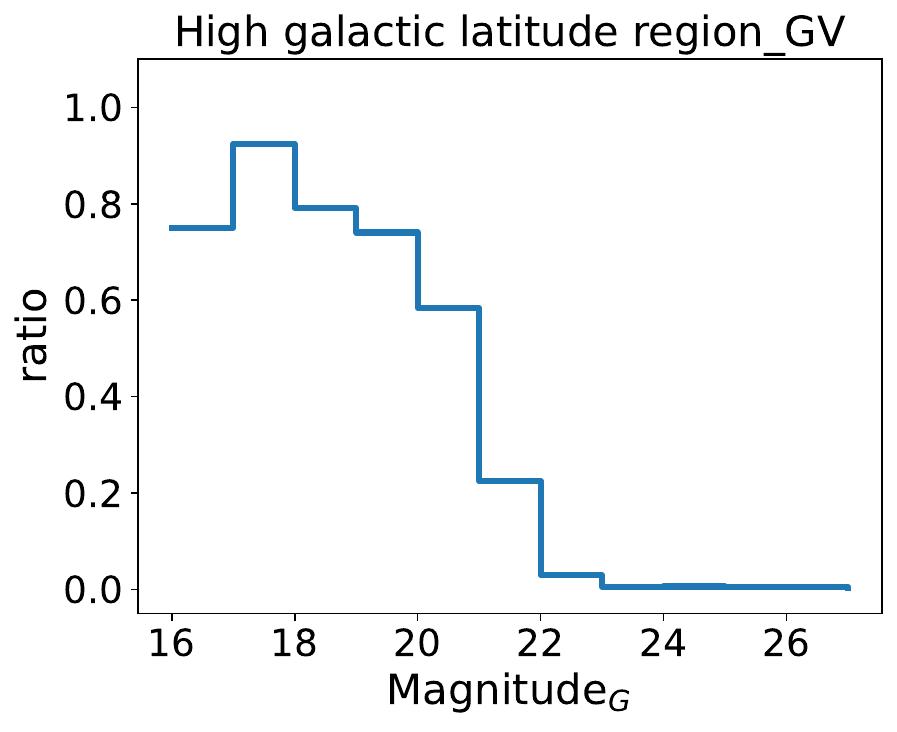}{0.32\textwidth}{(h)}
          \hspace{-5pt}
          \fig{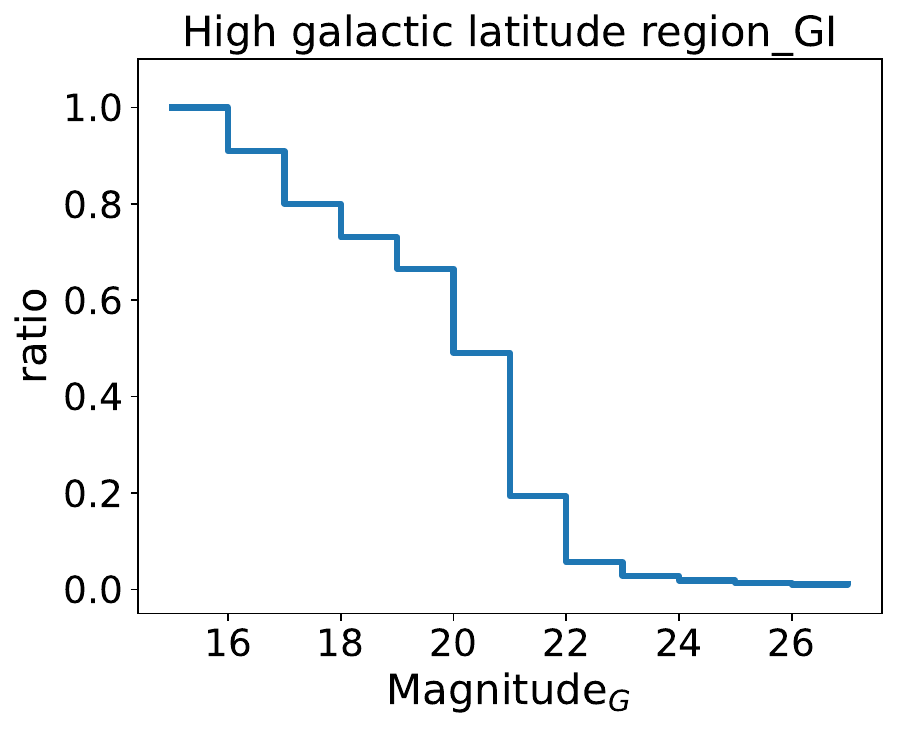}{0.32\textwidth}{(i)}
}
\caption{The detection rates of different magnitudes in the GU, GI, and GV bands for high galactic latitude regions, as well as low-density and medium-density regions of the Galactic and nearby galaxies. \label{fig:predict}}
\end{figure*}

\begin{figure*}
\gridline{
  \fig{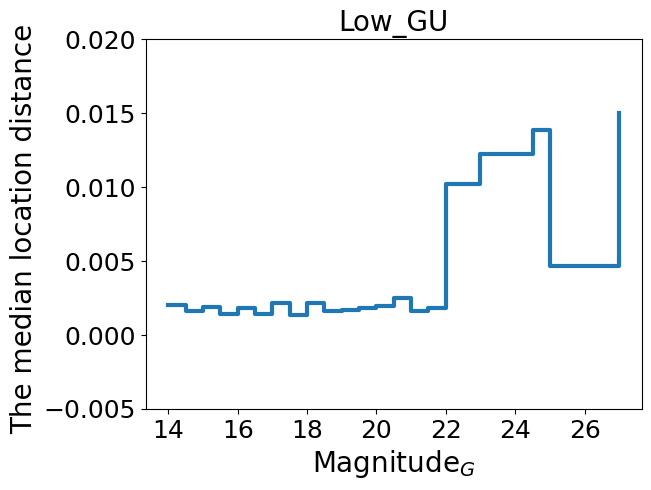}{0.32\textwidth}{(a)}
  \hspace{-5pt} 
  \fig{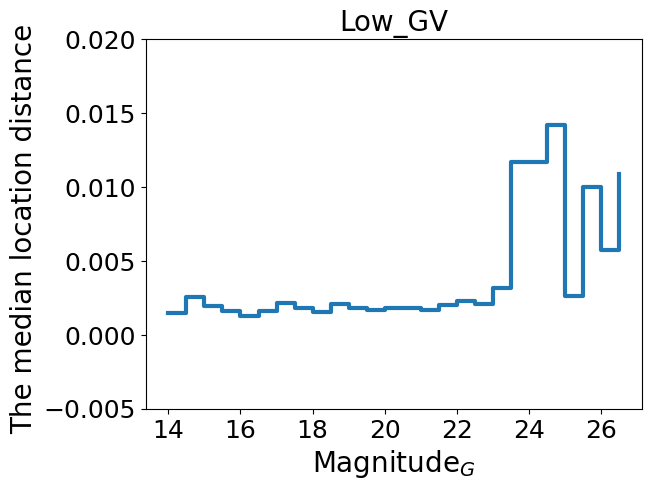}{0.32\textwidth}{(b)}
  \hspace{-5pt}
  \fig{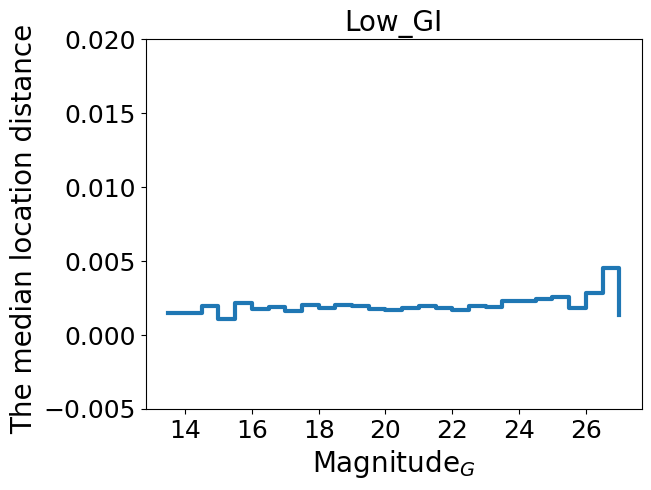}{0.32\textwidth}{(c)}
}
\gridline{
  \fig{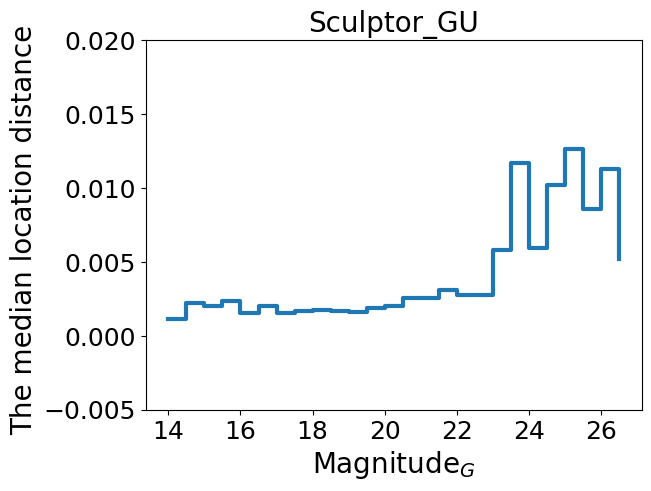}{0.32\textwidth}{(d)}
  \hspace{-5pt}
  \fig{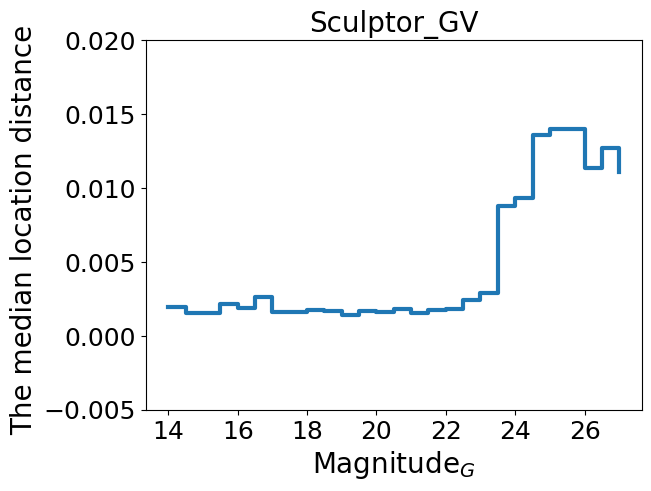}{0.32\textwidth}{(e)}
  \hspace{-5pt}
  \fig{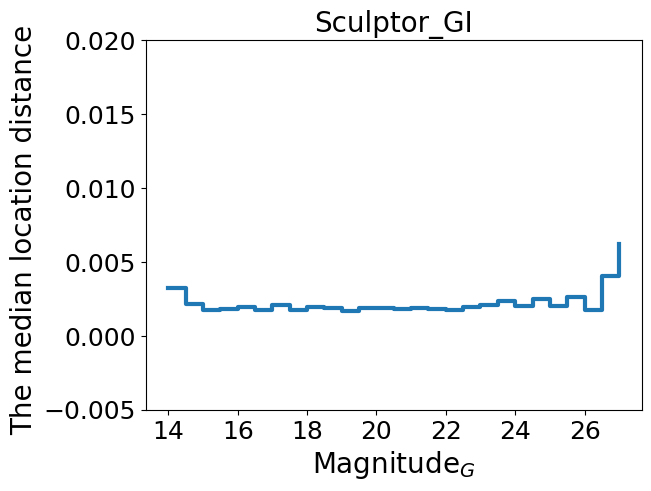}{0.32\textwidth}{(f)}
}
\gridline{\fig{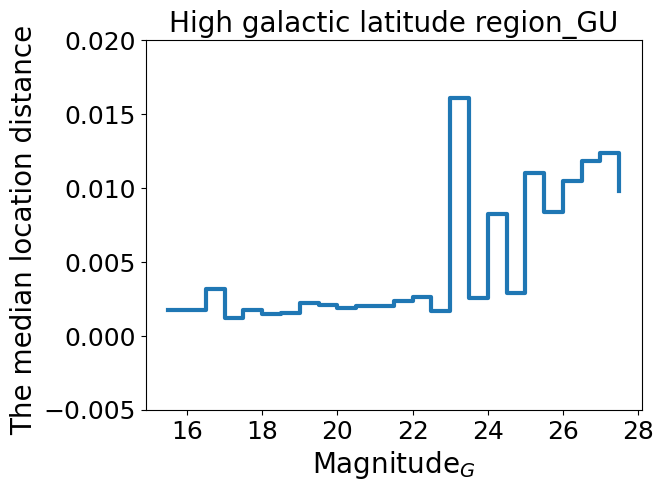}{0.32\textwidth}{(g)}
          \hspace{-5pt}
          \fig{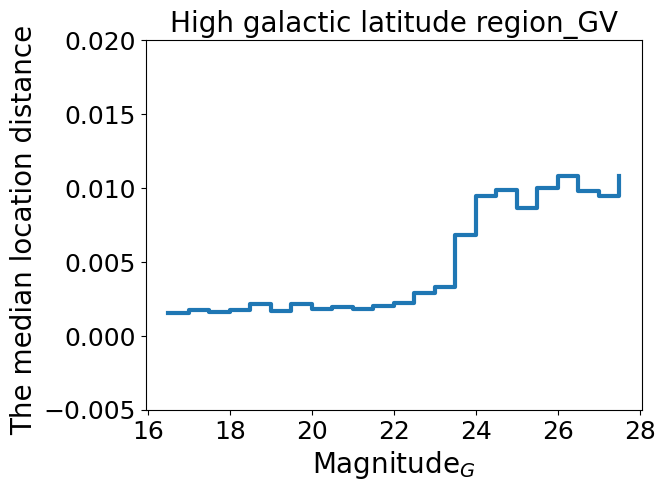}{0.32\textwidth}{(h)}
          \hspace{-5pt}
          \fig{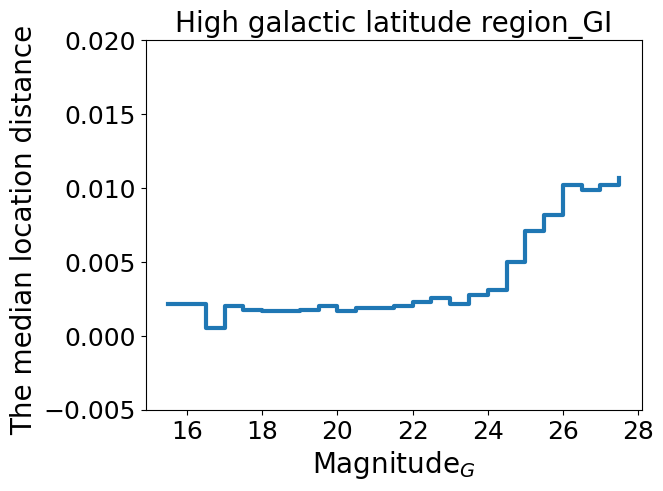}{0.32\textwidth}{(i)}
}
\caption{The median location distance of the zero-order images between the catalog and the YOLO output.
The median location distance is measured in normalized relative coordinates. Given an image size of 640$\times$640 pixels, a distance of 0.01 indicates that the positions of the zero-order images differ by 6.4 pixels between the catalog and the model output.\label{fig:location}}
\end{figure*}

Figure \ref{fig:detected_example} presents the schematic diagrams of slitless spectroscopy image detection. 
To accurately assess the detection rates for different magnitudes in the test set, we matched the coordinates of the zeroth-order images with the source coordinates from the star catalog. 
The star catalog provides the coordinates from direct imaging. 
Using the following dispersion function \citep{Zhang2025slitless,Wei2025overview}, the position coordinates of the zeroth-order images and spectral lines were determined:
\begin{align}
  \Delta y & (x,y,\Delta x)=\nonumber \\
  & (b_{00} + b_{01}x + b_{02}y + b_{03}x^2 + b_{04}xy + b_{05}y^2) + \nonumber \\
  & \Delta x(b_{10} + b_{11}x + b_{12}y + b_{13}x^2 + b_{14}xy + b_{15}y^2)
\end{align}
the position coordinates of the zero-order images and spectral lines can be obtained. 
Here, $x$ and $y$ represent the coordinates from direct imaging, while $x+\Delta x$ and $y+\Delta y$ correspond to the coordinates of the zeroth-order image and first-order spectrum in the detected image. 
Figure \ref{fig:predict} shows the detection rates of sources at different magnitudes in the GU, GI, and GV bands. 
More details can be found in Tables \ref{tab:cat_vs_predict_low}, \ref{tab:cat_vs_predict_Sculptor}, and \ref{tab:cat_vs_predict_HGL}.
Excluding the low-SNR images in the GU band, the model achieved detection rates exceeding 80\% for targets brighter than 21 mag in the GI and GV bands within medium-density regions of the Galactic and nearby galaxies. In low-density regions of these fields, detection rates remained above 80\% for targets brighter than 20 mag in the GI and GV bands. 
For the high galactic latitude region, the model achieved detection rates exceeding 70\% for targets brighter than 18 mag in the GI and GV bands.
And the median location distance of the zero-order images between the catalog and the YOLO output is shown in Figure \ref{fig:location}.
The detection rate and the median location distance presented here were obtained by matching the coordinates of the zero-order images output by the model with the coordinates of the zero-order images provided in the catalog (specifically, by converting the coordinates of the direct imaging given in the catalog using Equation (5)). 
We did not evaluate the detection rate or median location error for spectral lines because their positions and bounding boxes, 
derived from the star catalog and Equation (5), are affected by field distortions. 
These distortions vary across the image, lack a precise analytical description, and are instead approximated through interpolation of optical simulation data, 
with maximum pixel offsets reaching up to 200 pixels \citep{Zhang2025slitless,Wei2025overview}.
Therefore, we cannot obtain accurate magnitude values and positional coordinates for the spectral lines as we can for the zero-order images.

\begin{deluxetable*}{llllllllll}
\tablecaption{The number of cataloged and detected sources in low-density region. \label{tab:cat_vs_predict_low}}
\tablewidth{\textwidth}
\tablehead{
\colhead{Mag} & 
\multicolumn{3}{c}{GU} & 
\multicolumn{3}{c}{GI} & 
\multicolumn{3}{c}{GV} \\
\colhead{} & 
\colhead{Cat} & \colhead{Pre} & \colhead{Rates} &
\colhead{Cat} & \colhead{Pre} & \colhead{Rates} &
\colhead{Cat} & \colhead{Pre} & \colhead{Rates} 
}
\startdata
13 & 0 & 0 & -- & 2 & 2 & 100\% & 0 & 0 & --\\
14 & 72 & 60 & 83.33\% & 44 & 40 & 90.91\% & 42 & 40 & 95.24\%\\
15 & 116 & 104 & 89.66\% & 134 & 124 & 92.54\% & 120 & 114 & 95.00\%\\
16 & 200 & 165 & 82.5\% & 198 & 172 & 86.87\% & 240 & 214 & 89.17\%\\
17 & 298 & 227 & 76.17\% & 304 & 276 & 90.79\% & 320 & 276 & 86.25\%\\
18 & 452 & 261 & 57.74\% & 474 & 431 & 90.93\% & 374 & 338 & 90.37\%\\
19 & 608 & 212 & 34.87\% & 572 & 516 & 90.21\% & 574 & 534 & 93.03\%\\
20 & 726 & 132 & 18.18\% & 810 & 724 & 89.38\% & 762 & 680 & 89.24\%\\
21 & 1300 & 38 & 2.92\% & 1218 & 964 & 79.15\% & 1162 & 647 & 55.68\%\\
22 & 1716 & 6 & 0.35\% & 1868 & 921 & 49.30\% & 1802 & 162 & 8.99\%\\
23 & 0 & 0 & -- & 1856 & 225 & 12.12\% & 1768 & 3 & 0.17\%\\
24 & 2116 & 4 & 0.19\% & 1982 & 59 & 2.98\% & 2084 & 2 & 0.1\%\\
25 & 0 & 0 & -- & 2440 & 30 & 1.23\% & 2380 & 12 & 0.5\%\\
26 & 3074 & 1 & 0.03\% & 3140 & 17 & 0.54\% & 2916 & 5 & 0.17\%\\
\enddata
\end{deluxetable*}

\begin{deluxetable*}{llllllllll}
\tablecaption{The number of cataloged and detected sources in medium-density region. \label{tab:cat_vs_predict_Sculptor}}
\tablewidth{0pt}
\tablehead{
\colhead{Mag} & 
\multicolumn{3}{c}{GU} & 
\multicolumn{3}{c}{GI} & 
\multicolumn{3}{c}{GV} \\
\colhead{} & 
\colhead{Cat} & \colhead{Pre} & \colhead{Rates} &
\colhead{Cat} & \colhead{Pre} & \colhead{Rates} &
\colhead{Cat} & \colhead{Pre} & \colhead{Rates} 
}
\startdata
14 & 70 & 71 & 100\% & 48 & 48 & 100\% & 52 & 52 & 100\% \\
15 & 72 & 72 & 100\% & 88 & 82 & 93.18\% & 80 & 72 & 90.0\% \\
16 & 148 & 130 & 87.84\% & 136 & 129 & 94.85\% & 140 & 131 & 93.57\% \\
17 & 212 & 190 & 89.62\% & 178 & 154 & 86.52\% & 180 & 156 & 86.67\% \\
18 & 968 & 849 & 87.71\% & 318 & 294 & 92.45\% & 622 & 560 & 90.03\% \\
19 & 2070 & 1376 & 66.47\% & 552 & 512 & 92.75\% & 1284 & 1152 & 89.72\% \\
20 & 2268 & 483 & 21.3\% & 680 & 623 & 91.62\% & 1474 & 1332 & 90.37\% \\
21 & 7690 & 130 & 1.69\% & 1484 & 1299 & 87.53\% & 4564 & 3931 & 86.13\% \\
22 & 832 & 10 & 1.2\% & 832 & 428 & 51.44\% & 764 & 268 & 35.08\% \\
23 & 998 & 8 & 0.8\% & 956 & 162 & 16.95\% & 918 & 22 & 2.4\% \\
24 & 1128 & 11 & 0.98\% & 1254 & 70 & 5.58\% & 1086 & 20 & 1.84\% \\
25 & 1628 & 9 & 0.55\% & 1590 & 49 & 3.08\% & 1496 & 5 & 0.33\% \\
26 & 1918 & 9 & 0.47\% & 1954 & 20 & 1.02\% & 1760 & 21 & 1.19\% \\
\enddata
\end{deluxetable*}

\begin{deluxetable*}{llllllllll}
\tablecaption{The number of cataloged and detected sources in the high Galactic latitude region. \label{tab:cat_vs_predict_HGL}}
\tablewidth{0pt}
\tablehead{
\colhead{Mag} & 
\multicolumn{3}{c}{GU} & 
\multicolumn{3}{c}{GI} & 
\multicolumn{3}{c}{GV} \\
\colhead{} & 
\colhead{Cat} & \colhead{Pre} & \colhead{Rates} &
\colhead{Cat} & \colhead{Pre} & \colhead{Rates} &
\colhead{Cat} & \colhead{Pre} & \colhead{Rates} 
}
\startdata
15 & 1 & 1 & 100\% & 1 & 1 & 100\% & 0 & 0 & -- \\
16 & 3 & 2 & 66.67\% & 11 & 10 & 90.91\% & 4 & 3 & 75.0\% \\
17 & 45 & 25 & 55.56\% & 65 & 52 & 80.0\% & 66 & 61 & 92.42\% \\
18 & 80 & 37 & 46.25\% & 204 & 149 & 73.04\% & 153 & 121 & 79.08\% \\
19 & 214 & 68 & 31.78\% & 508 & 338 & 66.54\% & 320 & 237 & 74.06\% \\
20 & 515 & 109 & 21.17\% & 1125 & 552 & 49.07\% & 685 & 400 & 58.39\% \\
21 & 1629 & 62 & 3.81\% & 2396 & 463 & 19.32\% & 1629 & 366 & 22.47\% \\
22 & 4073 & 9 & 0.22\% & 4727 & 268 & 5.67\% & 4153 & 122 & 2.94\% \\
23 & 6955 & 7 & 0.1\% & 7820 & 220 & 2.81\% & 7068 & 41 & 0.58\% \\
24 & 9448 & 15 & 0.16\% & 11371 & 213 & 1.87\% & 9669 & 59 & 0.61\% \\
25 & 12899 & 21 & 0.16\% & 16327 & 222 & 1.36\% & 13641 & 79 & 0.58\% \\
26 & 19132 & 29 & 0.15\% & 25416 & 259 & 1.02\% & 20375 & 103 & 0.51\% \\
27 & 13609 & 25 & 0.18\% & 18165 & 193 & 1.06\% & 14636 & 63 & 0.43\% \\
\enddata
\end{deluxetable*}

In the CSST data processing pipeline, the preprocessing stage incorporates cosmic ray removal procedures. However, complete elimination cannot be guaranteed. 
Consequently, our training set intentionally includes some images with cosmic rays. 
These cosmic rays exhibit morphological similarities to zero-order images in slitless spectral data (though with different PSF profiles), making them potentially susceptible to confusion with actual zero-order signals. 
Furthermore, we have conducted comparative evaluations on the test set to analyze the impact of cosmic rays on model performance specifically regarding zero-order image detection (see Table \ref{tab:CM_discussion}).
We can observe that the presence of cosmic rays slightly degrades the model's performance.

\begin{deluxetable*}{lcccccc}
\tablecaption{The Impact of Cosmic Ray on Model Performance Across Different Bands \label{tab:CM_discussion}}
\tablewidth{0pt}
\tablehead{
\colhead{Band} & 
\multicolumn{3}{c}{With CR} & 
\multicolumn{3}{c}{Without CR} \\
\colhead{} & 
\colhead{Precision} & \colhead{Recall} & \colhead{F1-Score} & 
\colhead{Precision} & \colhead{Recall} & \colhead{F1-Score} 
}
\startdata 
GU    & 0.73      & 0.23   & 0.35     & 0.95      & 0.29   & 0.44     \\
GV    & 0.85      & 0.82   & 0.84     & 0.98      & 0.86   & 0.92     \\
GI    & 0.84      & 0.87   & 0.85     & 0.96      & 0.88   & 0.92     \\
\enddata
\begin{tablenotes}
\item[] Note: 
\begin{enumerate}
  \item \textit{With CR} represents images containing cosmic rays, and \textit{Without CR} represents images without cosmic rays.
  \item The recall in the GU band is very low because the GU band images have a very low SNR, making it difficult for the model to identify sources.
\end{enumerate}
\end{tablenotes}
\end{deluxetable*}

\section{DISCUSSION} \label{sec:discussion}

In this section, we discussed the performance of our source detection model for CSST slitless spectroscopy. 
We discussed dataset-related considerations, evaluated the impact of image segmentation strategies, identified key challenges such as low-SNR regions and extended galaxy morphology that lead to missed detections, and assessed the model's robustness against a variable ratio of zeroth-order images to spectra. 
We also compared the performance differences between our method and traditional approaches.

Our dataset comprises three distinct sky regions—high galactic latitude fields (containing both galaxies and stars), 
medium-density fields at the Galactic and nearby galaxies (containing only stars), 
and low-density fields at the Galactic and nearby galaxies (containing only stars). 
The simulated data for low-density fields at the Galactic and nearby galaxies represent the most fundamental and easily identifiable scenario. 
Beyond this, we aimed to enable the model to handle regions with significant spectral crowding, which led us to incorporate simulated data from medium-density fields at the Galactic and nearby galaxies. 
Finally, to ensure the model can also identify galaxies, we included simulated data from high galactic latitude fields, as these regions contain a substantial population of galaxies.
Beyond stars and galaxies, rarer objects such as nebulae and active galactic nuclei may also be present—objects not considered in the current study. 
Although our model does not include AGNs, since our work focuses on the estimation of source numbers and counts, whether it is an AGN or a typical galaxy, it actually manifests as an extended source spectrum in slitless spectroscopy. This does not affect our learning outcomes, but in real observations, such a model still needs to be tested. 
Future work will need to further refine the dataset by incorporating galaxies across different sky regions, additional rare object types, and a wider range of density conditions.

During training and testing, each slitless spectroscopy image was divided into 16 equal sections by resampling them to 640$\times$640 pixels. 
This approach can cause some spectra to be truncated at the segment boundaries, potentially leading to partial recognition (e.g., only half of a spectrum) 
when the segmented images are reassembled into a complete image along the junction lines. 
To address this issue, some researchers have adopted a method of horizontal slicing followed by vertical stretching to avoid such boundary artifacts \citep{Zhou_2025}.
The authors first slice the original image into ten horizontal strips with a 62-pixel overlap between neighboring strips, preventing targets from being truncated exactly at the edges. Each strip is then stretched vertically to a square format to meet YOLO’s input requirement. Since the spectra are elongated horizontally, this horizontal slicing preserves their full extent and thereby eliminates boundary artifacts.
However, stretching cropped images into square dimensions can introduce geometric distortions, potentially leading to errors in subsequent spectral extraction or morphological analysis. In the case of zero-order images, cropping and resizing may be particularly unsuitable, as their inherently small size increases the risk of losing critical details. Future work should therefore consider adapting cropping strategies to the specific scientific objectives of each analysis.

In slitless spectroscopy images, in addition to the first-order spectra, higher-order spectra such as second-order spectra occasionally appear. 
However, the first-order spectra generally have much higher energy than the higher-order ones. 
Our goal is to detect the zero-order images and first-order spectra of stars and galaxies.
Typically, a single ``source" manifests as one zero-order image and one first-order spectrum in the slitless spectroscopic images. We need to count the number of sources (i.e., the number of zero-order images) and determine the positions of both the zero-order images and their corresponding first-order spectra generated by these sources.
When using CVAT for annotation, we primarily label zero-order images and first-order spectra, without annotating spectral lines of other orders. 
For most sources, the majority of energy is concentrated in the first-order spectra, while spectral lines of other orders are typically not visible. Although brighter sources may exhibit visible spectral lines of other orders, these are usually distinct from first-order spectra—spectral lines of other orders tend to be broader and easier to distinguish.
Nevertheless, occasional confusion between first-order and higher-order spectra may have occurred, leading to annotations that include not only first-order spectra but also a small number of second-order spectra. A similar issue may arise during detection as well. 
Therefore, when calculating the number of sources and detection rates, we used the detected zeroth-order images as the basis for the statistics.

The energy of zeroth-order images accounts for only about 10\% of the total incident energy \citep{Zhan2021}, 
which results in a lower SNR for zeroth-order images compared to spectral lines. 
The lower SNR can be attributed to the fact that the signal strength of zeroth-order images is weaker, making them more susceptible to noise interference. 
Zeroth-order images, though characterized by lower energy and SNR, are indispensable for wavelength calibration as they provide reliable reference positions. 
Wavelength calibration is a vital process in spectroscopy, as it ensures the accurate determination of the wavelengths of detected spectral lines. 
Without zeroth-order images, the detected spectral lines would lack the necessary reference for precise wavelength calibration, 
rendering them essentially useless for further analysis and interpretation.

\begin{figure*}
\gridline{\fig{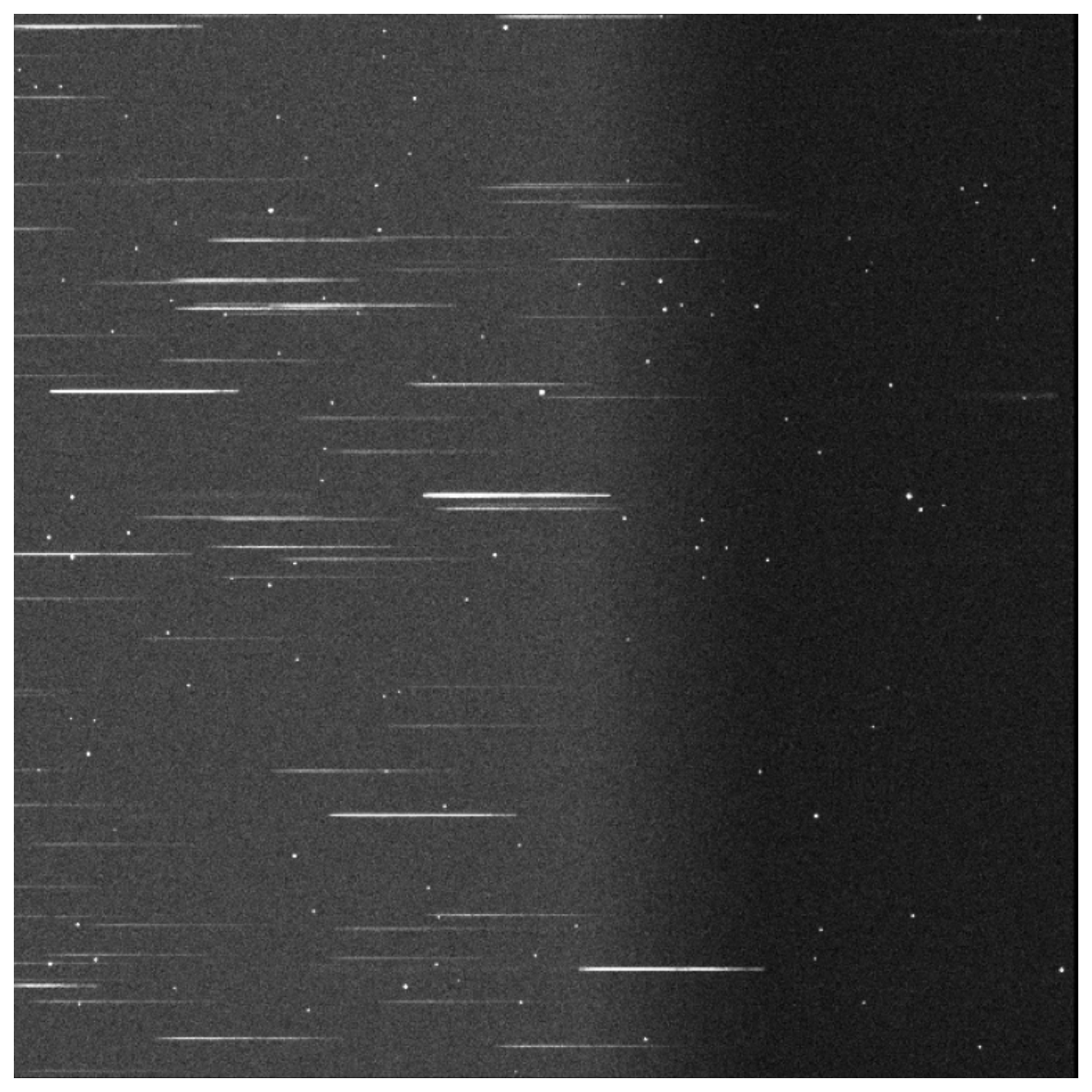}{0.5\textwidth}{(a)}
          \fig{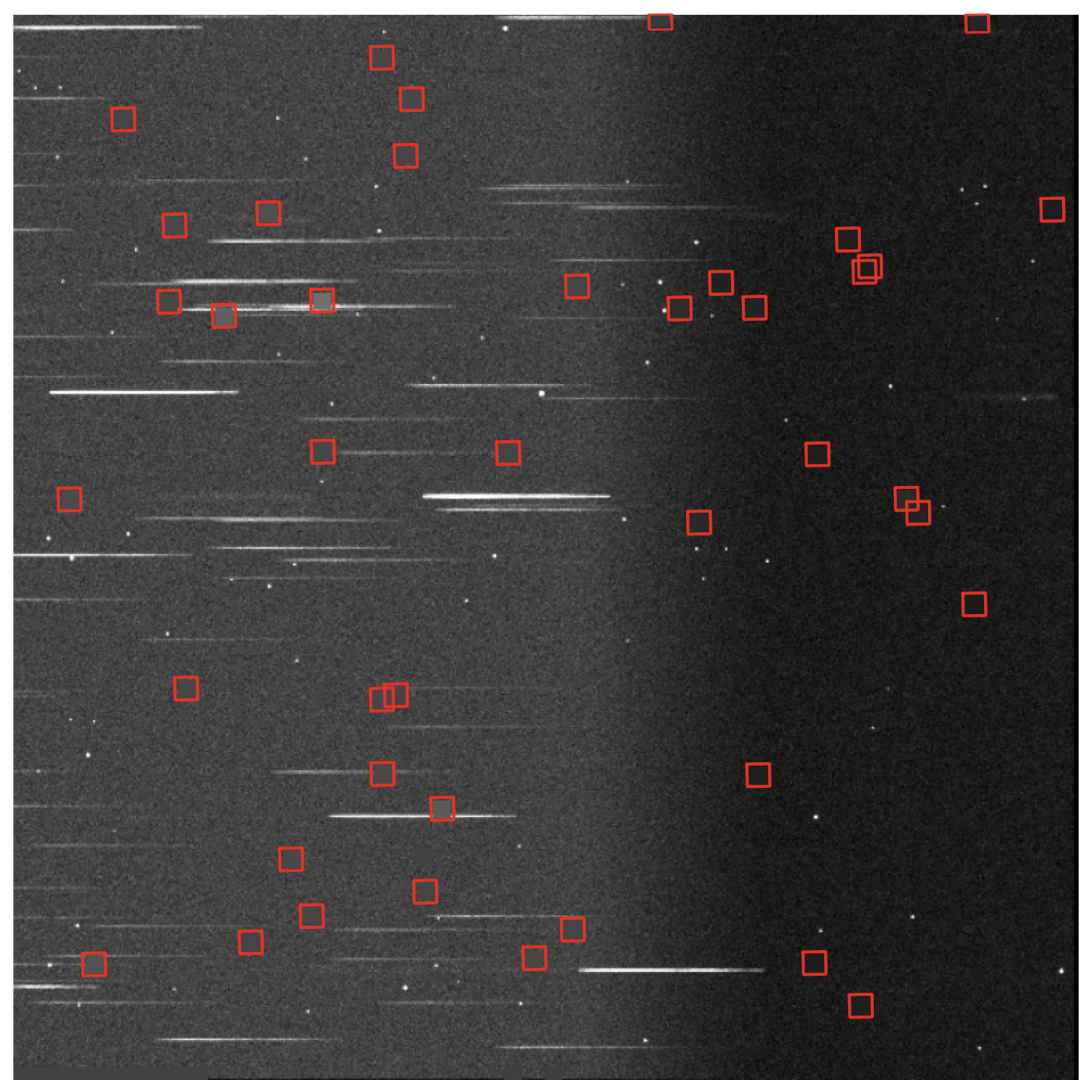}{0.5\textwidth}{(b)}
          }
\gridline{\fig{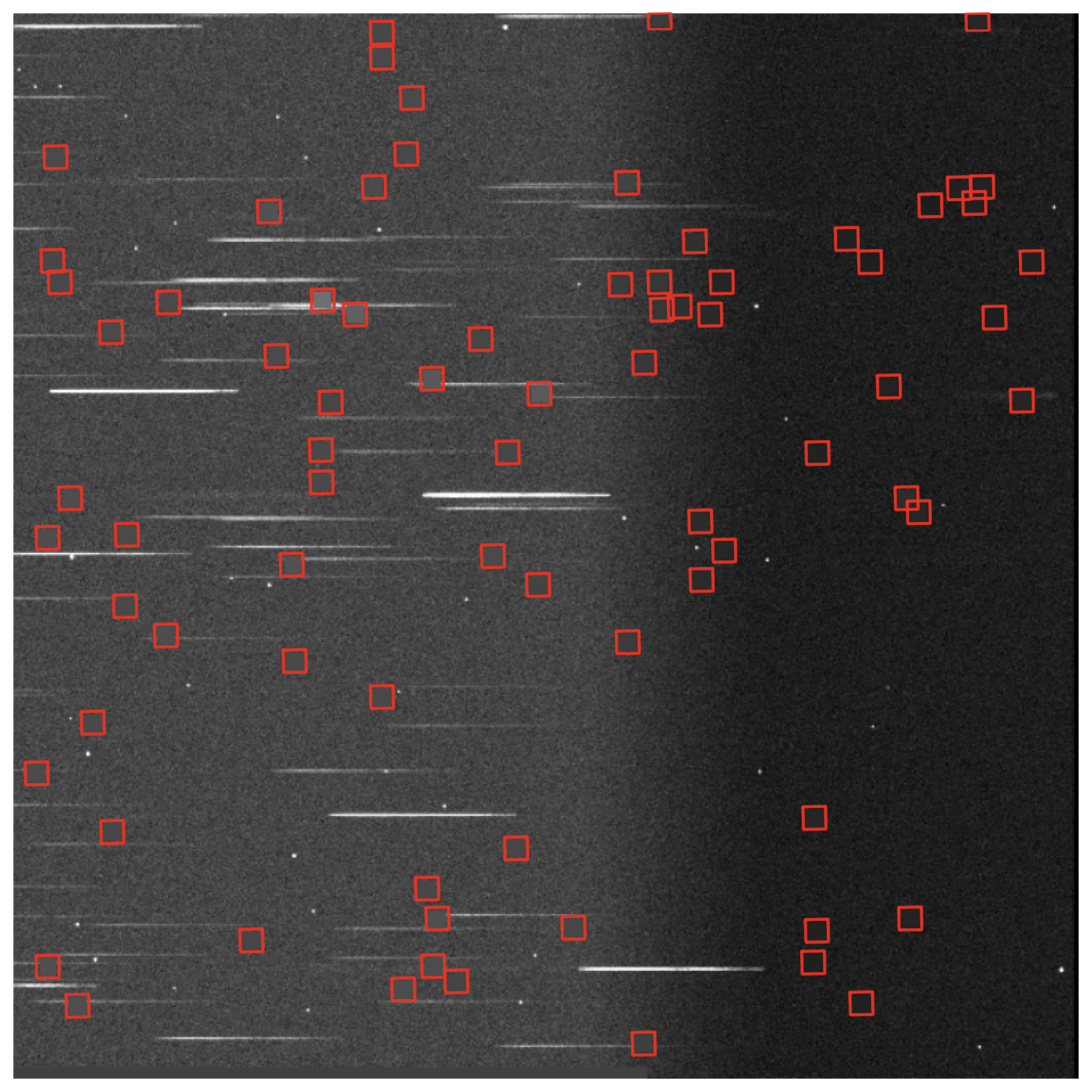}{0.5\textwidth}{(c)}
          \fig{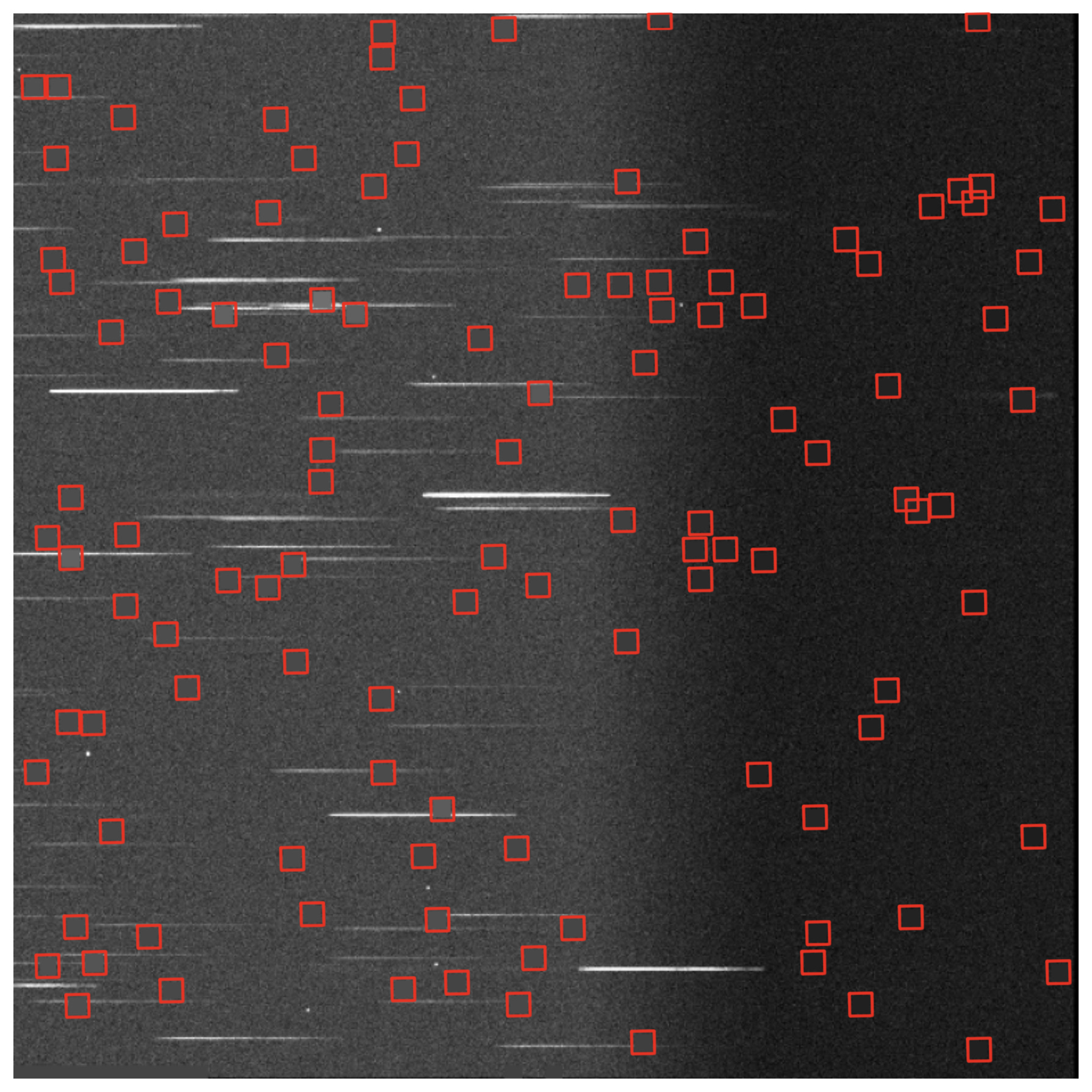}{0.5\textwidth}{(d)}
          }
\caption{Schematic diagrams of slitless spectral images with varying masked ratios applied to the zero-order image: (a) masked ratio = 0; (b) masked ratio = 0.3; (c) masked ratio = 0.6; (d) masked ratio = 0.9. The masked ratio indicates the fraction of the zero-order image that is obscured, while the spectral lines remain fully visible.
The red bounding boxes in the figure indicate the positions of the mask region.
There are fragmentation risks where 1D spectra are split by masks, but in real images, the fragmentation risks, as illustrated in (d), do not occur.
\label{fig:mask_image}
}\end{figure*}

We need to count the number of sources (i.e., the statistical count of zero-order images), but the low energy of zeroth-order images may render some of them undetectable in slitless spectral images, resulting in variability in the ratio of visible zeroth-order images to spectral lines.
To evaluate the model's detection performance of zero-order images under different ratios of zeroth-order to spectral components, we selected a GV-band slitless spectral image from a medium-density Galactic and nearby galaxy region for testing. 
The image was divided into 16 equal segments. For each sub-image, we randomly selected multiple sources and masked their corresponding zeroth-order images 
by replacing the affected pixels with the median values of surrounding pixels (Figure \ref{fig:mask_image}), followed by source detection analysis.
The surrounding pixels are defined by dilating the original target bounding box—which is derived by applying Equation (5) to convert the direct imaging coordinates from the catalog file—with a 1-pixel kernel, thus creating an expanded region that encompasses both the original bounding box surrounding the point target and its immediate one-pixel border.
Figure \ref{fig:mask_predict} shows the detection rates of zero-order images brighter than the 22 magnitude at various masked ratios. 
The results indicate that lowering the ratio of zeroth-order images to spectral lines has little effect on the detection performance of zero-order images, demonstrating the model's robustness and stability under varying conditions.
However, in practical applications, it is still essential to strike a balance between the number of zeroth-order images and spectral lines to ensure reliable wavelength calibration.

\begin{figure}[ht!]
\centering
\includegraphics[width=0.8\textwidth]{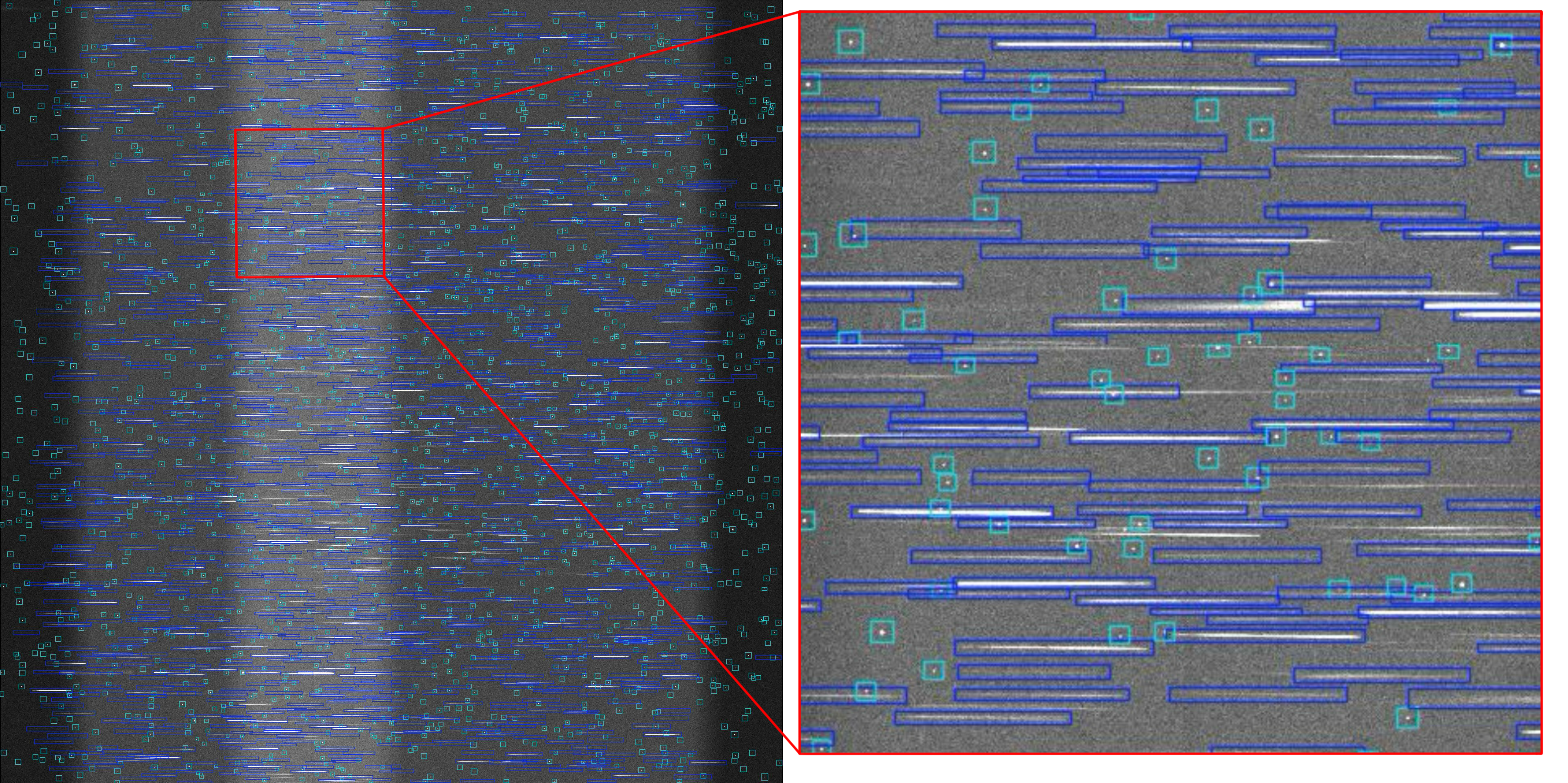}
\caption{The missed detection in slitless spectroscopic images of medium-density regions within the Galaxy and nearby galaxies.\label{fig:missed}
}\end{figure}

Each detector used by the CSST for slitless spectroscopic observations is paired with two gratings that disperse light in opposite directions. 
As a result, the SNR of targets tends to be lower at the grating junctions in the final slitless spectroscopic image. 
Visual inspection of the detection results reveals that, most missed detections occur at these junctions, 
especially for slitless spectroscopic images of medium-density regions within the Galactic and nearby galaxies, as illustrated in Figure \ref{fig:missed}.
And the missed detections account for approximately 12.8\% of the total visual inspection.

\begin{figure*}
\gridline{\fig{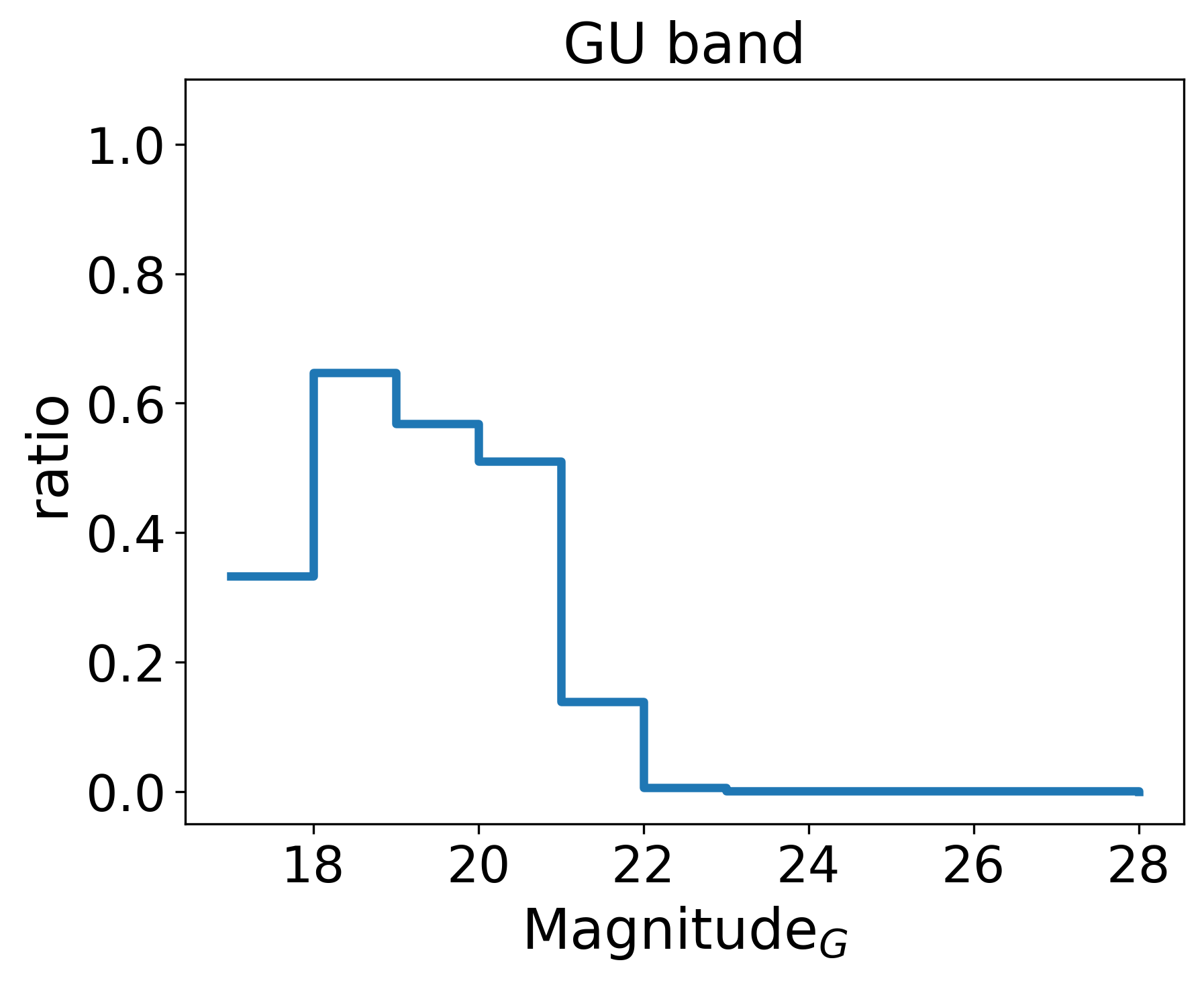}{0.32\textwidth}{(a)}
          \hspace{-5pt}
          \fig{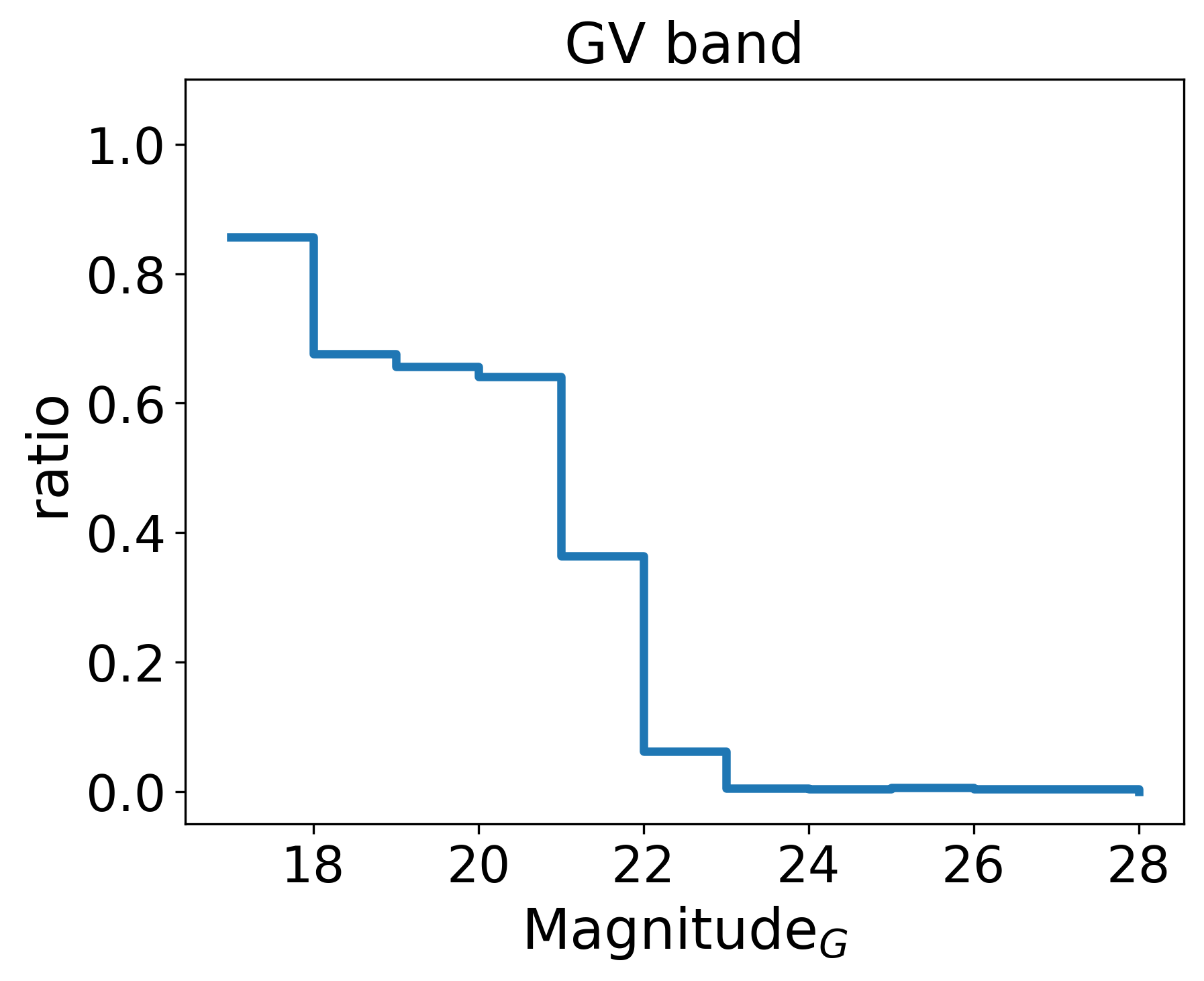}{0.32\textwidth}{(b)}
          \hspace{-5pt}
          \fig{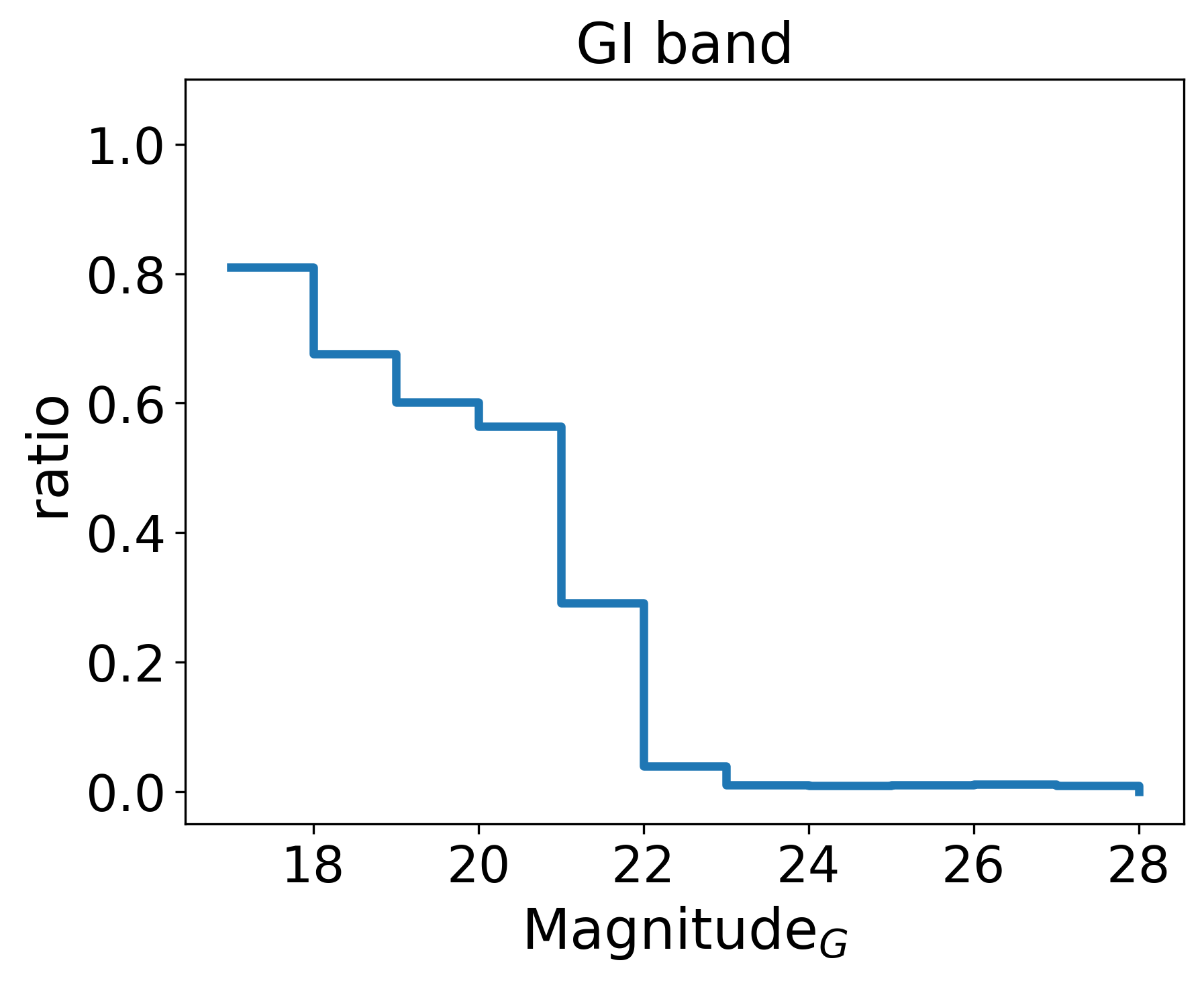}{0.32\textwidth}{(c)}
}
\caption{The detection rates of different magnitudes in the GU, GI, and GV bands for galaxy. \label{fig:galaxy}}
\end{figure*}

\begin{figure}[ht!]
\centering
\includegraphics[width=0.5\textwidth]{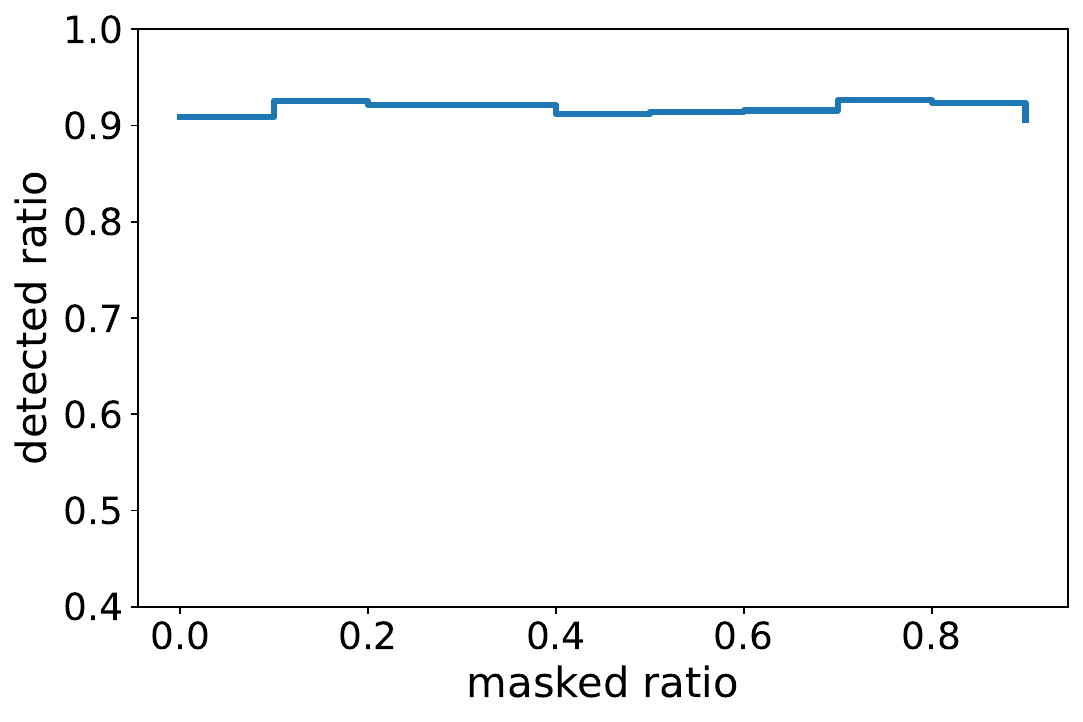}
\caption{Detection rates of targets brighter than the 22 magnitude at various masked ratios for zero-order images. 
The masked ratio indicates the proportion of the zero-order image that is covered, with the spectral lines remaining fully visible.
\label{fig:mask_predict}
}\end{figure}

In slitless spectroscopic images of high galactic latitude regions, galaxies constitute a large proportion of the observed sources. 
In the test set, the ratio of the number of stars to galaxies with magnitude values less than 22 is approximately 0.072.
Unlike stars, galaxies exhibit more diffuse light distributions across the field of view. According to the SNR formula \citep{1981SPIE..290...28M,1992ASPC...23..130G}:
\begin{equation}
  \frac{S}{N} = \frac{S_\star}{\sqrt{S_\star + n_{\text{p}}\cdot\bigl(1+\frac{n_{\text{p}}}{n_{\text{s}}}\bigr)\cdot(S_S+t\cdot dc+R^2+G^2\sigma_f^2)}}
\end{equation}
where $S_\star$ denotes the number of signal electrons; $S_S$ is the average number of sky background electrons per pixel; $n_{\text{p}}$ is the number of pixels used to measure the signal; $n_{\text{s}}$ is the number of pixels used to estimate the background; $t$ represents the exposure time; $dc$ is the dark current; $R$ is the readout noise; $G$ is the gain; and $\sigma_f$ is the quantization error, modeled as an independent uniform random variable in the range 
[-0.5, 0.5],
galaxies with the same flux as stars typically have lower SNRs due to their extended profiles (i.e., a larger $n_{\text{p}}$), leading to reduced visibility in the images, as shown in Figure \ref{fig:detected_example} (c). 
As a result, they are more difficult to detect, leading to lower detection rates at equivalent magnitudes.
We separately examined the model's detection rate for galaxies, as shown in Figure \ref{fig:galaxy}.

For comparative analysis, we also employed the traditional source detection tool Source Extractor alongside the YOLO model, 
with the results summarized in Table \ref{tab:SExtractior}. 
The algorithm of Source Extractor tends to generate a significant number of false detections, yielding far more sources than present in the input catalog.
Source Extractor requires manual parameter tuning, and its detection performance is highly sensitive to these settings, limiting its ability to adapt automatically to diverse datasets. In contrast, deep learning-based approaches, such as YOLO, offer greater robustness and scalability, making them more suitable for large-scale data processing pipelines.

\begin{deluxetable}{lcccccccccccc}
\tablecaption{The results of Source Extractor. \label{tab:SExtractior}}
\tablewidth{0pt}
\tablehead{
\colhead{Region} & \multicolumn{3}{c}{GU} & \multicolumn{3}{c}{GV} & \multicolumn{3}{c}{GI} \\
\colhead{} & 
\colhead{Catalog} & \colhead{SE} & \colhead{YOLO} &
\colhead{Catalog} & \colhead{SE} & \colhead{YOLO} &
\colhead{Catalog} & \colhead{SE} & \colhead{YOLO}
}
\startdata
Low & 2469 & 1725 & 1161 & 2432 & 4616 & 2196 & 2538 & 6418 & 2285 \\
Sculptor & 13498 & 13242 & 3301 & 8396 & 10014 & 7386 & 3484 & 6304 & 3141 \\
HGL & 129 & 9713 & 65 & 223 & 12309 & 185 & 281 & 14728 & 212 \\
\enddata
\begin{tablenotes}
  \item Note: ``Low" represents the low-density region in the Galactic and nearby galaxies, ``Sculptor" represents the medium-density region in the Galactic and nearby galaxies, and ``HGL" represents the high galactic latitude region. The table displays the number of targets in each waveband and sky region, respectively. ``Catalog" represents the number of sources in the catalog, ``SE" represents the number detected by the Source Extractor method, and ``YOLO" represents the number detected by the method used in this study.
\end{tablenotes}
\end{deluxetable}

\section{CONCLUSION} \label{sec:conclution}

This paper presents a source detection model for slitless spectral images based on the YOLOv8 algorithm. 
To enhance the detection of small and densely packed sources, a common challenge in astronomical images, we applied a simple cropping strategy. 
This process divided 1,560 full-frame images into 24,960 sub-images for training and validation, substantially improving performance on crowded targets.
And the main results are summarized as follows:
\begin{enumerate}
  \item The model achieved precise detection on the validation set, with precision/recall of 88.6\%/90.4\% for spectral lines and 87.0\%/80.8\% for zeroth-order images.
  \item On the test set, it maintained $>$80\% detection for sources brighter than 21 mag (medium-density regions) and 20 mag (low-density regions) in the Galactic and nearby galaxy regions, and $>$70\% for sources brighter than 18 mag in high Galactic latitude regions.
  \item Compared to traditional source detection methods, the YOLO approach offers superior performance and a higher level of automation.
\end{enumerate}

The model developed in this study is capable of outputting both the number of sources in the image and their positional coordinates. 
In future work, we plan to explore the possibility of performing spectral extraction based on these coordinates.

\begin{acknowledgments}
     This work was supported by the Strategic Priority Research Program of the Chinese Academy of Sciences, Grant No. XDB0550100. 
     J.L is supported by the National Natural Science Foundation of China (NSFC; grant No. 12273027).
     Man I Lam is supported by the National Natural Science Foundation of China (NSFC; grant No. 12373048).
\end{acknowledgments}

\bibliography{Source-file}{}

\begin{thebibliography}{}
\expandafter\ifx\csname natexlab\endcsname\relax\def\natexlab#1{#1}\fi
\providecommand{\url}[1]{\href{#1}{#1}}
\providecommand{\dodoi}[1]{doi:~\href{http://doi.org/#1}{\nolinkurl{#1}}}
\providecommand{\doeprint}[1]{\href{http://ascl.net/#1}{\nolinkurl{http://ascl.net/#1}}}
\providecommand{\doarXiv}[1]{\href{https://arxiv.org/abs/#1}{\nolinkurl{https://arxiv.org/abs/#1}}}

\bibitem[{A. {Anthore} \& D. {Cornu}(2024){Anthore} \&
  {Cornu}}]{2024eas..conf.1350A}
{Anthore}, A., \& {Cornu}, D. 2024, \bibinfo{title}{{Galaxy detection: Prospect
  for unveiling SKA precursor data with deep learning},} in EAS2024, European
  Astronomical Society Annual Meeting, 1350.
\newblock \url{https://ui.adsabs.harvard.edu/abs/2024eas..conf.1350A}

\bibitem[{E. {Ba{\~n}ados} {et~al.}(2025){Ba{\~n}ados}, {Le Brun},
  {Belladitta}, {Momcheva}, {Stern}, {Wolf}, {Ezziati}, {Mortlock}, {Humphrey},
  {Smart}, {Casewell}, {P{\'e}rez-Garrido}, {Goldman}, {Mart{\'\i}n},
  {Mohandasan}, {Reyl{\'e}}, {Dominguez-Tagle}, {Copin}, {Lusso}, {Matsuoka},
  {McCarthy}, {Ricci}, {Rix}, {Rottgering}, {Schindler}, {Weaver}, {Allaoui},
  {Bedrine}, {Castellano}, {Chabaud}, {Daste}, {Dufresne}, {Gracia-Carpio},
  {K{\"u}mmel}, {Moresco}, {Scodeggio}, {Surace}, {Vibert}, {Balestra},
  {Bonnefoi}, {Caillat}, {Cogato}, {Costille}, {Dusini}, {Ferriol},
  {Franceschi}, {Gillard}, {Jahnke}, {Le Mignant}, {Ligori}, {Medinaceli},
  {Morgante}, {Passalacqua}, {Paterson}, {Pires}, {Sirignano}, {Andika},
  {Atek}, {Barrado}, {Bisogni}, {Conselice}, {Dannerbauer}, {Decarli}, {Dole},
  {Dupuy}, {Feltre}, {Fotopoulou}, {Gillis}, {Lopez Lopez}, {Onoue},
  {Rodighiero}, {Sedighi}, {Shankar}, {Siudek}, {Spinoglio}, {Vergani},
  {Vietri}, {Walter}, {Zamorani}, {Zapatero Osorio}, {Zhang}, {Bethermin},
  {Aghanim}, {Altieri}, {Amara}, {Andreon}, {Baccigalupi}, {Baldi}, {Bardelli},
  {Basset}, {Battaglia}, {Biviano}, {Bonchi}, {Bonino}, {Branchini}, {Brescia},
  {Brinchmann}, {Camera}, {Capobianco}, {Carbone}, {Carretero}, {Casas},
  {Castignani}, {Cavuoti}, {Cimatti}, {Colodro-Conde}, {Congedo}, {Conversi},
  {Courbin}, {Courtois}, {Cropper}, {Cuby}, {Da Silva}, {Degaudenzi}, {De
  Lucia}, {Giorgio}, {Dolding}, {Dubath}, {Duncan}, {Dupac}, {Ealet}, {Farina},
  {Faustini}, {Fourmanoit}, {Frailis}, {Galeotta}, {George}, {Giocoli},
  {Granett}, {Grazian}, {Grupp}, {Guzzo}, {Haugan}, {Hoar}, {Hoekstra},
  {Holmes}, {Hook}, {Hormuth}, {Hornstrup}, {Hudelot}, {Jhabvala}, {Joachimi},
  {Keih{\"a}nen}, {Kermiche}, {Kubik}, {Kuijken}, {Kunz}, {Kurki-Suonio},
  {Lilje}, {Lindholm}, {Lloro}, {Mainetti}, {Maino}, {Maiorano}, {Mansutti},
  {Marggraf}, {Markovic}, {Martinelli}, {Martinet}, {Marulli}, {Massey}, {Mei},
  {Mellier}, {Meneghetti}, {Merlin}, {Meylan}, {Mora}, {Moscardini},
  {Neissner}, {Niemi}, {Nightingale}, {Padilla}, {Paltani}, {Pasian},
  {Pedersen}, {Percival}, {Pettorino}, {Polenta}, {Poncet}, {Popa}, {Pozzetti},
  {Raison}, {Rebolo}, {Renzi}, {Rhodes}, {Riccio}, {Romelli}, {Roncarelli},
  {Rossetti}, {Saglia}, {Sakr}, {Sapone}, {Sartoris}, {Schewtschenko},
  {Schirmer}, {Schneider}, {Schrabback}, {Secroun}, {Sefusatti}, \&
  {Seidel}}]{2025MNRAS.tmp.1224B}
{Ba{\~n}ados}, E., {Le Brun}, V., {Belladitta}, S., {et~al.} 2025,
  \bibinfo{title}{{Euclid: The potential of slitless infrared spectroscopy: A z
  = 5.4 quasar and new ultracool dwarfs<SUP></SUP>},} \mnras,
  \dodoi{10.1093/mnras/staf1274}

\bibitem[{K. {Boyett} {et~al.}(2022){Boyett}, {Mascia}, {Pentericci},
  {Leethochawalit}, {Trenti}, {Brammer}, {Roberts-Borsani}, {Strait}, {Treu},
  {Bradac}, {Glazebrook}, {Acebron}, {Bergamini}, {Calabr{\`o}}, {Castellano},
  {Fontana}, {Grillo}, {Henry}, {Jones}, {Marchesini}, {Mason}, {Mercurio},
  {Morishita}, {Nanayakkara}, {Rosati}, {Scarlata}, {Vanzella}, {Vulcani},
  {Wang}, \& {Willott}}]{2022ApJ...940L..52B}
{Boyett}, K., {Mascia}, S., {Pentericci}, L., {et~al.} 2022,
  \bibinfo{title}{{Early Results from GLASS-JWST. VI. Extreme Rest-optical
  Equivalent Widths Detected in NIRISS Wide Field Slitless Spectroscopy},}
  \apjl, 940, L52, \dodoi{10.3847/2041-8213/ac9f17}

\bibitem[{G. {Brammer}(2019){Brammer}}]{2019ascl.soft05001B}
{Brammer}, G. 2019, {Grizli: Grism redshift and line analysis software},,
  Astrophysics Source Code Library, record ascl:1905.001
  \url{https://ui.adsabs.harvard.edu/abs/2019ascl.soft05001B}

\bibitem[{E. {Carley}(2020){Carley}}]{2020EGUGA..22.5109C}
{Carley}, E. 2020, \bibinfo{title}{{Using supervised machine learning to
  automatically detect type II and III solar radio bursts},} in EGU General
  Assembly Conference Abstracts, EGU General Assembly Conference Abstracts,
  5109, \dodoi{10.5194/egusphere-egu2020-5109}

\bibitem[{R.~G. {Clowes} {et~al.}(1980){Clowes}, {Emerson}, {Smith}, {Wallace},
  {Cannon}, {Savage}, \& {Boksenberg}}]{1980MNRAS.193..415C}
{Clowes}, R.~G., {Emerson}, D., {Smith}, M.~G., {et~al.} 1980,
  \bibinfo{title}{{The use of objective prism plates from the UK Schmidt
  Telescope for low resolution spectrophotometry of quasars},} \mnras, 193,
  415, \dodoi{10.1093/mnras/193.3.415}

\bibitem[{Y. Copin {et~al.}(2025)Copin, Fumana, Mancini, Appleton, Chary,
  Conseil, Faisst, Hemmati, Masters, Scarlata, Scodeggio, Alavi, Carle,
  Casenove, Contini, Das, Gillard, Herzog, \& Scott}]{article}
Copin, Y., Fumana, M., Mancini, C., {et~al.} 2025, \bibinfo{title}{Euclid Quick
  Data Release (Q1). From spectrograms to spectra: The SIR spectroscopic
  processing function,} Astronomy \& Astrophysics,
  \dodoi{10.1051/0004-6361/202554627}

\bibitem[{D. Cornu {et~al.}(2024)Cornu, Salomé, Semelin, Marchal, Freundlich,
  Aicardi, Lu, Sainton, Mertens, Combes, \& Tasse}]{Cornu_2024}
Cornu, D., Salomé, P., Semelin, B., {et~al.} 2024,
  \bibinfo{title}{YOLO-CIANNA: Galaxy detection with deep learning in radio
  data: I. A new YOLO-inspired source detection method applied to the SKAO
  SDC1,} Astronomy \& amp; Astrophysics, 690, A211,
  \dodoi{10.1051/0004-6361/202449548}

\bibitem[{C. Cortes \& V.~N. Vapnik(1995)Cortes \&
  Vapnik}]{Cortes1995SupportVectorN}
Cortes, C., \& Vapnik, V.~N. 1995, \bibinfo{title}{Support-Vector Networks,}
  Machine Learning, 20, 273.
\newblock \url{https://api.semanticscholar.org/CorpusID:52874011}

\bibitem[{N. Dalal \& B. Triggs(2005)Dalal \& Triggs}]{1467360}
Dalal, N., \& Triggs, B. 2005, \bibinfo{title}{Histograms of oriented gradients
  for human detection,} in 2005 IEEE Computer Society Conference on Computer
  Vision and Pattern Recognition (CVPR'05), Vol.~1, 886--893 vol. 1,
  \dodoi{10.1109/CVPR.2005.177}

\bibitem[{R. {Doyon} {et~al.}(2023){Doyon}, {Willott}, {Hutchings},
  {Sivaramakrishnan}, {Albert}, {Lafreni{\`e}re}, {Rowlands}, {Bego{\~n}a
  Vila}, {Martel}, {LaMassa}, {Aldridge}, {Artigau}, {Cameron}, {Chayer},
  {Cook}, {Cooper}, {Darveau-Bernier}, {Dupuis}, {Earnshaw}, {Espinoza},
  {Filippazzo}, {Fullerton}, {Gaudreau}, {Gawlik}, {Goudfrooij}, {Haley},
  {Kammerer}, {Kendall}, {Lambros}, {Ignat}, {Maszkiewicz}, {McColgan},
  {Morishita}, {Ouellette}, {Pacifici}, {Philippi}, {Radica}, {Ravindranath},
  {Rowe}, {Roy}, {Roy}, {Saad}, {Sohn}, {Talens}, {Touahri}, {Thatte},
  {Taylor}, {Vandal}, {Volk}, {Wander}, {Warner}, {Zheng}, {Zhou}, {Abraham},
  {Beaulieu}, {Benneke}, {Ferrarese}, {Jayawardhana}, {Johnstone},
  {Kaltenegger}, {Meyer}, {Pipher}, {Rameau}, {Rieke}, {Salhi}, \&
  {Sawicki}}]{2023PASP..135i8001D}
{Doyon}, R., {Willott}, C.~J., {Hutchings}, J.~B., {et~al.} 2023,
  \bibinfo{title}{{The Near Infrared Imager and Slitless Spectrograph for the
  James Webb Space Telescope. I. Instrument Overview and In-flight
  Performance},} \pasp, 135, 098001, \dodoi{10.1088/1538-3873/acd41b}

\bibitem[{D. {Engels} {et~al.}(1988){Engels}, {Groote}, {Hagen}, \&
  {Reimrs}}]{1988ASPC....2..143E}
{Engels}, D., {Groote}, D., {Hagen}, H.~J., \& {Reimrs}, D. 1988,
  \bibinfo{title}{{The Hamburg Quasar Survey},} in Astronomical Society of the
  Pacific Conference Series, Vol.~2, Optical Surveys for Quasars, ed.
  P.~{Osmer}, M.~M. {Phillips}, R.~{Green}, \& C.~{Foltz}, 143.
\newblock \url{https://ui.adsabs.harvard.edu/abs/1988ASPC....2..143E}

\bibitem[{D. {Engels} {et~al.}(1998){Engels}, {Hagen}, {Cordis}, {Koehler},
  {Wisotzki}, \& {Reimers}}]{1998A&AS..128..507E}
{Engels}, D., {Hagen}, H.~J., {Cordis}, L., {et~al.} 1998, \bibinfo{title}{{The
  Hamburg Quasar Survey. II. A first list of 121 quasars},} \aaps, 128, 507,
  \dodoi{10.1051/aas:1998390}

\bibitem[{ {Euclid Collaboration} {et~al.}(2025{\natexlab{a}}){Euclid
  Collaboration}, {Mellier}, {Abdurro'uf}, {Acevedo Barroso}, {Ach{\'u}carro},
  {Adamek}, {Adam}, {Addison}, {Aghanim}, {Aguena}, {Ajani}, {Akrami},
  {Al-Bahlawan}, {Alavi}, {Albuquerque}, {Alestas}, {Alguero}, {Allaoui},
  {Allen}, {Allevato}, {Alonso-Tetilla}, {Altieri}, {Alvarez-Candal}, {Alvi},
  {Amara}, {Amendola}, {Amiaux}, {Andika}, {Andreon}, {Andrews}, {Angora},
  {Angulo}, {Annibali}, {Anselmi}, {Anselmi}, {Arcari}, {Archidiacono},
  {Aric{\`o}}, {Arnaud}, {Arnouts}, {Asgari}, {Asorey}, {Atayde}, {Atek},
  {Atrio-Barandela}, {Aubert}, {Aubourg}, {Auphan}, {Auricchio}, {Aussel},
  {Aussel}, {Avelino}, {Avgoustidis}, {Avila}, {Awan}, {Azzollini},
  {Baccigalupi}, {Bachelet}, {Bacon}, {Baes}, {Bagley}, {Bahr-Kalus},
  {Balaguera-Antolinez}, {Balbinot}, {Balcells}, {Baldi}, {Baldry}, {Balestra},
  {Ballardini}, {Ballester}, {Balogh}, {Ba{\~n}ados}, {Barbier}, {Bardelli},
  {Baron}, {Barreiro}, {Barrena}, {Barriere}, {Barros}, {Barthelemy},
  {Bartolo}, {Basset}, {Battaglia}, {Battisti}, {Baugh}, {Baumont},
  {Bazzanini}, {Beaulieu}, {Beckmann}, {Belikov}, {Bel}, {Bellagamba}, {Bella},
  {Bellini}, {Benabed}, {Bender}, {Benevento}, {Bennett}, {Benson},
  {Bergamini}, {Bermejo-Climent}, {Bernardeau}, {Bertacca}, {Berthe},
  {Berthier}, {Bethermin}, {Beutler}, {Bevillon}, {Bhargava}, {Bhatawdekar},
  {Bianchi}, {Bisigello}, {Biviano}, {Blake}, {Blanchard}, {Blazek}, {Blot},
  {Bosco}, {Bodendorf}, {Boenke}, {B{\"o}hringer}, {Boldrini}, {Bolzonella},
  {Bonchi}, {Bonici}, {Bonino}, {Bonino}, {Bonvin}, {Bon}, {Booth}, {Borgani},
  {Borlaff}, {Borsato}, {Bose}, {Botticella}, {Boucaud}, {Bouche}, {Boucher},
  {Boutigny}, {Bouvard}, {Bouwens}, {Bouy}, {Bowler}, {Bozza}, {Bozzo},
  {Branchini}, {Brando}, {Brau-Nogue}, {Brekke}, {Bremer}, {Brescia}, {Breton},
  {Brinchmann}, {Brinckmann}, {Brockley-Blatt}, {Brodwin}, {Brouard}, {Brown},
  {Bruton}, {Bucko}, {Buddelmeijer}, {Buenadicha}, {Buitrago}, {Burger},
  {Burigana}, {Busillo}, {Busonero}, {Cabanac}, {Cabayol-Garcia}, {Cagliari},
  {Caillat}, {Caillat}, {Calabrese}, {Calabro}, {Calderone}, {Calura}, {Camacho
  Quevedo}, {Camera}, {Campos}, {Ca{\~n}as-Herrera}, {Candini}, {Cantiello},
  {Capobianco}, {Cappellaro}, {Cappelluti}, {Cappi}, {Caputi}, {Cara},
  {Carbone}, {Cardone}, {Carella}, {Carlberg}, {Carle}, {Carminati}, {Caro},
  {Carrasco}, {Carretero}, {Carrilho}, {Carron Duque}, \&
  {Carry}}]{2025A&A...697A...1E}
{Euclid Collaboration}, {Mellier}, Y., {Abdurro'uf}, {et~al.}
  2025{\natexlab{a}}, \bibinfo{title}{{Euclid: I. Overview of the Euclid
  mission},} \aap, 697, A1, \dodoi{10.1051/0004-6361/202450810}

\bibitem[{ {Euclid Collaboration} {et~al.}(2025{\natexlab{b}}){Euclid
  Collaboration}, {Aussel}, {Tereno}, {Schirmer}, {Alguero}, {Altieri},
  {Balbinot}, {de Boer}, {Casenove}, {Corcho-Caballero}, {Furusawa},
  {Furusawa}, {Hudson}, {Jahnke}, {Libet}, {Macias-Perez}, {Masoumzadeh},
  {Mohr}, {Odier}, {Scott}, {Vassallo}, {Verdoes Kleijn}, {Zacchei}, {Aghanim},
  {Amara}, {Andreon}, {Auricchio}, {Awan}, {Azzollini}, {Baccigalupi}, {Baldi},
  {Balestra}, {Bardelli}, {Basset}, {Battaglia}, {Belikov}, {Bender},
  {Biviano}, {Bonchi}, {Bonino}, {Branchini}, {Brescia}, {Brinchmann},
  {Camera}, {Ca{\~n}as-Herrera}, {Capobianco}, {Carbone}, {Cardone},
  {Carretero}, {Casas}, {Castander}, {Castellano}, {Castignani}, {Cavuoti},
  {Chambers}, {Cimatti}, {Colodro-Conde}, {Congedo}, {Conselice}, {Conversi},
  {Copin}, {Courbin}, {Courtois}, {Cropper}, {Cuby}, {Da Silva}, {da Silva},
  {Degaudenzi}, {de Jong}, {De Lucia}, {Di Giorgio}, {Dinis}, {Dolding},
  {Dole}, {Douspis}, {Dubath}, {Duncan}, {Dupac}, {Dusini}, {Ealet},
  {Escoffier}, {Fabricius}, {Farina}, {Farinelli}, {Faustini}, {Ferriol},
  {Fotopoulou}, {Fourmanoit}, {Frailis}, {Franceschi}, {Franzetti}, {Galeotta},
  {George}, {Gillard}, {Gillis}, {Giocoli}, {G{\'o}mez-Alvarez},
  {Gracia-Carpio}, {Granett}, {Grazian}, {Grupp}, {Guzzo}, {Gwyn}, {Haugan},
  {Herent}, {Hoar}, {Hoekstra}, {Holliman}, {Holmes}, {Hook}, {Hormuth},
  {Hornstrup}, {Hudelot}, {Ili{\'c}}, {Jhabvala}, {Joachimi}, {Keih{\"a}nen},
  {Kermiche}, {Kiessling}, {Kubik}, {Kuijken}, {K{\"u}mmel}, {Kunz},
  {Kurki-Suonio}, {Lahav}, {Le Boulc'h}, {Le Brun}, {Le Mignant}, {Liebing},
  {Ligori}, {Lilje}, {Lindholm}, {Lloro}, {Mainetti}, {Maino}, {Maiorano},
  {Mansutti}, {Marcin}, {Marggraf}, {Markovic}, {Martinelli}, {Martinet},
  {Marulli}, {Massey}, {Maurogordato}, {McCracken}, {Medinaceli}, {Mei},
  {Melchior}, {Mellier}, {Meneghetti}, {Merlin}, {Meylan}, {Mora}, {Moresco},
  {Morris}, {Moscardini}, {Mourre}, {Nakajima}, {Neissner}, {Nichol}, {Niemi},
  {Nightingale}, {Nutma}, {Padilla}, {Paltani}, {Pasian}, {Peacock},
  {Pedersen}, {Percival}, {Pettorino}, {Pires}, {Polenta}, {Pollack}, {Poncet},
  {Popa}, {Pozzetti}, {Racca}, {Raison}, {Rebolo}, {Renzi}, {Rhodes}, {Riccio},
  {Rix}, {Romelli}, {Roncarelli}, {Rossetti}, {Rusholme}, {Saglia}, {Sakr},
  {S{\'a}nchez}, {Sapone}, {Sartoris}, {Sauvage}, {Schewtschenko}, {Schneider},
  {Scodeggio}, {Secroun}, {Sefusatti}, \& {Seidel}}]{2025arXiv250315302E}
{Euclid Collaboration}, {Aussel}, H., {Tereno}, I., {et~al.}
  2025{\natexlab{b}}, \bibinfo{title}{{Euclid Quick Data Release (Q1) -- Data
  release overview},} arXiv e-prints, arXiv:2503.15302,
  \dodoi{10.48550/arXiv.2503.15302}

\bibitem[{ {Euclid Collaboration} {et~al.}(2025{\natexlab{c}}){Euclid
  Collaboration}, {Jahnke}, {Gillard}, {Schirmer}, {Ealet}, {Maciaszek},
  {Prieto}, {Barbier}, {Bonoli}, {Corcione}, {Dusini}, {Grupp}, {Hormuth},
  {Ligori}, {Martin}, {Morgante}, {Padilla}, {Toledo-Moreo}, {Trifoglio},
  {Valenziano}, {Bender}, {Castander}, {Garilli}, {Lilje}, {Rix}, {Andersen},
  {Auricchio}, {Balestra}, {Barriere}, {Battaglia}, {Berthe}, {Bodendorf},
  {Boenke}, {Bon}, {Bonnefoi}, {Caillat}, {Capobianco}, {Carle}, {Casas},
  {Cho}, {Costille}, {Ducret}, {Ferriol}, {Franceschi}, {Gimenez}, {Holmes},
  {Hornstrup}, {Jhabvala}, {Kohley}, {Kubik}, {Laureijs}, {Le Mignant},
  {Lloro}, {Medinaceli}, {Mellier}, {Polenta}, {Racca}, {Renzi}, {Salvignol},
  {Secroun}, {Seidel}, {Seiffert}, {Sirignano}, {Sirri}, {Strada}, {Smadja},
  {Stanco}, {Wachter}, {Anselmi}, {Borsato}, {Caillat}, {Cogato},
  {Colodro-Conde}, {Crouzet}, {Conforti}, {D'Alessandro}, {Copin},
  {Cuillandre}, {Davies}, {Davini}, {Derosa}, {Diaz}, {Di Domizio}, {Di
  Ferdinando}, {Farinelli}, {Ferrari}, {Fornari}, {Gabarra}, {Garcia},
  {Gutierrez}, {Giacomini}, {Lagier}, {Gianotti}, {Krause}, {Madrid},
  {Laudisio}, {Macias-Perez}, {Naletto}, {Niclas}, {Marpaud}, {Mauri}, {da
  Silva}, {Passalacqua}, {Paterson}, {Patrizii}, {Risso}, {Solheim},
  {Scodeggio}, {Stassi}, {Steinwagner}, {Tenti}, {Testera}, {Travaglini},
  {Tosi}, {Troja}, {Tubio}, {Valieri}, {Vescovi}, {Ventura}, {Aghanim},
  {Altieri}, {Amara}, {Amiaux}, {Andreon}, {Appleton}, {Aussel}, {Baccigalupi},
  {Baldi}, {Bardelli}, {Basset}, {Bonchi}, {Bonino}, {Branchini}, {Brescia},
  {Brinchmann}, {Camera}, {Carbone}, {Cardone}, {Carretero}, {Casas},
  {Castellano}, {Castignani}, {Cavuoti}, {Chabaud}, {Cimatti}, {Congedo},
  {Conselice}, {Conversi}, {Courbin}, {Courtois}, {Crocce}, {Cropper}, {Cuby},
  {Da Silva}, {Degaudenzi}, {De Lucia}, {Di Giorgio}, {Dinis}, {Douspis},
  {Dubath}, {Duncan}, {Dupac}, {Fabricius}, {Farina}, {Farrens}, {Faustini},
  {Fosalba}, {Fotopoulou}, {Fourmanoit}, {Frailis}, {Franzetti}, {Galeotta},
  {George}, {Gillis}, {Giocoli}, {G{\'o}mez-Alvarez}, {Granett}, {Grazian},
  {Guzzo}, {Hailey}, {Haugan}, {Hoar}, {Hoekstra}, {Hook}, {Hudelot},
  {Ili{\'c}}, {Joachimi}, {Keih{\"a}nen}, {Kermiche}, {Kiessling}, {Kilbinger},
  {Kitching}, {K{\"u}mmel}, {Kunz}, {Kurki-Suonio}, {Lahav}, {Liebing},
  {Lindholm}, {Lorenzo Alvarez}, \& {Mainetti}}]{2025A&A...697A...3E}
{Euclid Collaboration}, {Jahnke}, K., {Gillard}, W., {et~al.}
  2025{\natexlab{c}}, \bibinfo{title}{{Euclid: III. The NISP Instrument},}
  \aap, 697, A3, \dodoi{10.1051/0004-6361/202450786}

\bibitem[{P.~F. Felzenszwalb {et~al.}(2010)Felzenszwalb, Girshick, McAllester,
  \& Ramanan}]{5255236}
Felzenszwalb, P.~F., Girshick, R.~B., McAllester, D., \& Ramanan, D. 2010,
  \bibinfo{title}{Object Detection with Discriminatively Trained Part-Based
  Models,} IEEE Transactions on Pattern Analysis and Machine Intelligence, 32,
  1627, \dodoi{10.1109/TPAMI.2009.167}

\bibitem[{Y. Freund \& R.~E. Schapire(1997)Freund \& Schapire}]{FREUND1997119}
Freund, Y., \& Schapire, R.~E. 1997, \bibinfo{title}{A Decision-Theoretic
  Generalization of On-Line Learning and an Application to Boosting,} Journal
  of Computer and System Sciences, 55, 119,
  \dodoi{https://doi.org/10.1006/jcss.1997.1504}

\bibitem[{R. Girshick(2015)Girshick}]{10.1109/ICCV.2015.169}
Girshick, R. 2015, \bibinfo{title}{Fast R-CNN,} in Proceedings of the 2015 IEEE
  International Conference on Computer Vision (ICCV), ICCV '15 (USA: IEEE
  Computer Society), 1440–1448, \dodoi{10.1109/ICCV.2015.169}

\bibitem[{R. Girshick {et~al.}(2014)Girshick, Donahue, Darrell, \&
  Malik}]{6909475}
Girshick, R., Donahue, J., Darrell, T., \& Malik, J. 2014, \bibinfo{title}{Rich
  Feature Hierarchies for Accurate Object Detection and Semantic Segmentation,}
  in 2014 IEEE Conference on Computer Vision and Pattern Recognition, 580--587,
  \dodoi{10.1109/CVPR.2014.81}

\bibitem[{T.~P. {Greene} {et~al.}(2016){Greene}, {Chu}, {Egami}, {Hodapp},
  {Kelly}, {Leisenring}, {Rieke}, {Robberto}, {Schlawin}, \&
  {Stansberry}}]{2016SPIE.9904E..0EG}
{Greene}, T.~P., {Chu}, L., {Egami}, E., {et~al.} 2016,
  \bibinfo{title}{{Slitless spectroscopy with the James Webb Space Telescope
  Near-Infrared Camera (JWST NIRCam)},} in Society of Photo-Optical
  Instrumentation Engineers (SPIE) Conference Series, Vol. 9904, Space
  Telescopes and Instrumentation 2016: Optical, Infrared, and Millimeter Wave,
  ed. H.~A. {MacEwen}, G.~G. {Fazio}, M.~{Lystrup}, N.~{Batalha}, N.~{Siegler},
  \& E.~C. {Tong}, 99040E, \dodoi{10.1117/12.2231347}

\bibitem[{K. {Grishin} {et~al.}(2025){Grishin}, {Mei}, {Ilic}, {Aguena},
  {Boutigny}, {Paturel}, \& {LSST Dark Energy Science
  Collaboration}}]{2025A&A...695A.246G}
{Grishin}, K., {Mei}, S., {Ilic}, S., {et~al.} 2025, \bibinfo{title}{{YOLO-CL
  cluster detection in the Rubin/LSST DC2 simulations},} \aap, 695, A246,
  \dodoi{10.1051/0004-6361/202452119}

\bibitem[{C.~A. {Gullixson}(1992){Gullixson}}]{1992ASPC...23..130G}
{Gullixson}, C.~A. 1992, \bibinfo{title}{{Two Dimensional Imagery},} in
  Astronomical Society of the Pacific Conference Series, Vol.~23, Astronomical
  CCD Observing and Reduction Techniques, ed. S.~B. {Howell}, 130.
\newblock \url{https://ui.adsabs.harvard.edu/abs/1992ASPC...23..130G}

\bibitem[{H.~J. {Hagen} {et~al.}(1995){Hagen}, {Groote}, {Engels}, \&
  {Reimers}}]{1995A&AS..111..195H}
{Hagen}, H.~J., {Groote}, D., {Engels}, D., \& {Reimers}, D. 1995,
  \bibinfo{title}{{The Hamburg Quasar Survey. I. Schmidt observations and plate
  digitization.},} \aaps, 111, 195.
\newblock \url{https://ui.adsabs.harvard.edu/abs/1995A&AS..111..195H}

\bibitem[{K. He {et~al.}(2017)He, Gkioxari, Dollár, \& Girshick}]{8237584}
He, K., Gkioxari, G., Dollár, P., \& Girshick, R. 2017, \bibinfo{title}{Mask
  R-CNN,} in 2017 IEEE International Conference on Computer Vision (ICCV),
  2980--2988, \dodoi{10.1109/ICCV.2017.322}

\bibitem[{Z. He {et~al.}(2021)He, Qiu, Luo, Shi, Kong, \& Jiang}]{9624090}
He, Z., Qiu, B., Luo, A.-L., {et~al.} 2021, \bibinfo{title}{Deep learning
  applications based on SDSS photometric data: detection and classification of
  sources,} Monthly Notices of the Royal Astronomical Society, 508, 2039,
  \dodoi{10.1093/mnras/stab2243}

\bibitem[{L. Jiao {et~al.}(2019)Jiao, Zhang, Liu, Yang, Li, Feng, \&
  Qu}]{8825470}
Jiao, L., Zhang, F., Liu, F., {et~al.} 2019, \bibinfo{title}{A Survey of Deep
  Learning-Based Object Detection,} IEEE Access, 7, 128837,
  \dodoi{10.1109/ACCESS.2019.2939201}

\bibitem[{M. Kümmel {et~al.}(2009)Kümmel, Walsh, Pirzkal, Kuntschner, \&
  Pasquali}]{K_mmel_2009}
Kümmel, M., Walsh, J.~R., Pirzkal, N., Kuntschner, H., \& Pasquali, A. 2009,
  \bibinfo{title}{The Slitless Spectroscopy Data Extraction Software aXe,}
  Publications of the Astronomical Society of the Pacific, 121, 59–72,
  \dodoi{10.1086/596715}

\bibitem[{R. Lienhart \& J. Maydt(2002)Lienhart \& Maydt}]{1038171}
Lienhart, R., \& Maydt, J. 2002, \bibinfo{title}{An extended set of Haar-like
  features for rapid object detection,} in Proceedings. International
  Conference on Image Processing, Vol.~1, I--I,
  \dodoi{10.1109/ICIP.2002.1038171}

\bibitem[{T.-Y. Lin {et~al.}(2017)Lin, Goyal, Girshick, He, \&
  Dollár}]{8237586}
Lin, T.-Y., Goyal, P., Girshick, R., He, K., \& Dollár, P. 2017,
  \bibinfo{title}{Focal Loss for Dense Object Detection,} in 2017 IEEE
  International Conference on Computer Vision (ICCV), 2999--3007,
  \dodoi{10.1109/ICCV.2017.324}

\bibitem[{W. Liu {et~al.}(2016)Liu, Anguelov, Erhan, Szegedy, Reed, Fu, \&
  Berg}]{10.1007/978-3-319-46448-0_2}
Liu, W., Anguelov, D., Erhan, D., {et~al.} 2016, \bibinfo{title}{SSD: Single
  Shot MultiBox Detector,} in Computer Vision -- ECCV 2016, ed. B.~Leibe,
  J.~Matas, N.~Sebe, \& M.~Welling (Springer International Publishing).
\newblock \url{https://link.springer.com/chapter/10.1007/978-3-319-46448-0_2}

\bibitem[{D.~G. Lowe(2004)Lowe}]{2004Distinctive}
Lowe, D.~G. 2004, \bibinfo{title}{Distinctive Image Features from
  Scale-Invariant Keypoints,} International Journal of Computer Vision, 60, 91.
\newblock
  \url{https://link.springer.com/article/10.1023/B:VISI.0000029664.99615.94}

\bibitem[{T. {Morishita} {et~al.}(2019){Morishita}, {Abramson}, {Treu},
  {Brammer}, {Jones}, {Kelly}, {Stiavelli}, {Trenti}, {Vulcani}, \&
  {Wang}}]{2019ApJ...877..141M}
{Morishita}, T., {Abramson}, L.~E., {Treu}, T., {et~al.} 2019,
  \bibinfo{title}{{Massive Dead Galaxies at z {\ensuremath{\sim}} 2 with HST
  Grism Spectroscopy. I. Star Formation Histories and Metallicity Enrichment},}
  \apj, 877, 141, \dodoi{10.3847/1538-4357/ab1d53}

\bibitem[{L. {Mortara} \& A. {Fowler}(1981){Mortara} \&
  {Fowler}}]{1981SPIE..290...28M}
{Mortara}, L., \& {Fowler}, A. 1981, \bibinfo{title}{{Evaluations of
  Charge-Coupled Device / CCD / Performance for Astronomical Use},} in Society
  of Photo-Optical Instrumentation Engineers (SPIE) Conference Series, Vol.
  290, Society of Photo-Optical Instrumentation Engineers (SPIE) Conference
  Series, ed. J.~C. {Geary} \& D.~W. {Latham}, 28, \dodoi{10.1117/12.965833}

\bibitem[{A. {Naufal} {et~al.}(2024){Naufal}, {Koyama}, {D'Eugenio},
  {Dannerbauer}, {Shimakawa}, {P{\'e}rez-Mart{\'\i}nez}, {Kodama}, {Zhang}, \&
  {Daikuhara}}]{2024ApJ...977...58N}
{Naufal}, A., {Koyama}, Y., {D'Eugenio}, C., {et~al.} 2024,
  \bibinfo{title}{{Revealing the Quiescent Galaxy Population in the Spiderweb
  Protocluster at z = 2.16 with Deep HST/WFC3 Slitless Spectroscopy},} \apj,
  977, 58, \dodoi{10.3847/1538-4357/ad8dcf}

\bibitem[{P.~A. {Oesch} {et~al.}(2016){Oesch}, {Brammer}, {van Dokkum},
  {Illingworth}, {Bouwens}, {Labb{\'e}}, {Franx}, {Momcheva}, {Ashby}, {Fazio},
  {Gonzalez}, {Holden}, {Magee}, {Skelton}, {Smit}, {Spitler}, {Trenti}, \&
  {Willner}}]{2016ApJ...819..129O}
{Oesch}, P.~A., {Brammer}, G., {van Dokkum}, P.~G., {et~al.} 2016,
  \bibinfo{title}{{A Remarkably Luminous Galaxy at z=11.1 Measured with Hubble
  Space Telescope Grism Spectroscopy},} \apj, 819, 129,
  \dodoi{10.3847/0004-637X/819/2/129}

\bibitem[{A. Pasquali {et~al.}(2006)Pasquali, Pirzkal, Larsen, Walsh, \&
  Kümmel}]{22b385bc-245f-39bd-a48e-ccaaab6d04eb}
Pasquali, A., Pirzkal, N., Larsen, S., Walsh, J., \& Kümmel, M. 2006,
  \bibinfo{title}{Slitless Grism Spectroscopy with the Hubble Space Telescope
  Advanced Camera for Surveys,} Publications of the Astronomical Society of the
  Pacific, 118, 270.
\newblock \url{http://www.jstor.org/stable/10.1086/498731}

\bibitem[{C. {Piaulet-Ghorayeb} {et~al.}(2024){Piaulet-Ghorayeb}, {Benneke},
  {Radica}, {Raul}, {Coulombe}, {Ahrer}, {Kubyshkina}, {Howard},
  {Krissansen-Totton}, {MacDonald}, {Roy}, {Louca}, {Christie},
  {Fournier-Tondreau}, {Allart}, {Miguel}, {Schlichting}, {Welbanks},
  {Cadieux}, {Dorn}, {Evans-Soma}, {Fortney}, {Pierrehumbert},
  {Lafreni{\`e}re}, {Acu{\~n}a}, {Komacek}, {Innes}, {Beatty}, {Cloutier},
  {Doyon}, {Gagnebin}, {Gapp}, \& {Knutson}}]{2024ApJ...974L..10P}
{Piaulet-Ghorayeb}, C., {Benneke}, B., {Radica}, M., {et~al.} 2024,
  \bibinfo{title}{{JWST/NIRISS Reveals the Water-rich ``Steam World''
  Atmosphere of GJ 9827 d},} \apjl, 974, L10, \dodoi{10.3847/2041-8213/ad6f00}

\bibitem[{J. Redmon {et~al.}(2016)Redmon, Divvala, Girshick, \&
  Farhadi}]{7780460}
Redmon, J., Divvala, S., Girshick, R., \& Farhadi, A. 2016, \bibinfo{title}{You
  Only Look Once: Unified, Real-Time Object Detection,} in 2016 IEEE Conference
  on Computer Vision and Pattern Recognition (CVPR), 779--788,
  \dodoi{10.1109/CVPR.2016.91}

\bibitem[{D. {Reimers} {et~al.}(1996){Reimers}, {Koehler}, \&
  {Wisotzki}}]{1996A&AS..115..235R}
{Reimers}, D., {Koehler}, T., \& {Wisotzki}, L. 1996, \bibinfo{title}{{The
  Hamburg/ESO survey for bright QSOs. II. Follow-up spectroscopy of 160 quasars
  and Seyferts.},} \aaps, 115, 235.
\newblock \url{https://ui.adsabs.harvard.edu/abs/1996A&AS..115..235R}

\bibitem[{D. {Reimers} \& L. {Wisotzki}(1997){Reimers} \&
  {Wisotzki}}]{1997Msngr..88...14R}
{Reimers}, D., \& {Wisotzki}, L. 1997, \bibinfo{title}{{The Hamburg/ESO
  Survey.},} The Messenger, 88, 14.
\newblock \url{https://ui.adsabs.harvard.edu/abs/1997Msngr..88...14R}

\bibitem[{S. Ren {et~al.}(2017)Ren, He, Girshick, \& Sun}]{7485869}
Ren, S., He, K., Girshick, R., \& Sun, J. 2017, \bibinfo{title}{Faster R-CNN:
  Towards Real-Time Object Detection with Region Proposal Networks,} IEEE
  Transactions on Pattern Analysis and Machine Intelligence, 39, 1137,
  \dodoi{10.1109/TPAMI.2016.2577031}

\bibitem[{R.~E. {Ryan} {et~al.}(2018{\natexlab{a}}){Ryan}, {Casertano}, \&
  {Pirzkal}}]{2018wfc..rept...13R}
{Ryan}, R.~E., {Casertano}, S., \& {Pirzkal}, N. 2018{\natexlab{a}}, {Linear
  Reconstruction of Grism Spectroscopy I. Simulation and Extraction Examples},,
  Instrument Science Report WFC3 2018-13, 18 pages
  \url{https://ui.adsabs.harvard.edu/abs/2018wfc..rept...13R}

\bibitem[{R.~E. {Ryan} {et~al.}(2018{\natexlab{b}}){Ryan}, {Casertano}, \&
  {Pirzkal}}]{2018PASP..130c4501R}
{Ryan}, Jr., R.~E., {Casertano}, S., \& {Pirzkal}, N. 2018{\natexlab{b}},
  \bibinfo{title}{{Linear: A Novel Algorithm for Reconstructing Slitless
  Spectroscopy from HST/WFC3},} \pasp, 130, 034501,
  \dodoi{10.1088/1538-3873/aaa53e}

\bibitem[{B. {Sekachev} {et~al.}(2019){Sekachev}, {Manovich}, \&
  {Zhavoronkov}}]{2019zndo...3497106S}
{Sekachev}, B., {Manovich}, N., \& {Zhavoronkov}, A. 2019, {Computer Vision
  Annotation Tool}, 0.5.1 Zenodo, \dodoi{10.5281/zenodo.3497106}

\bibitem[{F. {Sun} {et~al.}(2023){Sun}, {Egami}, {Pirzkal}, {Rieke}, {Baum},
  {Boyer}, {Boyett}, {Bunker}, {Cameron}, {Curti}, {Eisenstein}, {Gennaro},
  {Greene}, {Jaffe}, {Kelly}, {Koekemoer}, {Kumari}, {Maiolino}, {Maseda},
  {Perna}, {Rest}, {Robertson}, {Schlawin}, {Smit}, {Stansberry}, {Sunnquist},
  {Tacchella}, {Williams}, \& {Willmer}}]{2023ApJ...953...53S}
{Sun}, F., {Egami}, E., {Pirzkal}, N., {et~al.} 2023, \bibinfo{title}{{First
  Sample of H{\ensuremath{\alpha}}+[O III]{\ensuremath{\lambda}}5007 Line
  Emitters at z > 6 Through JWST/NIRCam Slitless Spectroscopy: Physical
  Properties and Line-luminosity Functions},} \apj, 953, 53,
  \dodoi{10.3847/1538-4357/acd53c}

\bibitem[{L. {Varela} {et~al.}(2019){Varela}, {Boucheron}, {Malone}, \&
  {Spurlock}}]{2019amos.confE..89V}
{Varela}, L., {Boucheron}, L., {Malone}, N., \& {Spurlock}, N. 2019,
  \bibinfo{title}{{Streak detection in wide field of view images using
  Convolutional Neural Networks (CNNs)},} in Advanced Maui Optical and Space
  Surveillance Technologies Conference, ed. S.~{Ryan}, 89.
\newblock \url{https://ui.adsabs.harvard.edu/abs/2019amos.confE..89V}

\bibitem[{F. {Wang} {et~al.}(2023){Wang}, {Hu}, \& {Gu}}]{2023SPIE12511E..1CW}
{Wang}, F., {Hu}, T., \& {Gu}, M. 2023, \bibinfo{title}{{Detection and
  classification of galaxy morphology based on YOLOv5},} in Society of
  Photo-Optical Instrumentation Engineers (SPIE) Conference Series, Vol. 12511,
  Third International Conference on Computer Vision and Data Mining (ICCVDM
  2022), ed. T.~{Zhang} \& T.~{Yang}, 125111C, \dodoi{10.1117/12.2659975}

\bibitem[{C. Wei {et~al.}(2023)Wei, Zhang, Liu, Fang, Meng, Ban, Luo, Tian, Li,
  Li, Li, Li, Qi, \& Li}]{csst_cycle6_sim}
Wei, C., Zhang, X., Liu, D., {et~al.} 2023, CSST Main Survey Simulation Data:
  Cycle 6 Data Product Description,, China Science and Technology Cloud
  (CSTCloud)
  \url{https://pan.cstcloud.cn/web/view_iframe.html?fid=585816359305371\&shareId=zRF7mu3ET5c}

\bibitem[{C. Wei {et~al.}(2024)Wei, Zhang, Liu, Fang, Meng, Ban, Luo, Tian, Li,
  Li, Li, Li, Qi, \& Li}]{csst_cycle9_sim}
Wei, C., Zhang, X., Liu, D., {et~al.} 2024, CSST Main Survey Simulation Data:
  Cycle 9 Data Product Description,, Kingsoft Docs
  \url{https://www.kdocs.cn/l/cmIkLt4cfzxw?from=docs}

\bibitem[{C.-L. Wei {et~al.}(2025)Wei, Li, Fang, Zhang, Luo, Tian, Liu, Meng,
  Ban, Li, Luo, Xian, Wang, Peng, Li, Li, Shao, Zhang, Tang, Chen, Qi, Cao,
  Shan, Nie, Wang, He, Luo, \& Liu}]{Wei2025overview}
Wei, C.-L., Li, G.-L., Fang, Y.-D., {et~al.} 2025, \bibinfo{title}{Mock
  Observations for the {CSST} Mission: Main Surveys - An Overview of Framework
  and Simulation Suite,} Research in Astronomy and Astrophysics, in press

\bibitem[{R. Wen {et~al.}(2024)Wen, Zheng, Han, Yang, Wang, Zou, Liu, Zhang,
  Zu, Shi, Gu, \& Wang}]{10.1093/mnras/stae157}
Wen, R., Zheng, X.~Z., Han, Y., {et~al.} 2024, \bibinfo{title}{CSST large-scale
  structure analysis pipeline: II. The CSST Emulator for Slitless
  Spectroscopy,} Monthly Notices of the Royal Astronomical Society, 528, 2770,
  \dodoi{10.1093/mnras/stae157}

\bibitem[{C.~J. Willott {et~al.}(2022)Willott, Doyon, Albert, Brammer, Dixon,
  Muzic, Ravindranath, Scholz, Abraham, Artigau, Bradač, Goudfrooij,
  Hutchings, Iyer, Jayawardhana, LaMassa, Martis, Meyer, Morishita, Mowla,
  Muzzin, Noirot, Pacifici, Rowlands, Sarrouh, Sawicki, Taylor, Volk, \&
  Zabl}]{Willott_2022}
Willott, C.~J., Doyon, R., Albert, L., {et~al.} 2022, \bibinfo{title}{The
  Near-infrared Imager and Slitless Spectrograph for the James Webb Space
  Telescope. II. Wide Field Slitless Spectroscopy,} Publications of the
  Astronomical Society of the Pacific, 134, 025002,
  \dodoi{10.1088/1538-3873/ac5158}

\bibitem[{L. {Wisotzki}(1994){Wisotzki}}]{1994IAUS..161..723W}
{Wisotzki}, L. 1994, \bibinfo{title}{{The Hamburg/ESO Survey for Bright QSOs -
  Slitless Spectroscopy at High Resolution},} in IAU Symposium, Vol. 161,
  Astronomy from Wide-Field Imaging, ed. H.~T. {MacGillivray}, 723.
\newblock \url{https://ui.adsabs.harvard.edu/abs/1994IAUS..161..723W}

\bibitem[{L. {Wisotzki} {et~al.}(1996){Wisotzki}, {Koehler}, {Groote}, \&
  {Reimers}}]{1996A&AS..115..227W}
{Wisotzki}, L., {Koehler}, T., {Groote}, D., \& {Reimers}, D. 1996,
  \bibinfo{title}{{The Hamburg/ESO survey for bright QSOs. I. Survey design and
  candidate selection procedure.},} \aaps, 115, 227.
\newblock \url{https://ui.adsabs.harvard.edu/abs/1996A&AS..115..227W}

\bibitem[{G. Worseck {et~al.}(2008)Worseck, Wisotzki, \& Selman}]{Worseck_2008}
Worseck, G., Wisotzki, L., \& Selman, F. 2008, \bibinfo{title}{A slitless
  spectroscopic survey for quasars near quasars,} Astronomy \& Astrophysics,
  487, 539–554, \dodoi{10.1051/0004-6361:200810157}

\bibitem[{Y. Xing {et~al.}(2023)Xing, Yi, Liang, Su, Du, He, Liu, Kong, Bu, \&
  Wu}]{Xing_2023}
Xing, Y., Yi, Z., Liang, Z., {et~al.} 2023, \bibinfo{title}{Edge-on
  Low-surface-brightness Galaxy Candidates Detected from SDSS Images Using
  YOLO,} The Astrophysical Journal Supplement Series, 269, 59,
  \dodoi{10.3847/1538-4365/ad0551}

\bibitem[{H. Zhan(2021)Zhan}]{Zhan2021}
Zhan, H. 2021, \bibinfo{title}{The wide-field multiband imaging and slitless
  spectroscopy survey to be carried out by the Survey Space Telescope of China
  Manned Space Program,} Chinese Science Bulletin, 66, 1290,
  \dodoi{10.1360/TB-2021-0016}

\bibitem[{X. Zhang {et~al.}(2025)Zhang, Fang, Wei, Li, Liu, Ji, Tian, Li, Meng,
  Chen, Wang, Wang, Liu, Hu, Li, Wei, \& Tang}]{Zhang2025slitless}
Zhang, X., Fang, Y.-D., Wei, C.-L., {et~al.} 2025, \bibinfo{title}{Mock
  Observations for the {CSST} Mission: Main Surveys - the Slitless Spectroscopy
  Simulation,} Research in Astronomy and Astrophysics, in press

\bibitem[{Q. Zhao {et~al.}(2019)Zhao, Sheng, Wang, Tang, Chen, Cai, \&
  Ling}]{10.1609/aaai.v33i01.33019259}
Zhao, Q., Sheng, T., Wang, Y., {et~al.} 2019, \bibinfo{title}{M2Det: a
  single-shot object detector based on Multi-Level Feature Pyramid Network,} in
  Proceedings of the Thirty-Third AAAI Conference on Artificial Intelligence
  and Thirty-First Innovative Applications of Artificial Intelligence
  Conference and Ninth AAAI Symposium on Educational Advances in Artificial
  Intelligence, AAAI'19/IAAI'19/EAAI'19 (AAAI Press),
  \dodoi{10.1609/aaai.v33i01.33019259}

\bibitem[{Z.-Q. Zhao {et~al.}(2019)Zhao, Zheng, Xu, \& Wu}]{8627998}
Zhao, Z.-Q., Zheng, P., Xu, S.-T., \& Wu, X. 2019, \bibinfo{title}{Object
  Detection With Deep Learning: A Review,} IEEE Transactions on Neural Networks
  and Learning Systems, 30, 3212, \dodoi{10.1109/TNNLS.2018.2876865}

\bibitem[{Y. Zhou {et~al.}(2025)Zhou, Liu, Tian, Zhang, \& Li}]{Zhou_2025}
Zhou, Y., Liu, C., Tian, H., Zhang, X., \& Li, N. 2025, \bibinfo{title}{CSST
  Slitless Spectra: Target Detection and Classification with Yolo,} The
  Astronomical Journal, 170, 256, \dodoi{10.3847/1538-3881/adee1c}

\end{thebibliography}
\bibliographystyle{aasjournalv7}



\end{document}